\documentclass[runningheads]{svmult}

\usepackage{makeidx}   
\usepackage{graphicx}  
\usepackage{subeqnar}  
\usepackage{multicol}  
\usepackage{physprbb}  
\makeindex             

\def\mb#1{\setbox0=\hbox{$#1$}\kern-.025em\copy0\kern-\wd0
\kern-0.05em\copy0\kern-\wd0\kern-.025em\raise.0233em\box0}

\begin{document}
\title*{Statistical mechanics of two-dimensional vortices and stellar systems}
\toctitle{Statistical mechanics of two-dimensional vortices and stellar systems}
%
%
\titlerunning{2D vortices and stellar systems}
%
\author{Pierre-Henri Chavanis}
\authorrunning{Pierre-Henri Chavanis\inst{1}}
%
%
\institute{Laboratoire de Physique Quantique,\\
Universit\'e Paul Sabatier,\\
 118, route de Narbonne\\ 31062 Toulouse, France }

\maketitle              

\begin{abstract}

The formation of large-scale vortices is an intriguing phenomenon in
two-dimensional turbulence. Such organization is observed in
large-scale oceanic or atmospheric flows, and can be reproduced in
laboratory experiments and numerical simulations. A general
explanation of this organization was first proposed by Onsager (1949)
by considering the statistical mechanics for a set of point vortices
in two-dimensional hydrodynamics. Similarly, the structure and the
organization of stellar systems (globular clusters, elliptical
galaxies,...) in astrophysics can be understood by developing a
statistical mechanics for a system of particles in gravitational
interaction as initiated by Chandrasekhar (1942). These statistical
mechanics turn out to be relatively similar and present the same
difficulties due to the unshielded long-range nature of the
interaction. This analogy concerns not only the equilibrium states,
i.e. the formation of large-scale structures, but also the relaxation
towards equilibrium and the statistics of fluctuations. We will discuss
these analogies in detail and also point out the specificities of each
system.

\end{abstract}

\section{Introduction}
\label{sec_introduction}

\begin{figure}
\centering
\includegraphics[angle=0,width=0.8\textwidth]{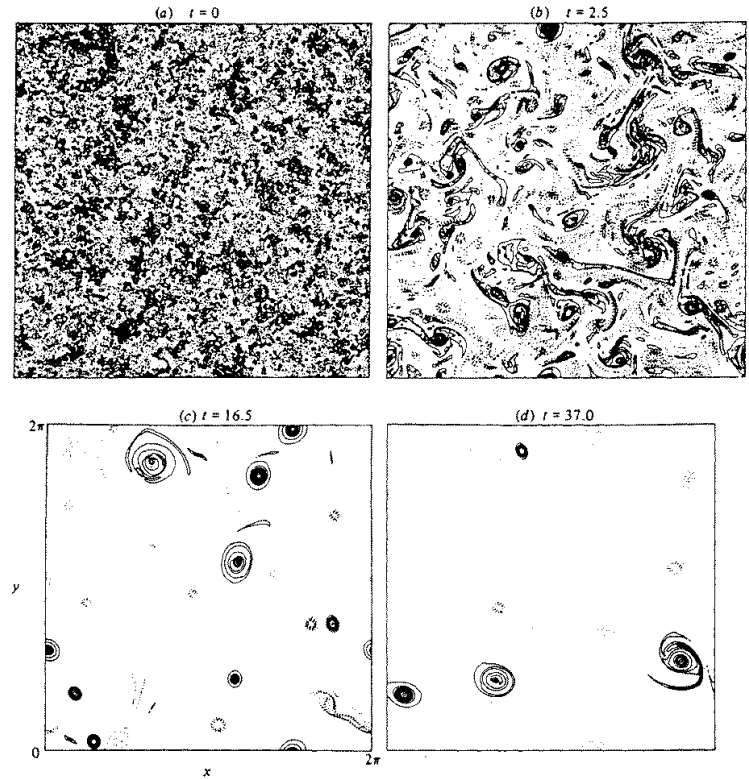}
\caption[]{Self-organization of two-dimensional turbulent flows into 
large-scale vortices \cite{williams}. These vortices are long-lived and 
dominate the dynamics.}
\label{McW}
\end{figure}

\begin{figure}
\centering
\includegraphics[angle=0,width=0.9\textwidth]{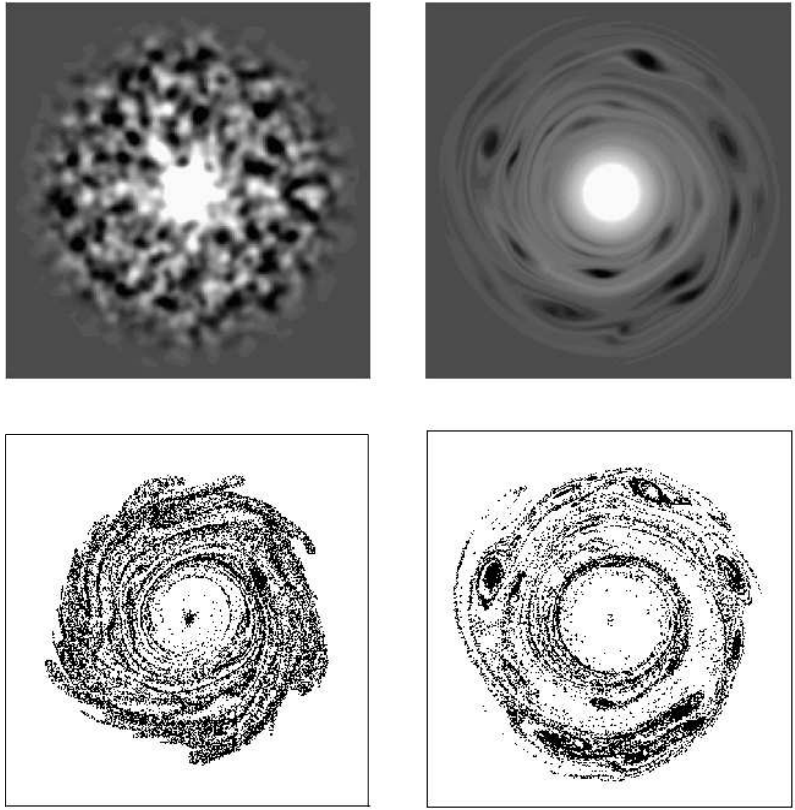}
\caption[]{A scenario of planet formation inside large-scale vortices 
presumably present in the Keplerian gaseous disk surrounding a star at
its birth. Starting from a random vorticity field, a series of
anticyclonic vortices appears spontaneously (upper panel). Due to the
Coriolis force and to the friction with the gas, these vortices can
efficiently trap dust particles passing nearby (lower pannel). The
local increase of dust concentration inside the vortices can initiate
the formation of planetesimals and planets by gravitational
instability. This numerical simulation is taken from \cite{bracco}.}
\label{planetes}
\end{figure}

Two-dimensional flows with high Reynolds numbers have the striking
property of organizing spontaneously into \index{Coherent structures}
coherent structures (the vortices) which dominate the dynamics
\cite{williams} (see Fig. \ref{McW}). The robustness of 
Jupiter's Great Red Spot, a huge vortex persisting for more than three
centuries in a turbulent shear between two zonal jets, is probably
related to this general property. Some other coherent structures like
dipoles (pairs of cyclone/anticyclone) and sometimes tripoles have
been found in atmospheric or oceanic systems and can persist during
several days or weeks responsible for atmospheric blocking. Some
astrophysicists invoke the existence of organized vortices in the
gaseous component of disk galaxies in relation with the emission of
spiral density waves
\cite{nezlin}. It has also been proposed that planetary formation
might have begun inside persistent gaseous vortices born out of the
protoplanetary nebula
\cite{barge,tanga,bracco,godon,cplanetes} (see Fig. \ref{planetes}). 
As a result, hydrodynamical vortices occur in a wide variety of
geophysical or astrophysical situations and their robustness demands a
general understanding.

\begin{figure}
\centering
\includegraphics[angle=0,width=0.55\textwidth]{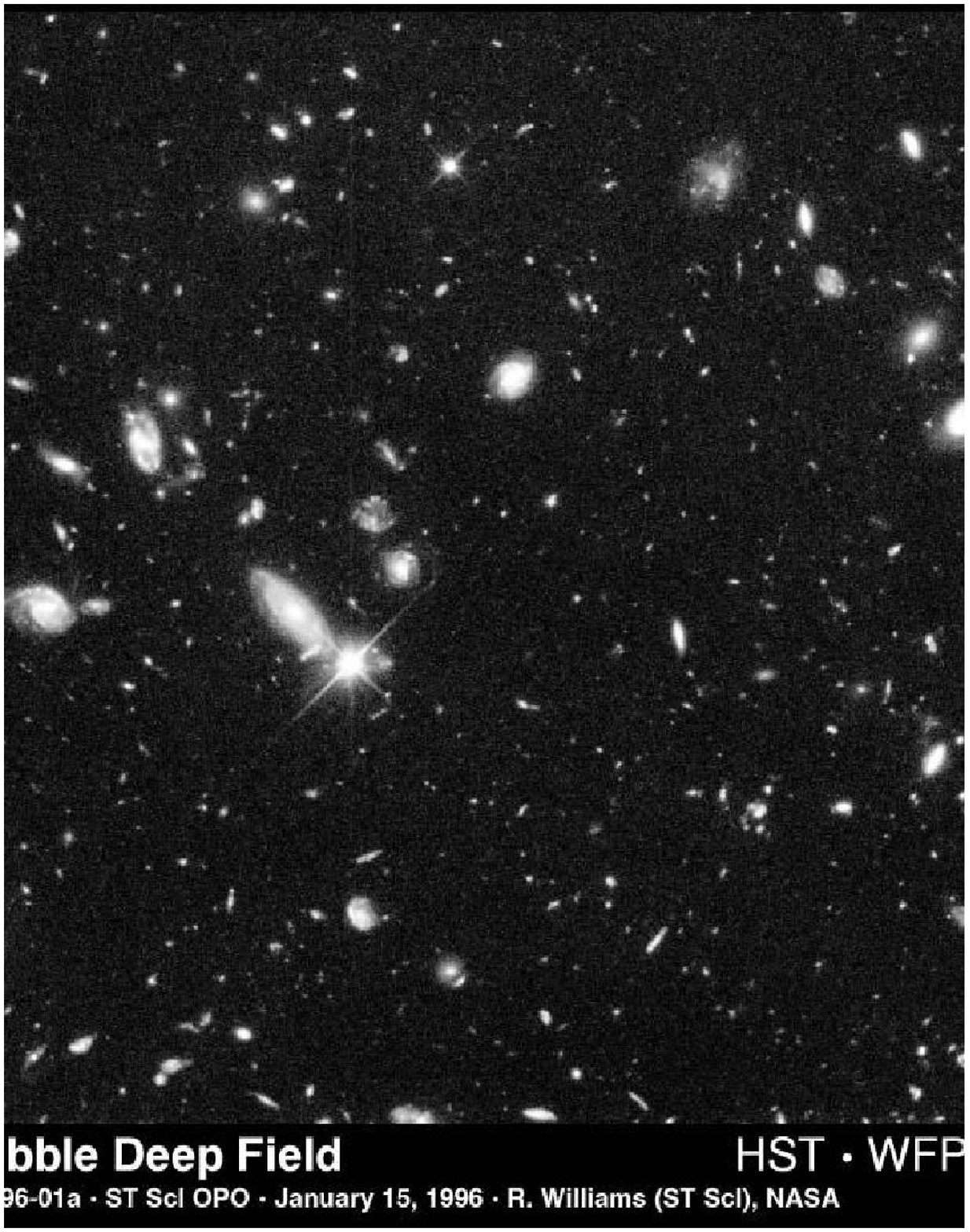}
\caption[]{Large-scale structures in the universe as observed with 
the Hubble space telescope. The analogy with Fig. \ref{McW} is
striking and will be discussed in detail in this paper.}
\label{Hubble}
\end{figure}

Similarly, it is striking to observe that self-gravitating systems
\index{Astrophysics} follow a kind of organization despite 
the diversity of their initial conditions and their environement
\cite{bt} (see Fig. \ref{Hubble}). This organization is illustrated by
morphological classification schemes such as the Hubble sequence for
galaxies and by simple rules which govern the structure of individual
self-gravitating systems. For example, elliptical galaxies display a
quasi-universal luminosity profile described by de Vaucouleur's
$R^{1/4}$ law and most of globular clusters are well-fitted by the
Michie-King model. On the other hand, the flat rotation curves of
spiral galaxies can be explained by the presence of a dark matter halo
with a density profile decreasing as $r^{-2}$ at large distances. The
fractal nature of the interstellar medium and the large scale
structures of the universe also display some form of organization.

The question that naturally emerges is what determines the particular
configuration to which a self-gravitating system or a large-scale
vortex settles. It is possible that their actual configuration
crucially depends on the conditions that prevail at their birth and on
the details of their evolution. However, in view of their apparent
regularity, it is tempting to investigate whether their organization
can be favoured by some fundamental physical principles like those of
thermodynamics and statistical physics. We ask therefore if the
actual states of self-gravitating systems in the universe and coherent
vortices in two-dimensional turbulent flows are not simply more
probable than any other possible configuration, i.e. if they cannot be
considered as {\it maximum entropy states}. This statistical mechanics
approach has been initiated by Onsager \cite{onsager} for a system of
point vortices and by Chandrasekhar \cite{chandra42} in the case of
self-gravitating systems.

It turns out that the statistical mechanics of two-dimensional
vortices and self-gravitating systems present a deep analogy despite
the very different physical nature of these systems. This analogy was
pointed out by Chavanis in \cite{cthese,cfloride,japon} and further
developed in
\cite{csr,drift,kin1,kin2,csire1,csire2}. In the following, we
will essentially discuss the statistical mechanics of 2D vortices and
refer to the review of Padmanabhan \cite{pad} (and his contribution in
this book) for more details about the statistical mechanics of
self-gravitating systems. We will see that the analogy between
two-dimensional vortices and (three-dimensional) self-gravitating
systems concerns not only the prediction of the equilibrium state,
i.e. the formation of large-scale structures, but also the statistics
of fluctuations and the relaxation towards equilibrium.

This paper is organized as follows. In Sec. \ref{sec_statmech}, we
discuss the statistical mechanics of point vortices introduced by
Onsager \cite{onsager} and further developed by Joyce \& Montgomery
\cite{jm} and Pointin \& Lundgren \cite{pl} among others (see a
complete list of references in the book of Newton \cite{newton}). We
discuss the existence of a thermodynamic limit in
Sec. \ref{sec_field} and make the connexion with field
theory. Statistical equilibrium states of axisymmetric flows are
obtained analytically in Sec. \ref{sec_disk}-\ref{sec_unbounded}. The
relation with equilibrium states of self-gravitating systems is shown
in Sec. \ref{sec_gravN}. In Sec. \ref{sec_fluc}, we discuss the
statistics of velocity fluctuations produced by a random distribution
of point vortices and use this stochastic approach to obtain an
estimate of the diffusion coefficient of point vortices. Application
to 2D decaying turbulence is considered in
Sec. \ref{sec_anomalous}. In Sec. \ref{sec_bath1}, we describe the
relaxation of a point vortex in a thermal bath and analyze this
relaxation in terms of a Fokker-Planck equation involving a diffusion
and a drift. In Sec. \ref{sec_kin}, we develop a more general kinetic
theory of point vortices. A new kinetic equation is obtained which
satisfies all conservation laws of the point vortex system and
increases the Boltzmann entropy (H-theorem). We mention the connexion
with the kinetic theory of stars developed by Chandrasekhar
\cite{chandra42}. In Sec. \ref{sec_violrelax}, we discuss the violent
relaxation of 2D vortices and stellar systems. We mention the analogy
between the Vlasov and the Euler equations and between the statistical
approach developed by Lynden-Bell \cite{lb} for collisionless stellar
systems and by Kuz'min \cite{kuzmin}, Miller
\cite{miller} and Robert
\& Sommeria \cite{rs1} for continuous vorticity fields. The concepts
of ``chaotic mixing'' and ``incomplete relaxation'' are discussed in
the light of a relaxation theory in Sec. \ref{sec_MEPP}. Application
of statistical mechanics to geophysical flows and Jupiter's Great Red
Spot are evocated in Sec. \ref{sec_del}.

\section{Statistical mechanics of point vortices in two-dimensional 
hydrodynamics}
\label{sec_statmech}

\subsection{Two-dimensional perfect flows}
\label{sec_perfect} 

The equations governing the dynamics of an invisicid flow are the equation of continuity and the Euler equation:
\begin{equation}
{\partial \rho\over\partial t}+\nabla (\rho{\bf u})=0,
\label{pf1}
\end{equation} 
\begin{equation}
{\partial {\bf u}\over\partial t}+({\bf u}\nabla){\bf u}=-{1\over\rho}\nabla p.
\label{pf2}
\end{equation} 
For an incompressible flow, the equation of continuity 
reduces to the condition 
\begin{equation}
\nabla {\bf u}=0.
\label{pf3}
\end{equation} 
If, in addition, the flow is two-dimensional, this last equation can be written
$\partial_{x}u+\partial_{y}v=0$, where $(u,v)$ are the components of the velocity. According to the Schwarz theorem, there exists a streamfunction $\psi$ such that $u=\partial_{y}\psi, v=-\partial_{x}\psi$, or, equivalently
\begin{equation}
{\bf u}=-{\bf z}\times\nabla\psi,
\label{pf4}
\end{equation} 
where ${\bf z}$ is a unit vector normal to the flow. The vorticity
\begin{equation}
{\mb\omega}=\nabla\times{\bf u}=\omega {\bf z}, \qquad {\rm with}\qquad \omega=\partial_{x}v-\partial_{y}u,
\label{pf5}
\end{equation} 
is directed along the vertical axis. According to the Stokes formula, the circulation of the velocity along a closed curve $(C)$ delimiting a domain area $(S)$ is
\begin{equation}
\Gamma=\oint_{(C)} {\bf u}\ d{\bf l}=\int_{(S)} \omega d^{2}{\bf r}.
\label{pf6}
\end{equation}

Taking  the curl of  Eq. (\ref{pf4}),  we find  that the  vorticity is
related to the stream function by a Poisson equation
\begin{equation}
\Delta\psi=-\omega,
\label{pf7}
\end{equation} 
where $\Delta=\partial^{2}_{xx}+\partial^{2}_{yy}$ is the Laplacian
operator. In an unbounded domain, this equation can be written in
integral form as
\begin{equation}
\psi({\bf r},t)=-{1\over 2\pi}\int \omega({\bf r}',t)
\ln |{\bf r}-{\bf r}'|d^{2}{\bf r}',
\label{pf8}
\end{equation} 
and the velocity field can be expressed in terms of the vorticity as
\begin{equation}
{\bf u}({\bf r},t)={1\over 2\pi}{\bf z}\times\int \omega({\bf
r}',t){{\bf r}-{\bf r}'\over |{\bf r}-{\bf r}'|^{2}}d^{2}{\bf r}'.
\label{pf9}
\end{equation} 
In a bounded domain, Eq. (\ref{pf8}) must be modified so as to take
into account vortex images. The impermeability condition implies that
$\psi$ is constant on the boundary and we shall take $\psi=0$ by
convention.  Taking the curl of Eq. (\ref{pf2}), the pressure term
disapears and the Euler equation becomes
\begin{equation}
{\partial\omega\over\partial t}+{\bf u} \nabla\omega=0.
\label{pf10}
\end{equation} 
This corresponds to the transport of the vorticity $\omega$ by the
velocity field ${\bf u}$.  It is easy to show that the flow conserves
the kinetic energy
\begin{equation}
E=\int {{\bf u}^{2}\over 2}  d^{2}{\bf r}. 
\label{pf10b}
\end{equation}
Using Eqs. (\ref{pf4})(\ref{pf7}), one has successively
\begin{equation}
E={1\over 2}\int (\nabla\psi)^{2} d^{2}{\bf r}={1\over 2}\int \psi
(-\Delta\psi) d^{2}{\bf r}={1\over 2}\int \omega\psi d^{2}{\bf r},
\label{pf11}
\end{equation}
where the second equality is obtained by a part integration with the
condition $\psi=0$ on the boundary.  Therefore, $E$ can be interpreted
either as the kinetic energy of the flow (see Eq. (\ref{pf10b})) or
as a potential energy of interaction between vortices (see Eq.
(\ref{pf11})).

\subsection{The point vortex gas}
\label{sec_gas} 

We shall consider the situation in which the velocity is created by a
collection of $N$ point vortices. In that case, the vorticity field
can be expressed as a sum of $\delta$-functions in the form
\begin{equation}
\omega ({\bf r},t)=\sum_{i=1}^{N}\gamma_{i}\delta ({\bf r}-{\bf r}_{i}(t)),
\label{pv1}
\end{equation}  
where ${\bf r}_{i}(t)$ denotes the position of point vortex $i$ at
time $t$ and $\gamma_{i}$ is its circulation. According to
Eqs. (\ref{pf9})(\ref{pv1}), the velocity of a point vortex is equal
to the sum of the velocities ${\bf V}(j\rightarrow i)$ produced by the
$N-1$ other vortices, i.e.
\begin{equation}
{\bf V}_{i}=\sum_{j\neq i}{\bf V}(j\rightarrow i)\qquad {\rm
with}\qquad {\bf V}(j\rightarrow i)=-{\gamma_{j}\over 2\pi}{\bf
z}\times {{\bf r}_{j}-{\bf r}_{i}\over |{\bf r}_{j}-{\bf r}_{i}|^{2}}.
\label{pv2}
\end{equation} 

As emphasized by Kirchhoff \cite{kirchhoff}, the above dynamics can be cast in a Hamiltonian form 
\begin{equation}
\gamma_{i} {d x_{i}\over dt}={\partial H\over\partial y_{i}},\qquad \gamma_{i} {d
y_{i}\over dt}=-{\partial H\over\partial x_{i}},
\label{pv3}
\end{equation} 
\begin{equation}
H=-{1\over 2\pi}\sum_{i<j}\gamma_{i}\gamma_{j}\ln |{\bf r}_{i}-{\bf r}_{j}|,
\label{pv4}
\end{equation} 
where the coordinates $(x,y)$ of the point vortices are canonically
conjugate.  These equations of motion still apply when the fluid is
restrained by boundaries, in which case the Hamiltonian (\ref{pv4}) is
modified so as to allow for vortex images, and may be constructed in
terms of Green's functions depending on the geometry of the
domain. Since $H$ is not explicitly time dependant, it is a constant
of the motion and it represents the ``potential'' energy of the point
vortices. The other conserved quantities are the angular momentum
$\sum_{i}\gamma_{i}r_{i}^{2}$ and the impulse $\sum_{i}\gamma_{i}{\bf
r}_{i}$.  Note that the Hamiltonian (\ref{pv4}) does not involve a
``kinetic'' energy of the point vortices in the usual sense (i.e., a
quadratic term $\sum_{i}{p_{i}^{2}\over 2m}$). This is related to the
particular circumstance that a point vortex is not a material
particle. Indeed, an isolated vortex remains at rest contrary to a
material particle which has a rectilinear motion due to its
inertia. Point vortices form therefore a very peculiar Hamiltonian
system. Note also that the Hamiltonian of point vortices can be either
positive or negative (in the case of vortices of different signs)
whereas the kinetic energy of the flow is necessarily positive. This
is clearly a drawback of the point vortex model.

\subsection{The microcanonical approach of Onsager (1949)}
\label{sec_onsager}

The statistical mechanics of point vortices was first considered by
Onsager \cite{onsager} who showed the existence of negative
temperature states \index{Negative temperatures} at which point
vortices of the same sign cluster into ``supervortices''. He could
therefore explain the formation of large, isolated vortices in nature.
This was a remarkable anticipation since observations were very scarce
at that time.

Let us consider a liquid enclosed by a boundary,
so that the vortices are confined to an area $A$. Since the
coordinates $(x,y)$ of the point vortices are canonically conjugate,
the phase space coincides with the configuration space and is {\it
finite}:
\begin{equation}
\int dx_{1}dy_{1}...dx_{N}dy_{N}=\biggl (\int dx dy\biggr )^{N}=A^{N}.
\label{on1}
\end{equation} 
This striking property contrasts with most classical Hamiltonian
systems considered in statistical mechanics which have unbounded phase
spaces due to the presence of a kinetic term in the Hamiltonian. 

As is usual in the microcanonical description of a system of $N$ particles, we
introduce the density of states 
\begin{equation}
g(E)=\int dx_{1}dy_{1}...dx_{N}dy_{N} \delta \biggl
(E-H(x_{1},y_{1},...,x_{N},y_{N})\biggr ),
\label{on2}
\end{equation} 
which gives the phase space volume per unit interaction energy $E$. The equilibrium $N$-body distribution of the system, satisfying the normalization condition $\int\mu({\bf r}_{1},...,{\bf r}_{N})d^{2}{\bf r}_{1}...d^{2}{\bf r}_{N}=1$, is given by 
\begin{equation}
\mu({\bf r}_{1},...,{\bf r}_{N})={1\over g(E)} \delta 
(E-H({\bf r}_{1},...,{\bf r}_{N}) ).
\label{on3}
\end{equation} 
This formula simply means that, in the microcanonical ensemble, all
accessible microstates (having the required energy $E$) are
equiprobable at statistical equilibrium.

The phase space volume which corresponds to energies $H({\bf
r}_{1},...,{\bf r}_{N})$ less than a given value $E$ can be written
\begin{equation}
\Phi(E)=\int_{E_{min}}^{E} g(E)dE.
\label{on4}
\end{equation}  
It increases monotonically from zero to $A^{N}$ when $E$ goes from $E_{min}$ to
$+\infty$. Therefore, $g(E)=d\Phi(E)/dE$ will have a maximum value at some
$E=E_{m}$, say, before decreasing to zero when $E\rightarrow +\infty$. 

In the microcanonical ensemble, the entropy and the temperature are defined by
\begin{equation}
S=\ln g(E), \qquad \beta={1\over T}={dS\over dE}.
\label{on5}
\end{equation} 
For $E>E_{m}$, $S(E)$ is a decreasing function of energy and
consequently the temperature is {\it negative}. Now, high energy
states $E\gg E_{m}$ are clearly those in which the vortices of the
same sign are crowded as close together as possible. For energies only
slighlty greater than $E_{m}$, the concentration will not be so
dramatic but there will be a tendency for the vortices to group
themselves together on a macroscopic scale and form ``clusters'' or
``supervortices''. By contrast, for $E<E_{m}$, the temperature is
positive and the vortices have the tendency to accumulate on the
boundary of the domain in order to decrease their energy. For a system
with positive and negative vortices, the negative temperature states,
achieved for relatively high energies, consist of two large
counter-rotating vortices physically well separated in the box. On the
contrary when $E\rightarrow -\infty$, the temperature is positive and
vortices of opposite circulation tend to pair off.

\subsection{The equation of state}
\label{sec_state} 

For a two-dimensional gas of particles interacting via a Coulombian 
potential in $\ln r$, the equation of state can be derived
exactly. Let us assume that the system is enclosed in a domain of
surface $V=R^{2}$. The density of states can be written 
\begin{equation}
g(E,V)=\int_{0}^{R}...\int_{0}^{R}\prod_{i=1}^{N}d^{2}{\bf r}_{i}\delta \biggl (E+{1\over 2\pi}\sum_{i<j}\gamma_{i}\gamma_{j}\ln |{\bf r}_{i}-{\bf r}_{j}|\biggr ).
\label{es1}
\end{equation}
Making the change of variable ${\bf x}={\bf r}/R$, we find that
$g(E,V)=V^{N}g(E',1)$ with $E'=E+{1\over 4\pi}\ln
V\sum_{i<j}\gamma_{i}\gamma_{j}$. Therefore, the entropy satisfies
$S(E,V)=N\ln V+S(E',1)$ and the pressure $P=T(\partial S/\partial
V)_{E}$ is exactly given by
\begin{equation}
P={N\over \beta V}\biggl (1+{\beta \over 4\pi N}\sum_{i<j}\gamma_{i}\gamma_{j} \biggr ).
\label{es2}
\end{equation}
If the vortices have the same circulation $\gamma$, we obtain
\begin{equation}
P={N\over \beta V}\biggl (1+{\gamma^{2} \over 8\pi}(N-1)\beta\biggr ).
\label{es3}
\end{equation}
In particular the pressure vanishes for 
\begin{equation}
\beta_{c}=-{8\pi\over (N-1)\gamma^{2}}.
\label{es4}
\end{equation}
We shall see in Sec. \ref{sec_disk} that this {\it negative} critical
inverse temperature is the minimum inverse temperature that the system
can achieve.

If, on the other hand, we consider a neutral system consisting of
$N/2$ vortices of circulation $\gamma$ and $N/2$ vortices of
circulation $-\gamma$, we find
\begin{equation}
P={N\over \beta V}\biggl (1-{\beta \gamma^{2}\over 8\pi}\biggr ).
\label{es5}
\end{equation}
This result is well-known is plasma physics \cite{salzberg}. The critical
temperature at which the pressure vanishes is now positive
\begin{equation}
\beta_{c}={8\pi\over\gamma^{2}},
\label{es6}
\end{equation}
and independant on the number of point vortices in the system. For
$\beta>\beta_{c}$ the pressure is negative so this range of
temperatures is forbidden. For simplicity, we have not taken into
account the contribution of images in the previous calculations; this
can slightly change the results but this should not alter the
existence of the critical inverse temperatures reported above.

\subsection{The mean-field approximation}
\label{sec_meanfield}

It is easy to show that the exact distribution of point vortices (\ref{pv1})
expressed in terms of $\delta$-functions is solution of the Euler
equation (\ref{pf10}). This is proved as follows. Taking the derivative of
Eq. (\ref{pv1}) with respect to time, we obtain
\begin{eqnarray}
{\partial\omega\over\partial t}=-\sum_{i=1}^{N}\gamma \nabla\delta({\bf
r}-{\bf r}_{i}(t)){\bf V}_{i}.
\label{mf1}
\end{eqnarray}
Since ${\bf V}_{i}={\bf u}({\bf r}_{i}(t),t)$, we can rewrite the foregoing
equation in the form
\begin{eqnarray}
{\partial\omega\over\partial t}=-\nabla\sum_{i=1}^{N}\gamma \delta({\bf
r}-{\bf r}_{i}(t)){\bf u}({\bf r},t).
\label{mf2}
\end{eqnarray}
Since the velocity is divergenceless, we obtain 
\begin{eqnarray}
{\partial\omega\over\partial t}=-{\bf u}({\bf
r},t)\nabla\sum_{i=1}^{N}\gamma \delta({\bf r}-{\bf r}_{i}(t))=-{\bf
u}\nabla\omega.
\label{mf3}
\end{eqnarray}
Therefore, in the point vortex model, the Euler equation (\ref{pf10})
contains exactly the same information as the Hamiltonian system
(\ref{pv3}).

This description in terms of $\delta$-functions, while being
technically correct, is useless for practical purposes because it
requires the knowledge of the exact trajectories of the point vortices
for an arbitrary initial condition or the solution of the Euler
equation (\ref{pf10}). When $N$ is large, this task is impossibly
difficult. Therefore, instead of the exact vorticity field expressed
in terms of $\delta$-functions, one is more interested by functions
which are smooth. For that reason, we introduce a smooth vorticity
field $\langle\omega\rangle({\bf r},t)$ which is proportional to the
average number of vortices contained in the cell $({\bf r},{\bf
r}+d{\bf r})$ at time $t$. This description requires that it is
possible to divide the domain in a large number of cells in such a way
that each cell is (a) large enough to contain a macroscopic number of
point vortices but (b) small enough for all the particles in the cell
can be assumed to possess the same average characteristics of the cell.

Formally, the average vorticity field is given by 
\begin{equation}
\langle\omega\rangle({\bf r},t)=\sum_{i=1}^{N}\gamma\langle\delta ({\bf r}-{\bf r}_{i}(t))\rangle,
\label{mf4}
\end{equation}
where the statistical average of a function $X({\bf r}_{1},...,{\bf r}_{N})$ is defined by
\begin{equation}
\langle X\rangle=\int \mu({\bf r}_{1},...,{\bf r}_{N},t)X({\bf r}_{1},...,{\bf r}_{N})d^{2}{\bf r}_{1}...d^{2}{\bf r}_{N},
\label{mf5}
\end{equation}
where $\mu({\bf r}_{1},...,{\bf r}_{N},t)$ is the $N$-body
distribution function of the system at time $t$. The average vorticity
can be rewritten
\begin{equation}
\langle\omega\rangle({\bf r},t)=N\gamma P({\bf r},t)=\gamma \langle n\rangle ({\bf r},t),
\label{mf6}
\end{equation}
where we have introduced the one-vortex distribution function
\begin{equation}
P({\bf r}_{1},t)=\int \mu({\bf r}_{1},...,{\bf r}_{N},t)d^{2}{\bf r}_{2}...d^{2}{\bf r}_{N},
\label{mf7}
\end{equation}
and the local vortex density $\langle n\rangle=N P({\bf r},t)$. In the
foregoing relations, we have implicitly used the fact that the
vortices are identical. The average energy of the system is given by
\begin{eqnarray}
E=\langle H\rangle=-{1\over 4\pi}\sum_{i\neq j}\gamma^{2}\langle \ln |{\bf r}_{i}-{\bf r}_{j}|\rangle\nonumber\\
=-{1\over 4\pi}N(N-1)\gamma^{2}\int g({\bf r},{\bf r}',t) \ln |{\bf r}-{\bf r}'|d^{2}{\bf r} d^{2}{\bf r}',
\label{mf8}
\end{eqnarray}
where 
\begin{equation}
g({\bf r}_{1},{\bf r}_{2},t)=\int \mu({\bf r}_{1},...,{\bf r}_{N},t)d^{3}{\bf r}_{2}...d^{2}{\bf r}_{N},
\label{mf9}
\end{equation}
is the two-body distribution function. In the mean-field approximation, which is exact in a properly defined thermodynamic limit with $N\rightarrow +\infty$ (see Sec. \ref{sec_field}), we have 
\begin{equation}
g({\bf r}_{1},{\bf r}_{2},t)= P({\bf r}_{1},t)P({\bf r}_{2},t).
\label{mf10}
\end{equation}
Accounting that $N(N-1)\simeq N^{2}$ for large $N$, the average energy takes the form
\begin{equation}
E=-{1\over 4\pi}\int \langle \omega\rangle ({\bf r},t) \langle \omega\rangle ({\bf r}',t) \ln |{\bf r}-{\bf r}'|d^{2}{\bf r} d^{2}{\bf r}'.
\label{mf11}
\end{equation}
Using Eq. (\ref{pf8}), the
expression (\ref{mf11}) for $E$ can be rewritten
\begin{equation}
E={1\over 2}\int\langle\omega\rangle \psi d^{2}{\bf r}=\int {\langle{\bf u}\rangle^{2}\over 2}  d^{2}{\bf r}.
\label{mf12}
\end{equation}
where $\psi$ is the streamfunction created by the average vorticity
$\langle\omega\rangle$ and where $\langle{\bf u}\rangle$ is the smooth
velocity field.

\subsection{The maximum entropy state}
\label{sec_entropie} 

We now wish to determine the equilibrium distribution of vortices 
following a statistical mechanics approach. Using Boltzmann procedure, 
we divide the domain in macrocells with area $\Delta$. Let $n_{i}$ denote the number of point vortices in the cell $\Delta_{i}$. We now decompose each macrocell into $\nu$ microcells with equal area. A {\it macrostate} is determined by the distribution $\lbrace n_{i}\rbrace$. Several configurations can lead to the same macrostate: each of them will be called a {\it microstate}. Using a combinatorial analysis, the number of microstates corresponding to the macrostate $\lbrace n_{i}\rbrace$  is
\begin{equation}
W(\lbrace n_{i}\rbrace)=N!\prod_{i}{\nu^{n_{i}}\over n_{i}!}.
\label{me1}
\end{equation}
The logarithm of this number defines the Boltzmann entropy. Using Stirling formula and considering the
continuum limit in which $\Delta,\nu\rightarrow 0$, we get the
classical formula
\begin{equation}
S=-N\int P({\bf r})\ln P({\bf r}) d^{2}{\bf r},
\label{me2}
\end{equation}
where $P({\bf r})$ is the density probability that a point vortex be
found in the surface element centered on ${\bf r}$.  At equilibrium, the
system is in the most probable macroscopic state, i.e. the state that
is the most represented at the microscopic level. This optimal state
is obtained by maximizing the Boltzmann entropy (\ref{me2}) at
fixed energy (\ref{mf12}) and vortex number $N$, or total circulation
\begin{equation}
\Gamma=N\gamma=\int\langle\omega\rangle d^{2}{\bf r}.
\label{me3}
\end{equation}

Writing the variational principle in the form
\begin{equation}
\delta S-\beta \delta E -\alpha\delta \Gamma=0,
\label{me4}
\end{equation}
where $\beta$ and $\alpha$ are Lagrange multipliers, it is readily
found that the maximum entropy state corresponds to the Boltzmann
distribution
\begin{equation}
\langle \omega\rangle=Ae^{-\beta\gamma\psi},
\label{me5}
\end{equation}
with inverse temperature $\beta$. We can account for the conservation
of angular momentum $L=\int \langle \omega \rangle r^{2} d^{2}{\bf r}$
(in a circular domain) and impulse $P=\int \langle \omega \rangle y
d^{2}{\bf r}$ (in a channel) by introducing appropriate Lagrange
multipliers $\Omega$ and $U$ for each of these constraints. In that
case, Eq. (\ref{me5}) remains valid provided that we replace the
streamfunction $\psi$ by the relative streamfunction $\psi'=\psi
+{\Omega\over 2}r^{2}-U y$. Substituting the Boltzmann relation
between $\langle\omega\rangle$ and $\psi$ in the Poisson equation
(\ref{pf7}), we obtain a differential equation for the streamfunction
\begin{equation}
-\Delta\psi=Ae^{-\beta\gamma\psi},
\label{me6}
\end{equation}
which determines the statistical equilibrium distribution of
vortices. This Boltzmann Poisson equation can be easily generalized to
include a spectrum of circulations among the vortices.

The combinatorial analysis presented in this section was
performed by Joyce \& Montgomery \cite{jm}. The mean-field equation
(\ref{me6}) was also obtained by Pointin \& Lundgren \cite{pl} from
the equilibrium BGK hierarchy.  In fact, this mean-field approach was
first developed by Onsager but his results were not published
(U. Frisch, private communication).

\subsection{Field theory and thermodynamic limit}
\label{sec_field}

Let us consider a collection of $N$ point vortices with equal
circulation $\gamma$ in a bounded domain of size $R$. We want to give
a rational to the mean-field approximation \index{Mean field models}
considered previously by defining a proper thermodynamic limit
\index{Thermodynamic limit} for point vortices. Rigorous results have
been established in
\cite{caglioti,eyink}. In the following, we present a less rigorous,
albeit equivalent, field theory approach inspired by the work of
\cite{horwitz,pad,vega4} for self-gravitating systems.

In the \index{Microcanonical Ensemble} microcanonical ensemble, the quantity of fundamental interest
is the density of states
\begin{equation}
g(E)=\int \delta \biggl (E-H({\bf r}_{1},...,{\bf r}_{N})\biggr )\prod_{i=1}^{N}d^{2}{\bf r}_{i},
\label{ft1}
\end{equation}
which is explicitly given by
\begin{equation}
g(E)=\int_{0}^{R}...\int_{0}^{R} \prod_{i=1}^{N}d^{2}{\bf r}_{i} \delta \biggl (E+{\gamma^{2}\over 2\pi}\sum_{i<j}\ln{|{\bf r}_{i}-{\bf r}_{j}|\over R}\biggr ),
\label{ft2}
\end{equation}
where the potential of interaction has been normalized by $R$. For
simplicity, we have ignored the contribution of the images but this
shall not affect the final results. We now introduce the change of
variables ${\bf x}={\bf r}/R$ and define the function
\begin{equation}
u({\bf x}_{1},...,{\bf x}_{N})={1\over N}\sum_{i<j}\ln|{\bf x}_{i}-{\bf x}_{j}|,
\label{ft3}
\end{equation}
and the dimensionless energy
\begin{equation}
\Lambda={2\pi E\over N^{2}\gamma^{2}}.
\label{ft4}
\end{equation}
In terms of these quantities, the density of states can be rewritten
\begin{equation}
g(E)={2\pi V^{N}\over N^{2}\gamma^{2}}\int_{0}^{1}...\int_{0}^{1} \prod_{i=1}^{N}d^{2}{\bf x}_{i} \delta \biggl (\Lambda+{1\over N}u({\bf x}_{1},...,{\bf x}_{N})\biggr ).
\label{ft5}
\end{equation}
The proper thermodynamic limit for a system of point vortices with
equal circulation in the microcanonical ensemble is such that
$N\rightarrow +\infty$ with fixed $\Lambda$. We see that the box size
$R$ does not enter in the normalized energy $\Lambda$. Therefore, the
thermodynamic limit corresponds to $N\rightarrow +\infty$ with
$\gamma\sim N^{-1}\rightarrow 0$ and $E\sim 1$. This is a very unusual
thermodynamic limit due to the non-extensivity of the system. Note
that the total circulation $\Gamma=N\gamma$ remains fixed in this
process.

For sufficiently large $N$, the density of states can be written
\begin{equation}
g(E)\simeq \int {\cal D}\rho \ e^{N S\lbrack \rho\rbrack}\ \delta \biggl(\Lambda-E\lbrack\rho\rbrack)\biggr )\delta\biggl (1-\int\rho({\bf r})d^{2}{\bf r}\biggr ),  
\label{ft6}
\end{equation}
with
\begin{equation}
S\lbrack \rho\rbrack=-\int \rho({\bf r})\ln\rho({\bf r})d^{2}{\bf r},  
\label{ft7}
\end{equation}
\begin{equation}
E\lbrack \rho\rbrack=-{1\over 2}\int \rho({\bf r})\rho({\bf r}')\ln|{\bf r}-{\bf r}'|d^{2}{\bf r}d^{2}{\bf r}'.  
\label{ft8}
\end{equation}
In the above formula, $g(E)$ has been expressed as a functional
integral over the macrostates recalling that $e^{N S\lbrack
\rho\rbrack}=W(\lbrace\rho\rbrace)$ gives the number of microstates
corresponding to the macrostate $\rho({\bf r})$ (see
Sec. \ref{sec_entropie}). The crucial point to realize is that the
vortex number $N$ appears explicitly in the exponential, all other
terms being of order unity. Therefore, at the thermodynamic limit
$N\rightarrow +\infty$, the functional integral is dominated by the
distribution $\rho_{*}({\bf r})$ which maximizes the Boltzmann entropy
(\ref{ft7}) under the constraints of fixed circulation and energy
brought by the $\delta$-functions. In this limit, the mean-field
approximation is {\it exact} and we have
\begin{equation}
g(E)=e^{N S\lbrack\rho_{*}\rbrack}, \qquad S(E)=N S\lbrack\rho_{*}\rbrack.
\label{ft9}
\end{equation}

In the \index{Canonical ensemble} canonical ensemble, the object of interest is the \index{Canonical partition function} partition function
\begin{equation}
Z(\beta)=\int e^{-\beta H({\bf r}_{1},...,{\bf r}_{N})}\prod_{i=1}^{N}d^{2}{\bf r}_{i},
\label{ft10}
\end{equation}
which is the normalization factor in the Gibbs measure
\begin{equation}
\mu({\bf r}_{1},...,{\bf r}_{N})={1\over Z} e^{-\beta H({\bf r}_{1},...,{\bf r}_{N})}.
\label{ft11}
\end{equation}
The free energy is defined by the relation
\begin{equation}
F(\beta)=-{1\over \beta}\ln Z.
\label{ft12}
\end{equation}
Using the notations introduced previously, we can rewrite the integral (\ref{ft10}) in the form
\begin{equation}
Z(\beta)=V^{N}\int_{0}^{N}...\int_{0}^{N}\prod_{i=1}^{N}d^{2}{\bf x}_{i} e^{\eta u({\bf x}_{1},...,{\bf x}_{N})},
\label{ft13}
\end{equation}
where 
\begin{equation}
\eta={\beta N\gamma^{2}\over 2\pi},
\label{ft14}
\end{equation}
is the normalized inverse temperature. It plays the role of the
``plasma parameter'' in plasma physics. It can be shown that the
partition function (\ref{ft13}) is convergent only for
$\eta>\eta_{c}=-4N/(N-1)$ which is consistent with the result
(\ref{es4}). The proper thermodynamic limit for a canonical distribution
of point vortices is such that $N\rightarrow +\infty$ with fixed $\eta$. In this limit, $\gamma\sim N^{-1}$ and $\beta\sim N$.  

The partition function is related to the density of states by the Laplace transform
\begin{equation}
Z(\beta)=\int_{-\infty}^{+\infty} dE \ g(E)e^{-\beta E}.
\label{ft15}
\end{equation}
Therefore, for large $N$ one has, using Eq. (\ref{ft6}),
\begin{equation}
Z(\beta)\simeq \int {\cal D}\rho \ e^{N J\lbrack \rho\rbrack}\delta\biggl (1-\int\rho({\bf r})d^{2}{\bf r}\biggr ),  
\label{ft16}
\end{equation}
with
\begin{equation}
J\lbrack \rho\rbrack=S\lbrack \rho\rbrack -\eta E\lbrack \rho\rbrack.
\label{ft17}
\end{equation}
At the thermodynamic limit, the functional integral is dominated by the distribution $\rho_{*}({\bf r})$ which maximizes the Massieu function  $J\lbrack \rho\rbrack$ under the constraint of a fixed circulation. In this limit, the mean-field approximation is exact and we have
\begin{equation}
Z(\beta)=e^{N J\lbrack\rho_{*}\rbrack}, \qquad F=E-TS.
\label{ft18}
\end{equation}
It can be noted that the critical points of entropy at fixed energy
and circulation (microcanonical description) and the critical points
of free energy at fixed temperature and circulation (canonical
description) coincide. This is not necessarily the case for the second
order variations of entropy and free energy. Therefore, a distribution
$\rho({\bf r})$ can be a maximum of $S[\rho]$ but a minimum (or a
saddle point) of $J[\rho]$. In that case, the equilibrium is stable in
the microcanonical ensemble but unstable in the canonical ensemble
(and the procedure leading to Eq. (\ref{ft18}) is clearly not
valid). When this situation happens, the ensembles are {\it
inequivalent} and phase transitions occur. This is the case, in
particular, for the gravitational problem (see Sec. \ref{sec_gravN}).

Finally, \index{Grandcanonical Ensemble} the grand canonical partition function \index{Grand partition function} is defined by
\begin{equation}
Z_{GC}=\sum_{N=0}^{+\infty}{z^{N}\over N!}\int \prod_{i=1}^{N}d^{2}{\bf r}_{i}e^{-\beta H_{N}({\bf r}_{1},...,{\bf r}_{N})},
\label{ft19}
\end{equation}
where $z$ is the fugacity. Using the Hubbard-Stratanovich transformation
\begin{equation}
\int {\cal D}\xi\ e^{-\int\lbrace {1\over 2}(\nabla\xi)^{2}-\rho({\bf r})\xi({\bf r})\rbrace d^{2}{\bf r}}=e^{-{1\over 2}\int \rho({\bf r})G({\bf r}-{\bf r}')\rho({\bf r}') d^{2}{\bf r}d^{2}{\bf r}'},
\label{ft20}
\end{equation}
where $G$ is the Green function of the Laplacian operator, we can rewrite the Boltzmann factor in the form
\begin{equation}
e^{-\beta H_{N}}=\int {\cal D}\xi \ e^{-\int {1\over 2}(\nabla\xi)^{2} d^{2}{\bf r}+\sqrt{-\beta}\sum_{i=1}^{N}\gamma \xi({\bf r}_{i})},
\label{ft21}
\end{equation}
where we have assumed $\beta<0$ which is the case of physical interest. Substituting this result in Eq. (\ref{ft19}) and introducing the notation $\phi({\bf r})=\sqrt{-\beta}\gamma\xi({\bf r})$, we can easily carry out the summation  on $N$ to obtain
\begin{equation}
Z_{GC}=\int {\cal D}\phi\ e^{-{1\over T_{eff}}\int d^{2}{\bf r}\lbrace {1\over 2}(\nabla\phi)^{2}-\mu^{2}e^{\phi}({\bf r})\rbrace},
\label{ft22}
\end{equation}
\begin{equation}
T_{eff}=-\beta\gamma^{2},\qquad \mu^{2}=-z\beta\gamma^{2}.
\label{ft23}
\end{equation}
Therefore, the grand partition function of the point vortex gas  corresponds to a Liouville field theory with an action
\begin{equation}
S\lbrack\phi\rbrack={1\over T_{eff}}\int  d^{2}{\bf r}\lbrace {1\over 2}(\nabla\phi)^{2}-\mu^{2}e^{\phi}({\bf r})\rbrace.
\label{ft24}
\end{equation}

While the previous description is formally correct if we {\it define}
$Z$ and $Z_{GC}$ by Eqs. (\ref{ft10}) and (\ref{ft19}), it must be
noted however that the canonical and grand canonical ensembles may not
have a physical meaning for point vortices. In particular, it is not
clear how one can impose a thermal bath at negative temperature. On
the other hand, the usual procedure to derive the canonical ensemble
from the microcanonical ensemble rests on a condition of additivity
\index{Additivity} which is clearly lacking in the present case.

\subsection{Axisymmetric equilibrium states in a disk}
\label{sec_disk} 

Let us consider a collection of $N$ point vortices with circulation
$\gamma$ confined within a disk of radius $R$. At statistical
equilibrium, the streamfunction $\psi$ is solution of the
Boltzmann-Poisson equation (\ref{me6}). If we work in a circular
domain, we must in principle account for the conservation of 
angular momentum. This can lead to bifurcations between axisymmetric
and off-axis solutions \cite{smith}. We shall, however, ignore this
constraint for the moment in order to obtain analytical expressions
for the thermodynamical parameters. This is a sufficient approximation
to illustrate the structure of the problem, which is our main concern
here.

\begin{figure}
\centering
\includegraphics[width=0.7\textwidth]{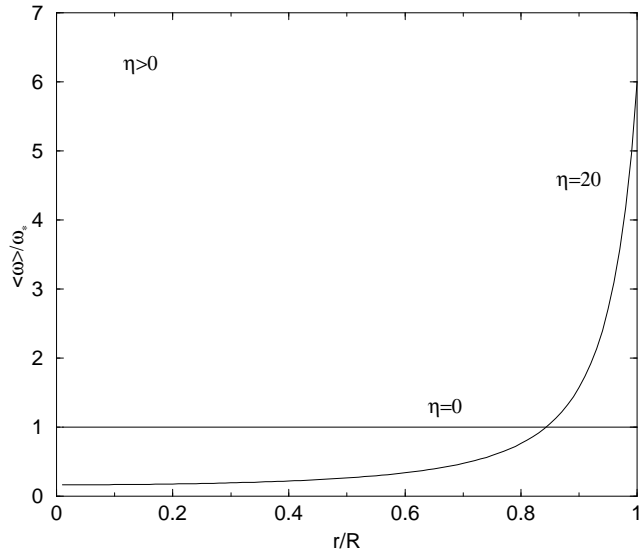}
\caption[]{Statistical equilibrium states of point vortices at positive temperatures ($\omega_{*}=\Gamma/\pi R^{2}$). The vortices are preferentially localized near the wall.}
\label{profilpos}
\end{figure}

\begin{figure}
\centering
\includegraphics[width=0.7\textwidth]{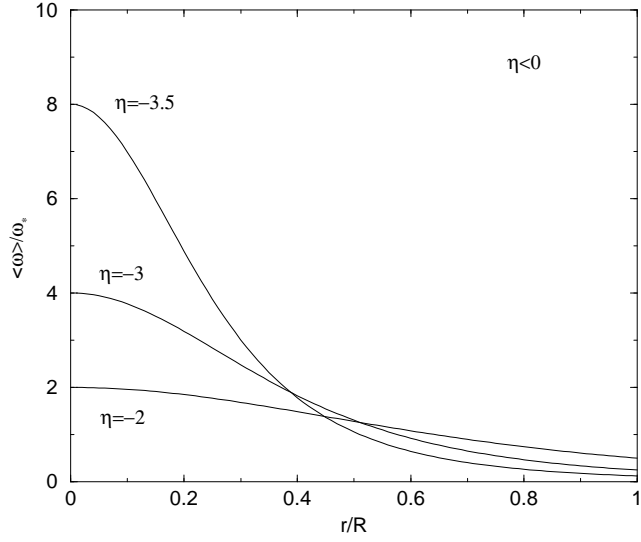}
\caption[]{Statistical equilibrium states of point vortices at negative temperatures showing a clustering. For $\eta=-4$, the vortices collapse at the center of the domain and the vorticity profile is a Dirac peak.}
\label{profilneg}
\end{figure}

If we confine our attention to axisymmetric solutions, the
Boltzmann-Poisson equation (\ref{me6}) can be written
\begin{equation}
{1\over r}{d\over dr}\biggl (r{d\psi\over dr}\biggr )=-\omega_{0}e^{-\beta\gamma(\psi-\psi_{0})}.
\label{ae1}
\end{equation} 
\begin{equation}
\psi'(0)=0, \qquad \psi(R)=0,
\label{ae2}
\end{equation} 
where $\psi_{0}$ and $\omega_{0}$ are the values of streamfunction and vorticity at $r=0$. Introducing the function $\phi=\beta\gamma(\psi-\psi_{0})$ and the dimensionless radial distance $\xi=(|\beta|\gamma\omega_{0})^{1/2}r$, Eq. (\ref{ae1}) can be reduced to the form
\begin{equation}
{1\over \xi}{d\over d\xi}\biggl (\xi{d\phi\over d\xi}\biggr )=\lambda e^{-\phi},
\label{ae3}
\end{equation} 
\begin{equation}
\phi(0)=\phi'(0)=0,
\label{ae4}
\end{equation} 
with $\lambda=1$ if $\beta<0$ and $\lambda=-1$ if $\beta>0$. It turns
out that this equation can be solved analytically as noticed by a
number of authors. With the change of variables
$t=\ln\xi$ and $\phi=2\ln\xi-z$, Eq. (\ref{ae3}) can be rewritten
\begin{equation}
{d^{2}z\over dt^{2}}=-\lambda e^{z}=-{d\over dz}(\lambda e^{z}).
\label{ae5}
\end{equation}
This corresponds to the motion of a ficticious particle in a potential $V(z)=\lambda e^{z}$. This equation is readily integrated and, returning to original variables,  we finally obtain
\begin{equation}
e^{-\phi}={1\over  (1+{\lambda\over 8}\xi^{2} )^{2}}.
\label{ae6}
\end{equation}
From the circulation theorem (\ref{pf6}) applied to an axisymmetric flow, we have
\begin{equation}
-{d\psi\over dr}(r)={\Gamma(r)\over 2\pi r}.
\label{ae7}
\end{equation}
where $\Gamma(r)=\int_{0}^{r}\omega(r')2\pi r' dr'$ is the circulation
within $r$. Taking $r=R$ and introducing the dimensionless variables
previously defined, we obtain
\begin{equation}
\eta\equiv {\beta\gamma\Gamma\over 2\pi}=-\alpha\phi'(\alpha),
\label{ae8}
\end{equation}
where $\alpha=(|\beta|\gamma\omega_{0})^{1/2}R$ is the value of $\xi$
at the box radius. Equation (\ref{ae8}) relates $\alpha$ to the
inverse temperature $\eta$. Using Eqs. (\ref{ae6}) and (\ref{ae8}),
the vorticity field can be written explicitly
\begin{equation}
\langle\omega\rangle(r)={4\Gamma\over \pi R^{2}(\eta+4)}{1\over  (1-{\eta\over \eta+4}{r^{2}\over R^{2}} )^{2}}.
\label{ae9}
\end{equation}
At positive temperatures ($\eta>0$), the vorticity is an increasing
function of the distance and the vortices tend to accumulate on the
boundary of the domain (Fig. \ref{profilpos}). On the contrary at
negative temperatures ($\eta<0$), the vorticity is a decreasing
function of the distance and the vortices tend to group themselves in
the core of the domain to form a ``supervortex''
(Fig. \ref{profilneg}). These results are consistent with Onsager's
prediction \cite{onsager}. We also confirm that statistical
equilibrium states only exist for $\eta>\eta_{c}=-4$, as previously
discussed. At this critical temperature, the
central vorticity becomes infinite and the solution tends to a Dirac
peak:
\begin{equation}
\langle\omega\rangle({\bf r})\rightarrow \Gamma\delta({\bf r}), \qquad {\rm for}\qquad \eta\rightarrow\eta_{c}=-4.
\label{ae10}
\end{equation}

\begin{figure}
\centering
\includegraphics[width=0.7\textwidth]{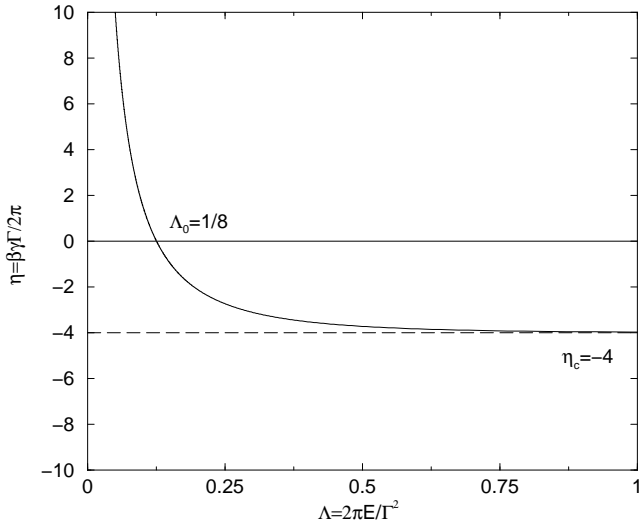}
\caption[]{Equilibrium phase diagram (caloric curve) \index{Caloric curve} for point vortices with equal circulation confined within a disk. For simplicity, the angular momentum has not been taken into account ($\Omega=0$).}
\label{etaLambda}
\end{figure}

The energy defined by Eq. (\ref{mf12}) can be written in the
dimensionless form
\begin{equation}
\Lambda\equiv {2\pi E\over \Gamma^{2}}={1\over 2\eta^{2}}\int_{0}^{\alpha}\phi'(\xi)^{2}\xi d\xi.
\label{ae11}
\end{equation}
The integral can be carried out explicitly using Eq. (\ref{ae6}). Eliminating $\alpha$ between Eqs. (\ref{ae11}) and (\ref{ae8}), we find that the temperature is related to the energy by the equation of state
\begin{equation}
\Lambda={1\over\eta}\biggl\lbrack {4\over\eta}\ln\biggl ({4\over 4+\eta}\biggr )+1\biggr \rbrack.
\label{ae12}
\end{equation}
The corresponding equilibrium phase diagram is represented in Fig. \ref{etaLambda}. We check explicitly that the energy becomes infinite when
$\eta\rightarrow \eta_{c}=-4$. The transition between positive and negative
temperatures occurs for $\Lambda_{0}=1/8$.

\begin{figure}
\centering
\includegraphics[width=0.7\textwidth]{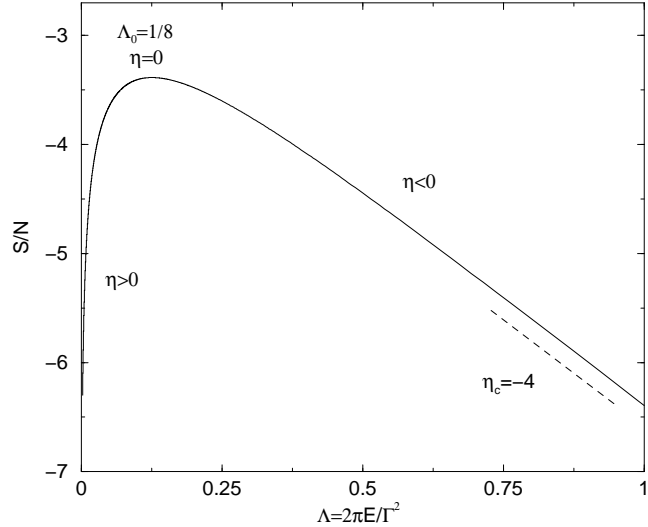}
\caption[]{Entropy vs energy plot for a system of point vortices with equal circulation confined within a disk.}
\label{sep}
\end{figure}

The entropy (\ref{me2}) can also be calculated easily from the above
results. Within an unimportant additive constant, it is given by
\begin{equation}
{S\over N}={8\over\eta}\ln 4-\biggl (1+{8\over \eta}\biggr )\ln (4+\eta).
\label{ae13}
\end{equation}
Using Eq. (\ref{ae12}), we can easily check that $dS/dE=\beta$, as
it should. For $\Lambda<\Lambda_{0}=1/8$, the entropy is an increasing
function of energy (positive temperatures) and for
$\Lambda>\Lambda_{0}$, it is a decreasing function of energy
(negative temperatures). For $\Lambda\rightarrow 0$, $S/N\sim
\ln\Lambda$ and for $\Lambda\rightarrow +\infty$, 
$S/N\sim -4\Lambda$ (Fig. \ref{sep}).

It is amusing to note that if we define a local pressure by the
relation $p({\bf r})=\langle\omega\rangle/\gamma\beta$ (which is
similar to the local equation of state $p=\rho T$ for an ideal gas of
material particles), we find that the pressure $P$ at the boundary of
the domain is exactly given by Eq. (\ref{es3}). Furthermore, using the
Boltzmann distribution (\ref{me5}), we easily check that $\nabla
p=-\langle\omega\rangle\nabla\psi$, which is similar to the equation
of hydrostatic equilibrium for a fluid in a gravitational field. It is
not clear whether these results bear more significance than is
apparent at first sight.

\subsection{Equilibrium states in an unbounded domain}
\label{sec_unbounded} 

In an infinite domain, it is necessary to take into account the
conservation of angular momentum, because this constraint determines
the typical size of the system. Furthermore, we shall assume that the
vortices have the same sign otherwise positive and negative vortices
will form dipoles and escape to infinity. Therefore, there is no
equilibrium state in an infinite domain for a neutral system of point
vortices.  When the conservation of angular momentum is taken into
account, the density of states is given by
\begin{equation}
g(E,L)=\int \delta \biggl (E-H({\bf r}_{1},...,{\bf r}_{N})\biggr )\ \delta\biggl (L-\sum_{i=1}^{N}\gamma r_{i}^{2}\biggr )\prod_{i=1}^{N}d^{2}{\bf r}_{i},
\label{ud1}
\end{equation}
and the angular velocity of the flow by
\begin{equation}
\Omega=2T\biggl ({\partial S\over\partial L}\biggr )_{E}.
\label{ud2}
\end{equation}
Using the same trick as in Sec. \ref{sec_state} with the change of
variables ${\bf x}={\bf r}/\sqrt{L}$, one finds the exact result
\begin{equation}
\Omega={2N\over\beta L}\biggl (1+{\beta N\gamma^{2}\over 8\pi}\biggr ).
\label{ud3}
\end{equation}
Therefore, the vorticity field is determined by the Boltzmann distribution
\begin{equation}
\langle\omega\rangle=Ae^{-\beta\gamma\psi'},
\label{ud4}
\end{equation}
where $\psi'$ is the relative streamfunction
\begin{equation}
\psi'\equiv \psi+{\Omega\over 2}r^{2}=\psi+{N\over 4\beta L}(4+\eta)r^{2}.
\label{ud7}
\end{equation}
For $\eta=0$, one has
\begin{equation}
\langle\omega\rangle={\gamma N\Gamma\over \pi L}e^{-{N \gamma\over L}r^{2}}.
\label{ud5}
\end{equation}
For large $r$, the asymptotic behavior of Eq. (\ref{ud4}) is
\begin{equation}
\langle\omega\rangle\sim {1\over r^{-\eta}}e^{-{N \gamma\over 4L}(4+\eta)r^{2}}\qquad (r\rightarrow +\infty),
\label{ud6}
\end{equation}
where we have used $\psi\sim -(\Gamma/2\pi)\ln r$ at large distances. From Eq. (\ref{ud6}), one sees that  $\eta\ge -4$ is required for the existence of an integrable solution.

Inserting the relation (\ref{ud4}) in the Poisson equation
(\ref{pf7}), we get
\begin{equation}
-\Delta\psi'=A e^{-{2\pi\eta\over N\gamma}\psi'}-{N^{2}\gamma^{2}\over 2\pi L\eta}(4+\eta).
\label{ud8}
\end{equation}
With the change of variables 
\begin{equation}
{\mb\xi}=\biggl ({\gamma N\over L}\biggr )^{1/2}{\bf r},\qquad \phi=\ln\biggl ({LA\over N^{2}\gamma^{2}}\biggr )-{2\pi\eta\over N\gamma}\psi',
\label{ud9}
\end{equation}
the Boltzmann-Poisson equation (\ref{ud8}) can be written
\begin{equation}
{d^{2}\phi\over d\xi^{2}}+{1\over\xi}{d\phi\over d\xi}=2\pi\eta e^{\phi}-(4+\eta),
\label{ud10}
\end{equation}
and the vorticity (\ref{ud4}) becomes
\begin{equation}
\langle\omega\rangle={N^{2}\gamma^{2}\over L}e^{\phi}.
\label{ud11}
\end{equation}
Equation (\ref{ud10}) has been obtained by Lundgren \& Pointin \cite{lp} from the equilibrium BGK hierarchy.  In the limit $\eta\rightarrow +\infty$, we have
\begin{equation}
\langle\omega\rangle= {\Gamma^{2}\over 2\pi L}, \qquad {\rm if} \quad r\le ({2L/\Gamma})^{1/2},
\label{ud12}
\end{equation}
and $\langle\omega\rangle=0$ otherwise. This vortex patch is the state
of minimum energy at fixed circulation and angular momentum. For
$\eta\rightarrow -4$, one has approximately
\begin{equation}
\langle\omega\rangle={N^{2}\gamma^{2}\over L}{A\over (1-A\pi\eta{\gamma N\over 4 L}r^{2})^{2}}e^{-{\gamma N\over 4 L}(4+\eta)r^{2}}.
\label{ud13}
\end{equation}
The first factor is an exact solution of Eq. (\ref{ud10}) with the second term on the right hand side neglected (see Sec. \ref{sec_disk}). The second factor is a correction for large $r$, in agreement with the asymptotic result expressed by Eq. (\ref{ud6}). The parameter $A$ tends to infinity as $\eta\rightarrow -4$ and is determined from the condition $\int \langle\omega\rangle d^{2}{\bf r}=\Gamma$ by the formula
\begin{equation}
\pi A+\ln(\pi A)=-C-\ln\bigl (1+{\eta\over 4}\bigr ),
\label{ud14}
\end{equation}
where $C=0.577...$ is the Euler constant. For $\eta=-4$, the vorticity profile is a Dirac peak and the energy tends to $+\infty$.

\subsection{The gravitational $N$-body problem}
\label{sec_gravN}

It is interesting to compare the previous results with those obtained
in the case of self-gravitating systems \cite{pad}. Formally, the
structure of the $N$-vortex problem \index{Dynamical system} shares
some analogies with \index{Astrophysics} the gravitational $N$-body
problem. The force by unit of mass experienced by a star is given by
\begin{equation}
{\bf F}_{i}=\sum_{j\neq i}{\bf F}(j\rightarrow i),\qquad {\bf F}(j\rightarrow i)=Gm {{\bf r}_{j}-{\bf r}_{i}\over |{\bf r}_{j}-{\bf r}_{i}|^{3}},
\label{gn1}
\end{equation}
where ${\bf F}(j\rightarrow i)$ is the force created by star $j$ on star $i$.
 The force can be written as the gradient ${\bf F}=-\nabla\Phi$ of a gravitational potential $\Phi$ which is related to the stellar density 
\begin{equation}
\rho({\bf r},t)=\sum_{i=1}^{N} m \delta({\bf r}-{\bf r}_{i}),
\label{gn2}
\end{equation} 
by the Poisson equation
\begin{equation}
\Delta\Phi=4\pi G\rho.
\label{gn3}
\end{equation} 
Furthermore, the equations of motion (Newton's equations) can be put in the Hamiltonian form
\begin{eqnarray}
{m {d{\bf r}_{i}\over dt}={\partial H\over\partial {\bf v}_{i}},\qquad m {d{\bf v}_{i}\over dt}=-{\partial H\over\partial {\bf r}_{i}}, }\nonumber\\
H={1\over 2}\sum_{i=1}^{N}mv_{i}^{2}-\sum_{i< j}{Gm^{2}\over |{\bf r}_{i}-{\bf r}_{j}|}.
\label{gn4}
\end{eqnarray}
In the analogy between stellar systems and two-dimensional vortices,
the star density $\rho$ plays the role of the vorticity $\omega$, the
force ${\bf F}$ the role of the velocity ${\bf V}$ and the
gravitational potential $\Phi$ the role of the streamfunction
$\psi$. The crucial point to realize is that, for the two systems, the
interaction is a long-range unshielded Coulombian interaction (in
$D=3$ or $D=2$ dimensions). This makes the connexion between point
vortices and stellar systems deeper than between point vortices and
electric charges for example. In particular, point vortices can
organize into large scale clusters, like stars in galaxies, while the
distribution of electric charges in a neutral plasma is uniform. There
are, on the other hand, fundamental differences between stars and
vortices. In particular, a star creates an acceleration while a vortex
creates a velocity. On the other hand, the gravitational interaction
is attractive and directed along the line joining the particles while
the interaction between vortices is rotational and perpendicular to
the line joining the vortices. 

\begin{figure}
\centering
\includegraphics[width=0.7\textwidth]{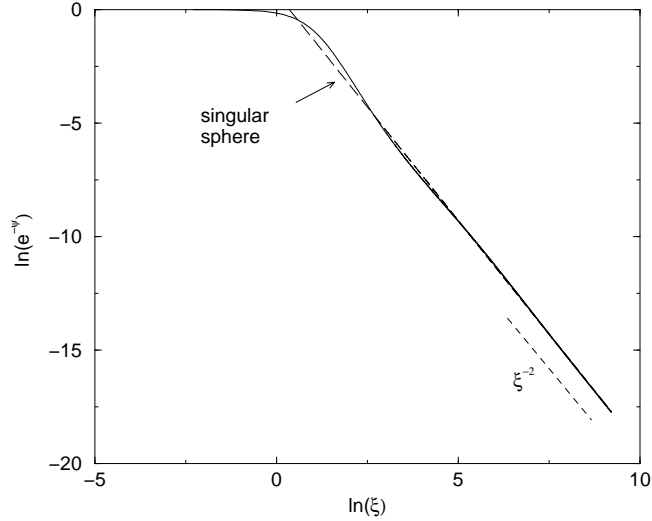}
\caption[]{Density profile of the self-gravitating gas at statistical equilibrium. The dashed line corresponds to the singular solution $\rho=1/2\pi G\beta r^{2}$.}
\label{densite}
\end{figure}

Despite these important differences,
the statistical mechanics of 2D vortices and stellar systems are
relatively similar. Like the point vortex gas, the self-gravitating
gas is described at statistical equilibrium by the Boltzmann
distribution
\begin{eqnarray}
\langle\rho\rangle=Ae^{-\beta\Phi},
\label{gn5}
\end{eqnarray} 
obtained by maximizing the Boltzmann entropy at fixed mass $M$ and
energy $E$ .  Its structure is therefore determined by solving the
Boltzmann-Poisson equation
\begin{eqnarray}
\Delta\Phi=4\pi G Ae^{-\beta\Phi},
\label{gn6}
\end{eqnarray} 
where $A$ and $\beta>0$ have to be related to $M$ and $E$. This
statistical mechanics approach has been developed principally for globular
clusters relaxing towards equilibrium via two-body encounters
\cite{bt}. It is clear that the Boltzmann-Poisson equation (\ref{gn6})
is similar to the Boltzmann-Poisson equation (\ref{me6}) for point
vortices at {\it negative} temperatures. The density profile
determined by these equations is a decreasing function of the
distance, which corresponds to a situation of {\it clustering} (see
Figs. \ref{profilneg} and \ref{densite}). The similarity of the
maximum entropy problem for stars and vortices, and the
Boltzmann-Poisson equations (\ref{gn6}) (\ref{me6}), is a first
manifestation of the formal analogy existing between these two
systems.

However, due to the different dimension of space ($D=3$ for stars
instead of $D=2$ for vortices), the mathematical problems differ in
the details. First of all, the density profile determined by the
Boltzmann-Poisson equation (\ref{gn6}) in $D=3$ decreases like
$r^{-2}$ at large distances leading to the so-called infinite mass
problem since $M=\int_{0}^{+\infty}\rho \ 4\pi r^{2}dr\rightarrow
+\infty$ \cite{chandra39}. There is no such problem for point vortices
in two dimensions: the vorticity decreases like $r^{-4}$, or even more
rapidly if the conservation of angular momentum is accounted for, and
the total circulation $\Gamma=\int_{0}^{+\infty}\omega \ 2\pi r dr$ is
finite.  The infinite mass problem implies that {\it no} statistical
equilibrium state exists for open star clusters, even in theory.  A
system of particles in gravitational interaction tends to evaporate so
that the final state is just two stars in Keplerian orbit. This
evaporation process has been clearly identified in the case of
globular clusters which gradually lose stars to the benefit of a
neighboring galaxy. In fact, the evaporation is so slow that we can
consider in a first approximation that the system passes by a
succession of quasiequilibrium states described by a truncated
isothermal distribution function (Michie-King model) \cite{bt}. This
justifies the statistical mechanics approach in that sense. Another
way of solving the infinite mass problem is to confine the system
within a box of radius $R$. However, even in that case, the notion of
equilibrium poses problem regarding what now happens at the center of
the configuration.

\begin{figure}
\centering
\includegraphics[width=0.7\textwidth]{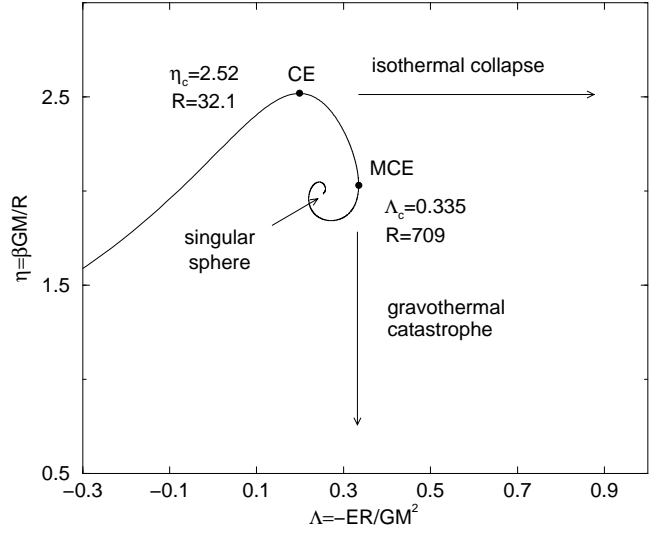}
\caption[]{Equilibrium phase diagram for self-gravitating systems confined within a box. For sufficiently low energy or temperature, there is no equilibrium state and the system undergoes gravitational collapse. }
\label{etalambdaS}
\end{figure}

\begin{figure}
\centering
\includegraphics[width=0.7\textwidth]{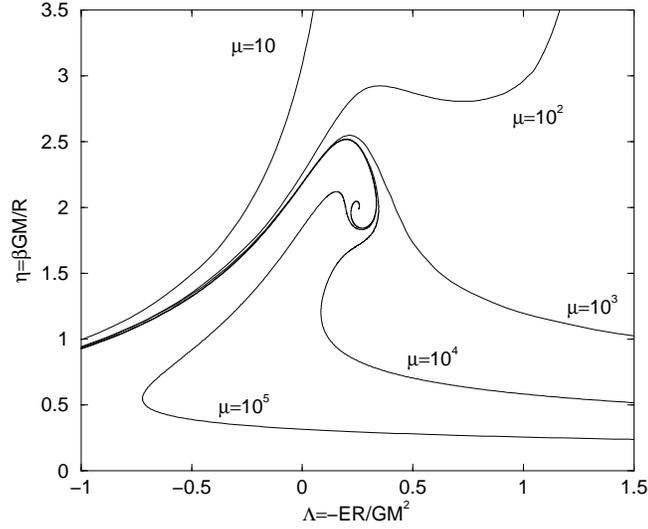}
\caption[]{Equilibrium phase diagram for self-gravitating fermions \cite{pt}. 
The degeneracy parameter $\mu$ plays the role of a small-scale cut-off
$\epsilon\sim 1/\mu$. For $\epsilon\rightarrow 0$, the classical
spiral of Fig. \ref{etalambdaS} is recovered.}
\label{etalambda}
\end{figure} 

The equilibrium phase diagram $(E,T)$ for bounded self-gravitating
systems is represented in Fig. \ref{etalambdaS}. The caloric curve
\index{Caloric curve} has a striking spiral behavior parametrized by
the density contrast ${\cal R}=\rho(0)/\rho(R)$ going from $1$
(homogeneous system) to $+\infty$ (singular sphere) as we proceed
along the spiral. There is no equilibrium state below
$E_{c}=-0.335GM^{2}/R$ or $T_{c}={GMm\over 2.52 kR}$
\cite{antonov,lbw}. In that case, the system is expected to collapse
indefinitely. This is called {\it gravothermal catastrophe}
\index{Gravothermal catastrophe} in the microcanonical ensemble (fixed
$E$) and {\it isothermal collapse} in the canonical ensemble (fixed
$T$). Dynamical models show that the collapse is self-similar and
develops a finite time singularity
\cite{penston,larson,cohn,lbe,kiess,crs}. However, although the
central density goes to $+\infty$, the shrinking of the core is so
rapid that the core mass goes to zero. Therefore, the singularity
contains no mass and this process alone cannot lead to a black hole.

Since the $T(E)$ curve has turning points, this
implies that the microcanonical and canonical ensembles are not
equivalent \index{Ensemble Inequivalence} and that phase transitions
will occur \cite{pad}. In the microcanonical ensemble, the series of
equilibria becomes unstable after the first turning point of energy
$(MCE)$ corresponding to a density contrast of $709$. At that point,
the solutions pass from local entropy maxima to saddle points.  In the
canonical ensemble, the series of equilibria becomes unstable after
the first turning point of temperature $(CE)$ corresponding to a
density contrast of $32.1$. At that point, the solutions pass from
minima of free energy ($F=E-TS$) to saddle points.  It can be noted
that the region of \index{Negative specific heat} negative specific
heats between $(CE)$ and $(MCE)$ is stable in the microcanonical
ensemble but unstable in the canonical ensemble, as expected on general
physical grounds. The thermodynamical stability of isothermal spheres
can be deduced from the topology of the $\beta-E$ curve by using the
turning point criterion of Katz
\cite{katz} who has extended Poincar\'e's theory of linear series of
equilibria.  The stability problem can also be reduced to the study of
an eigenvalue equation associated with the second order variations of
entropy or free energy as studied by Padmanabhan \cite{pad2} in the
microcanonical ensemble \index{Microcanonical Ensemble} and by Chavanis
\cite{chavacano} in the \index{Canonical ensemble} canonical ensemble. 
This study has been recently extended to other \index{Grandcanonical
Ensemble} statistical ensembles \cite{grand}: grand canonical, grand
microcanonical, isobaric.... The same stability limits as Katz are
obtained but this method provides in addition the form of the density
perturbation profiles that trigger the instability at the critical
points. It also enables one to show a clear equivalence between
thermodynamical stability in the canonical ensemble and dynamical
stability with respect to the Navier-Stokes equations (Jeans problem)
\cite{chavacano,grand}. These analytical methods can be extended to general
relativity \cite{relat}.  It must be stressed, however, that the statistical
equilibrium states of self-gravitating systems are at most
\index{Metastability} metastable: there is no {\it global} maximum of
entropy or free energy for a classical system of point masses in
gravitational interaction
\cite{antonov}.

\begin{figure}
\centering
\includegraphics[width=0.7\textwidth]{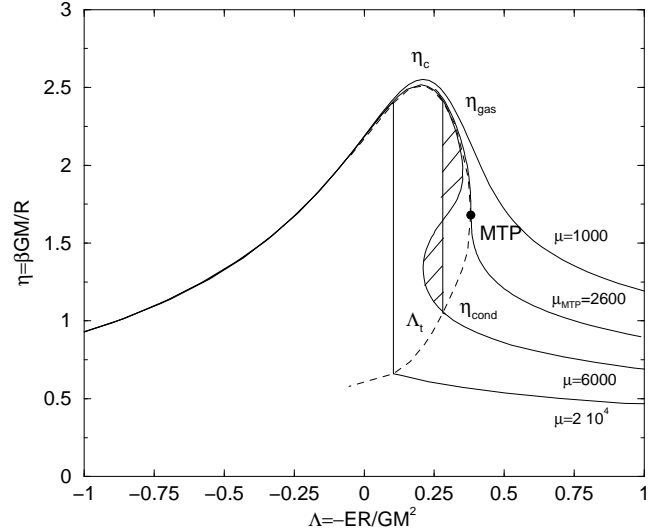}
\caption[]{Enlargement of the phase diagram near the tricritical point in the microcanonical ensemble. A priori, the phase transition should occur at the energy $E_{t}(\mu)$ at which the gaseous phase and the condensed phase have the same entropy. In fact, the entropic barrier played by the unstable solution on the wiggling branch is so hard to cross that the transition will not occur at $E_{t}$ but rather at, or near, $E_{c}$ the point of gravothermal catastrophe \cite{newkatz,ispolatov}.}
\label{tricritiquemicroP}
\end{figure}

\begin{figure}
\centering
\includegraphics[width=0.7\textwidth]{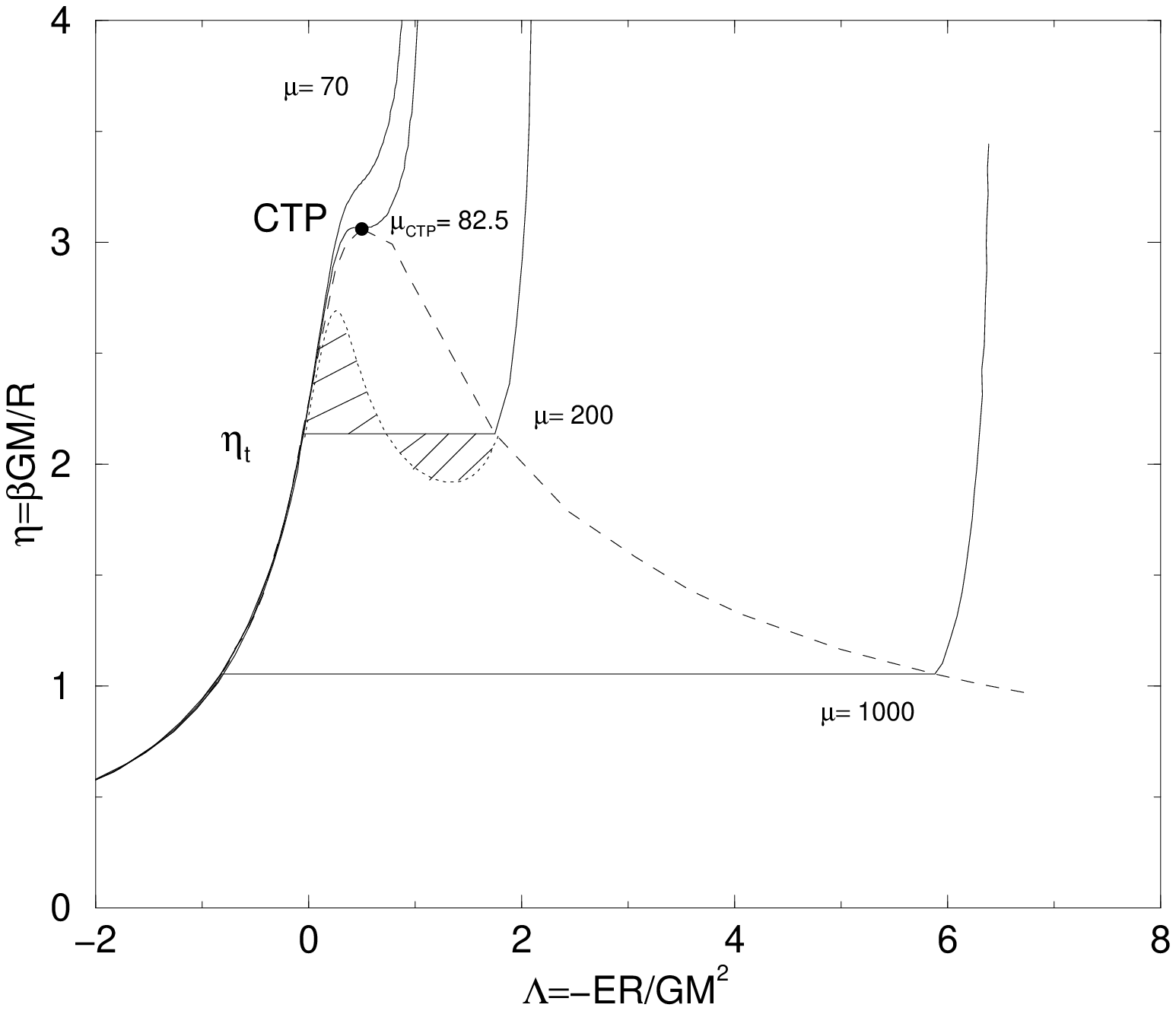}
\caption[]{Same as Fig. \ref{tricritiquemicroP} near the tricritical point in the canonical ensemble. The region of negative specific heats, allowed in the microcanonical ensemble, is replaced by a phase transition in the canonical ensemble. The notion of metastability \index{Metastability} is discussed in \cite{pt,ispolatov}.}
\label{tricritiqueP}
\end{figure}

Phase transitions in self-gravitating systems \index{Phase transition}
can be studied in detail by introducing a small-scale cut-off
$\epsilon$ in order to regularize the potential. This can be achieved
for example by considering a system of self-gravitating fermions (in
which case an effective repulsion is played by the Pauli exclusion
principle)
\cite{thirring,bilic,cs,pt,cape} or a hard spheres gas
\cite{aronson,pad,stahl}. Other forms of regularization are 
possible \cite{follana,millerB,ispolatov}. For these systems, there
can still be gravitational collapse but the core will cease to shrink
when it feels the influence of the cut-off. The result is the
formation of a compact object with a large mass: a ``fermion ball'' or
a hard spheres ``condensate''. The equilibrium phase diagram of
self-gravitating fermions is represented in Fig. \ref{etalambda} and
has been discussed at length by Chavanis \cite{pt} in the light of an
analytical model. The introduction of a small-scale cut-off has the
effect of unwinding the classical spiral of Fig. \ref{etalambdaS}. For
a small cut-off $\epsilon\ll 1$, the trace of the spiral is still
visible and the $T(E)$ curve is multivalued
(Fig. \ref{tricritiquemicroP}). This can lead to a gravitational first
order phase transition between a gaseous phase with an almost
homogeneous density profile (upper branch) and a condensed phase with
a core-halo structure (lower branch). At a critical cut-off value
$\epsilon=\epsilon_{MTP}$, the two phases merge and the gravitational
phase transition disapears.  This particular point is sometimes called
a microcanonical tricritical point (MTP).

For
$\epsilon>\epsilon_{MTP}$, the curve has the form of Fig.
\ref{tricritiqueP}. The $T(E)$ curve is now univalued so that the 
equilibrium states are always stable in the microcanonical ensemble
(they are global entropy maxima). In particular, the region of
\index{Negative specific heat} negative specific heats (leading to a
convex dip \index{Convex intruder} in the entropy vs energy plot
\cite{ispolatov}) is allowed in the microcanonical ensemble. By
contrast, the $E(T)$ curve is multivalued and this can lead to a
normal first order phase transition in the canonical ensemble. The
gaseous and condensed phases are thus connected by a Maxwell plateau
which replaces the region of negative specific heats.  At
$\epsilon=\epsilon_{CTP}$, the two phases merge, the specific heat
becomes infinite and the phase transition is second order. This
particular point is sometimes called a canonical tricritical point
(CTP). For $\epsilon>\epsilon_{CTP}$, the $T(E)$ curve is monotonic
and the specific heat is positive. Therefore, as the cut-off parameter
$\epsilon$ increases, the self-gravitating gas consecutively exhibits
gravitational first order, normal first order, second-order and no
phase transition at all
\cite{ispolatov}. A similar behavior exists for the Blume-Emery-Griffiths
(BEG) model with infinite range interaction
\cite{barre} and is probably representative of other systems
presenting inequivalence of ensembles (small systems or systems with
long-range interactions).

The equilibrium phase diagram of point vortices in two dimensions is
completely different (Fig. \ref{etaLambda}). Since there is no turning
point of energy or temperature, we can immediately infer that the
equilibria are always stable and that the microcanonical and canonical
ensembles are equivalent. In the microcanonical ensemble, an
equilibrium state exists for all values of energy.  Therefore, there
is no ``gravothermal catastrophe'' in two dimensions \cite{klb}. The
solutions of the Boltzmann-Poisson system for arbitrary dimension $D$
and the disappearance of the spiral as we approach the critical
dimension $D=2$ have been studied in \cite{scD} (note that the spiral
also disapears for $D\ge 10$).  There exists, on the other hand, a
critical inverse temperature $\eta_{c}=-4$ in two dimensions below
which there is no equilibrium state in the canonical ensemble. This
can lead to a situation of collapse \cite{scD} but, as indicated
previously, it is not clear whether the canonical ensemble makes sense
for a gas of point vortices.

\section{Statistics of velocity fluctuations arising from a random distribution of point vortices }
\label{sec_fluc}

\subsection{The marginal Gaussian distribution}
\label{sec_marginal}

The aim of equilibrium statistical mechanics is to predict the final
configuration of a system resulting from a complex evolution. We seek
now to develop a kinetic theory of point vortices to determine how the
system will reach this equilibrium configuration. The first problem to
consider is the characterization of the velocity fluctuations
experienced by a point vortex. These fluctuations are responsible for
a diffusion process which is one of the driving source of the
evolution. In this section, we describe a simple stochastic model
\cite{jimenez,min,weiss,chukbar,kuv,csire1,csire2,eff} from which we can
obtain an estimate of the diffusion coefficient of point vortices
\cite{csire1}. We follow the presentation of Chavanis \& Sire \cite{csire1,csire2}.

Let us consider a collection of $N$ point vortices randomly
distributed in a disk of radius $R$. We assume that the vortices have
a Poisson distribution, i.e. their positions are independent and
uniformly distributed in average over the entire domain. In terms of
statistical mechanics, this corresponds to an equilibrium state with
$\beta=0$ in a domain with no specific symmetry or $\beta\rightarrow
+\infty$ (i.e. a state of minimum energy) if the angular momentum is
conserved. We are particularly interested in the ``thermodynamic
limit'' in which the number of vortices and the size of the domain go
to infinity ($N\rightarrow\infty$, $R\rightarrow\infty$) in such a way
that the vortex density $n={N\over\pi R^{2}}$ remains finite. In this
limit, the Poisson distribution is stationary and is well-suited to
the analysis of the fluctuations. For simplicity, we assume that the
vortices have the same circulation $\gamma$. There is therefore a
solid rotation of the system $\langle {\bf V}\rangle={1\over 2}n\gamma
{\bf r}_{\perp}$. We shall work in a rotating frame of reference so as
to ignore this solid rotation. The generalization of our results to a
spectrum of circulations among the vortices is relatively
straightforward \cite{csire1}.

The velocity ${\bf V}$ occurring at the center $O$ of the domain is the 
sum of the velocities ${\bf \Phi}_{i}$ $(i=1,...,N)$ produced by
the $N$ vortices:
\begin{equation}
{\bf V}=\sum_{i=1}^{N}{\bf \Phi}_{i},\qquad {\bf \Phi}_{i}=-{\gamma\over 2\pi}{{\bf r}_{ \perp i} \over r_{i}^{2}} ,
\label{mg1}
\end{equation}
where ${\bf r}_{i}$ denotes the position of the $i^{th}$ vortex 
relative to the point under consideration and, by 
definition,  ${\bf r}_{\perp i}$ is the vector ${\bf r}_i$ rotated by 
$+{\pi/ 2}$. Since the vortices are randomly distributed, the velocity 
${\bf V}$ fluctuates. It is therefore of interest to study the statistics 
of these fluctuations, i.e. the probability $W({\bf V})d^{2}{\bf V}$ that 
${\bf V}$ occurs between ${\bf V}$ and ${\bf V}+d{\bf V}$. Basically, we have to  determine the distribution of a sum of random variables. If the variance of the individual velocity ${\bf \Phi}$ were finite, we could immediately apply the central limit theorem and deduce that the distribution of ${\bf V}$ is Gaussian. Alternatively, if the variance of ${\bf \Phi}$ were rapidly diverging (i.e., algebraically), the distribution of ${\bf V}$ would be a L\'evy law with an infinite variance and an algebraic tail. In the present case, the problem is intermediate between these two situations because the variance of the velocity created by a single vortex 
\begin{equation} 
\langle \Phi^{2}\rangle=\int_{|{\bf r}|=0}^{R}\tau({\bf
r})\Phi^{2}d^{2}{\bf r}=\int_{0}^{R}{1\over\pi R^{2}}{\gamma^{2}\over
4\pi^{2}r^{2}}2\pi r dr,
\label{mg2}
\end{equation}
diverges {\it slowly}, i.e. logarithmically. For that reason, the
distribution of ${\bf V}$ will be intermediate between Gaussian and L\'evy
laws.

The distribution $W_{N}({\bf V})$ of the velocity can be expressed as
\begin{equation}
W_{N}({\bf  V})=\int \prod_{i=1}^{N}\tau({\bf r}_{i})d^{2}{\bf
r}_{i}\delta\biggl ({\bf  V}-\sum_{i=1}^{N}{\bf \Phi}_{i}\biggr ),
\label{mg3}
\end{equation}
where $\tau({\bf r}_{i})d^{2}{\bf r}_{i}$ governs the probability of occurrence
of the $i$-th point vortex at position ${\bf r}_{i}$. In writing this
expression, we have assumed that the vortices are identical and uncorrelated.
Now, using a method originally due to Markov, we express the $\delta$-function
appearing in Eq. (\ref{mg3}) in terms of its Fourier transform
\begin{equation}
\delta({\bf x})={1\over (2\pi)^{2}}\int e^{-i {\mb \rho}{\bf x}}d^{2}{\mb \rho}.
\label{mg4}
\end{equation}
With this transformation, $W_{N}({\bf V})$ becomes
\begin{equation}
W_{N}({\bf V})={1\over 4\pi^{2}}\int A_{N}({\mb \rho})e^{-i {\mb \rho}{\bf
V}}d^{2}{\mb\rho}, 
\label{mg5}
\end{equation}
with
\begin{equation} 
A_{N}({\mb \rho})=\biggl (\int_{|{\bf r}|=0}^{R}e^{i {\mb \rho}{\bf \Phi}}\tau
({\bf r})d^{2}{\bf r}\biggr )^{N},
\label{mg6}
\end{equation}
where we have written
\begin{equation}
{\bf\Phi}=-{\gamma\over 2\pi}{{\bf r}_{\perp}\over r^{2}}.
\label{mg7}
\end{equation}
If we now suppose that the vortices are uniformly distributed  on average, then
\begin{equation}
\tau({\bf r})={1\over \pi R^{2}},
\label{mg8}
\end{equation}
and Eq. (\ref{mg6}) reduces to 
\begin{equation}
A_{N}({\mb \rho})=\Biggl ({1\over \pi R^{2}}\int_{|{\bf r}|=0}^{R} 
e^{i {\mb \rho}{\bf \Phi}}d^{2}{\bf r}\Biggr )^{N}.
\label{mg9}
\end{equation}
Since
\begin{equation}
{1\over \pi R^{2}}\int_{|{\bf r}|=0}^{R}d^{2}{\bf r}=1, 
\label{mg10}
\end{equation}
we can rewrite our expression for $A_{N}({\mb \rho})$ in the form  
\begin{equation}
A_{N}({\mb \rho})=\Biggl (1-{1\over \pi R^{2}}\int_{|{\bf r}|=0}^{R} 
(1-e^{i {\mb \rho}{\bf \Phi}})d^{2}{\bf r}\Biggr )^{N}.
\label{mg11}
\end{equation}
We now consider the limit in which the number of vortices and the size
of the domain go to infinity in such a way that the density remains
finite: $$\qquad N\rightarrow\infty,\quad R\rightarrow \infty, \quad
n={N\over \pi R^{2}} \quad {\rm finite}.$$ 
If the integral
occurring in Eq. (\ref{mg11}) increases less rapidly than $N$, then
\begin{equation}
A({\mb \rho})= e^{-n C({\mb \rho})},
\label{mg12}
\end{equation} 
with
\begin{equation}
C({\mb \rho})=\int_{|{\bf r}|=0}^{R} 
(1-e^{i {\mb \rho}{\bf \Phi}})d^{2}{\bf r}.
\label{mg13}
\end{equation} 
We have dropped the subscript $N$ to indicate that the limit
$N\rightarrow\infty$, in the previous sense,  has been taken. Note that $A({\mb
\rho})$ can still depend on $N$ through logarithmic factors, so that Eq.
(\ref{mg12}) must be considered as an equivalent of Eq. (\ref{mg6})
for large $N$, not a true limit. In fact, it is appropriate to
consider that $N\rightarrow +\infty$ but not $\ln N$ since in physical
situations the typical number of vortices does not exceed $10^{4}$.

The characteristic function $C({\mb \rho})$ can be calculated
explicitly by taking ${\bf \Phi}$ as a variable of integration instead
of ${\bf r}$ and transforming to polar coordinates \cite{csire1}. The
final result remains complicated but the following expression provides
a sufficient approximation for our purposes:
\begin{equation}
C({\mb \rho})={\gamma^{2}\rho^{2}\over 16\pi}\ln \biggl ( {4\pi N\over
n\gamma^{2}\rho^{2}}\biggr ).
\label{mg14}
\end{equation}
Since $C({\mb \rho})$ diverges weakly with $N$ (logarithmically), the
limiting process leading to formula (\ref{mg12}) is permissible.  The
velocity distribution $W({\bf V})$ is simply the Fourier transform of
$A({\mb\rho})$ with the expression (\ref{mg14}) for $C({\mb
\rho})$. This leads to the following distribution for the velocity 
fluctuations:
\begin{equation}
W({\bf  V})= {4\over  {n\gamma^{2}} \ln N} e^{-{4\pi\over n {\gamma^{2}}\ln N}
{V}^{2}}\quad (V\ll V_{crit}(N)),
\label{mg15}
\end{equation} 
\begin{equation}
W({\bf V})= {n\gamma^{2}\over 4 \pi^{2}V^{4}}\quad (V\gg V_{crit}(N)),
\label{mg16}
\end{equation}
\begin{equation}
V_{crit}(N)\sim \biggl ({n\gamma^{2}\over 4\pi}\ln N\biggr )^{1/2}\ln^{1/2}(\ln
N).
\label{mg17}
\end{equation}
For small fluctuations, the core of the distribution is Gaussian
as if the central limit theorem were applicable (this is due to the
quadratic behavior of $C(\rho)$). For sufficiently large values of
$V$, the velocity distribution $W({\bf V})$ decays algebraically as
for a L\'evy law (this is due to the logarithmic term in
$C(\rho)$). In the mathematical limit $\ln N\rightarrow +\infty$, the
algebraic tail is rejected to infinity and the distribution is purely
Gaussian. This is consitent with the generalized form of the central
limit theorem described by Ibragimov \& Linnik \cite{ibragimov} and
used by Min {\it et al.} \cite{min} and Weiss {\it et al.}
\cite{weiss}. However, the convergence is so slow with $N$ 
that the algebraic tail is always visible for point vortex systems
\cite{min,jimenez,weiss,kuv}. This algebraic tail arises because we are on
the frontier between Gaussian and L\'evy laws (see Fig. 1.1 of
Bouchaud \& Georges \cite{bouchaud}). Therefore, we proposed to call
Eqs. (\ref{mg15})-({\ref{mg17}) the {\it Marginal Gaussian distribution}
\cite{csire2}.

Since the distribution $W({\bf V})$ decreases like $V^{-4}$ for
$V\rightarrow\infty$, the variance of the velocity diverges
logarithmically.  Since $V=\gamma/2\pi r$, this corresponds to a
divergence at small scales ($r\rightarrow 0$).  On the other hand, if
we were to extend the Gaussian distribution (\ref{mg15}) for all values of $V$,
we would conclude that its variance
\begin{equation}
\langle V^{2}\rangle = {n {\gamma^{2}} \over 4\pi} \ln N, 
\label{mg18}
\end{equation}
diverges logarithmically when $N\rightarrow \infty$. This corresponds
to a divergence at large scales ($R\rightarrow +\infty$). We can
recover these results more directly by calculating the variance of
${\bf V}$ from Eq. (\ref{mg1}). Indeed,
\begin{equation}
\langle V^{2}\rangle =N\int_{L_{min}}^{R} {1\over \pi R^{2}}{\gamma^{2}\over 4\pi^{2} r^{2}}2\pi r dr={n\gamma^{2}\over 2\pi}\ln\biggl ({R\over L_{min}}\biggr ),
\label{mg19}
\end{equation}
where $L_{min}$ is a lower cut-off and $R$ an upper cut-off played here by the size of the domain.

We can give a physical interpretation of the algebraic tail of the
velocity distribution in terms of the nearest neighbor
approximation. Let us calculate the velocity distribution due to the
nearest neighbor. For that purpose, we must first determine the
probability $\tau_{n.n}(r)dr$ that the position of the nearest
neighbor occurs between $r$ and $r+dr$. Clearly, $\tau_{n.n}(r)dr$ is
equal to the probability that no vortices exist interior to $r$ times
the probability that a vortex (any) exists in the annulus between $r$
and $r+dr$. Therefore, it must satisfy an equation of the form
\begin{equation}
\tau_{n.n}(r)dr=\biggl (1-\int_{0}^{r}\tau_{n.n}(r')dr'\biggr ) n 2\pi r dr,
\label{mg20}
\end{equation}
where $n={N\over \pi R^{2}}$ denotes the mean density of vortices in the disk.
Differentiating Eq. (\ref{mg20}) with respect to $r$ we obtain 
\begin{equation}
{d\over dr}\biggl\lbrack{\tau_{n.n}(r)\over 2\pi n r}\biggl\rbrack
=-\tau_{n.n}(r). 
\label{mg21}
\end{equation} 
This equation is readily integrated with the condition
$\tau_{n.n}(r)\sim 2\pi n r$ as $r\rightarrow 0$, and we find
\begin{equation}
\tau_{n.n}(r)={2\pi n} r e^{-{\pi n}r^{2}}.
\label{mg22}
\end{equation}
This is the distribution of the nearest neighbor in a random distribution of
particles. From this formula, we can obtain the exact value for the ``average
distance'' $d$ between vortices. By definition,  
\begin{equation}
d=\int_{0}^{+\infty} \tau_{n.n}({ r})r d{ r}.
\label{mg23}
\end{equation}
Hence,
\begin{equation}
d={1\over 2\sqrt{n}}.
\label{mg24}
\end{equation}
If we assume that the velocity ${\bf V}$ is
entirely due to the nearest neighbor, then 
\begin{equation}
W_{n.n}({\bf V})d^{2}{\bf V}=\tau_{n.n}({\bf r})d^{2}{\bf r},
\label{mg25}
\end{equation}  
with
\begin{equation}
{\bf V}=-{\gamma\over 2\pi}{{\bf r}_{\perp}\over r^{2}}.
\label{mg26}
\end{equation}
Using Eq. (\ref{mg22}), we obtain
\begin{equation}
W_{n.n}({\bf V})={n\gamma^{2}\over 4\pi^{2} V^{4}} e^{-{n\gamma^{2}\over 4\pi V^{2}}}.
\label{mg27}
\end{equation}
The nearest neighbor approximation is expected to give relevant
results only for large values of the velocity. Thus, we can make the
additional approximation
\begin{equation}
W_{n.n}({\bf V})={n\gamma^{2}\over 4\pi^{2} V^{4}},
\label{mg28}
\end{equation}
in perfect agreement with Eq. (\ref{mg16}) valid for $V\gg V_{crit}$.
This shows that the algebraic tail of the velocity distribution is
essentially produced by the nearest neighbor as for a L\'evy law.

Note also that the typical velocity due to the nearest neighbor is
$V_{n.n}\sim \gamma/2\pi d$ where $d$ is the inter-vortex distance
(\ref{mg24}). Hence $V_{n.n}^{2}\sim n\gamma^{2}/\pi^{2}$. Comparing
this result with Eq. (\ref{mg18}), we see that the velocity due to all
the vortices is, { up to a logarithmic factor}, of the same order as
the velocity due to the nearest neighbor. We can say, in some sense,
that the velocity fluctuation is {marginally} dominated by the
nearest neighbor and that collective effects are responsible for
logarithmic corrections. In fact, we can show \cite{eff} that the
``effective velocity'' created by a vortex at a distance $r$ from the
point under consideration is given by
\begin{equation}
V_{eff}={\gamma\over 2\pi r}{1\over 1+r/\Lambda},
\label{mg29}
\end{equation}
where $\Lambda\sim d$ is of the order of the inter-vortex
distance. For $r\gg d$, the ``effective'' velocity decays as $r^{-2}$
instead of the ordinary $r^{-1}$ law recovered for $r\ll d$. This
result indicates that, in a statistical sense, the velocity produced
by a vortex is shielded by ``cooperative'' effects.

\subsection{The speed of fluctuations}
\label{sec_speed}

The function $W({\bf V})$  does not
provide us  with all the necessary information concerning the fluctuations of
${\bf V}$. An important aspect of the problem concerns the  {\it speed of
fluctuations}, i.e. the typical duration $T({ V})$ of  the velocity fluctuation
${\bf V}$. This requires the knowledge of  the bivariate probability $W({\bf
V},{\bf A}) d^{2}{\bf V}d^{2}{\bf A}$  to measure simultaneously  a velocity
${\bf V}$ with a rate of change
\begin{equation}
{\bf A}={d{\bf V}\over dt}=\sum_{i=1}^{N}{\mb\psi}_{i},
\label{sf1}
\end{equation}
\begin{equation}
{\mb{\psi}}_{i}=-{\gamma\over 2\pi}\biggl ( {{\bf v}_{\perp i}\over
r_{i}^{2}}-{2({\bf r}_{i} {\bf v}_{i}){\bf r}_{\perp i}\over
r_{i}^{4}}\biggr ),
\label{sf2}
\end{equation}
where ${\bf v}_{i}={d{\bf r}_{i}/ dt}$ is the velocity of vortex $i$. Then,
the duration $T({ V})$ can  be estimated by the formula
\begin{equation}
T({V})={|{\bf V}|\over\sqrt{\langle A^{2}\rangle_{\bf{V}}}},
\label{sf3}
\end{equation}
where
\begin{equation}
\langle {A^{2}}\rangle_{\bf V}={\int W({\bf V},{\bf A})A^{2}d^{2}{\bf A}\over
W({\bf V})},
\label{sf4}
\end{equation}
is the mean square acceleration associated with a velocity fluctuation   ${\bf
V}$. 

To determine the speed of fluctuations, we need therefore to calculate
the bivariate probability $W_{N}({\bf V},{\bf A})$ to measure
simultaneously a velocity ${\bf V}$ with a rate of change ${\bf
A}={d{\bf V}/dt}$. According to Eqs. (\ref{mg1}), (\ref{sf1}) and
(\ref{sf2}), ${\bf V}$ and ${\bf A}$ are the sum of $N$ random variables
${\bf \Phi}_{i}$ and ${\mb \psi}_{i}$ depending on the positions ${\bf
r}_{i}$ and on the velocities ${\bf v}_{i}$ of the point vortices. Unlike
material particles, the variables $\lbrace {\bf r}_{i}$, ${\bf
v}_{i}\rbrace$, for different $i$'s, are not independent because the
velocities of the vortices are determined by the configuration
$\lbrace {\bf r}_{i}\rbrace$ of the system as a whole (see Eq. (\ref{mg1})). 
To be able to solve the problem analytically, we shall make a
{\it decorrelation approximation} and treat $\lbrace {\bf
r}_{i}$, ${\bf v}_{i}\rbrace$ ($i=1,...,N$) as independent
variables governed by the distribution 
\begin{equation}
\tau({\bf r},{\bf v})={1\over \pi R^{2}}\times {4\over n\gamma^{2} \ln
N}e^{-{4\pi\over  n\gamma^{2}\ln N}v^{2}},
\label{sf5}
\end{equation}
resulting from Eqs. (\ref{mg8}) and (\ref{mg15}).  It is remarkable
that the distribution (\ref{sf5}) is formally equivalent to the
Maxwell-Boltzmann statistics of material particles at
equilibrium. Owing to this analogy, we can interpret the variance
\begin{equation}
\overline{v^{2}}={n\gamma^{2}\over 4\pi}\ln N,
\label{sf6}
\end{equation}    
as a kind of kinetic ``temperature'' of the point vortices. It should
not be confused with the temperature $\beta$ introduced in
Sec. \ref{sec_statmech} and measuring the {\it clustering} of the
vortices (in the present case, $\beta=0$ since the vortices are
uniformly distributed in average).

When this decorrelation hypothesis is implemented, a straightforward
generalization of the method used in Sec. \ref{sec_marginal} yields 
\begin{equation}
W_{N}({\bf V},{\bf A})={1\over 16\pi^{4}}\int A_{N}({\mb \rho},{\mb \sigma})
e^{-i ({\mb \rho}{\bf V}+{\mb \sigma}{\bf A})}d^{2}{\mb \rho}d^{2}{\mb \sigma},
\label{sf7}
\end{equation}    
with 
\begin{equation}
A_{N}({\mb \rho},{\mb \sigma})=\biggl (\int_{|{\bf r}|=0}^{R}\int_{|{\bf
v}|=0}^{+\infty} \tau({\bf r},{\bf v})e^{i({\mb \rho}{\bf \Phi}+{\mb
\sigma}{\mb\psi})}d^{2}{\bf r}d^{2}{\bf v}\biggr )^{N},
\label{sf8}
\end{equation}    
where we have defined
\begin{equation}
{\bf \Phi}=-{\gamma\over 2\pi}{{\bf r}_{\perp}\over r^{2}},
\label{sf9}
\end{equation}  
\begin{equation}
{\mb \psi}=-{\gamma\over 2\pi}\biggl ( {{\bf v}_{\perp }\over r_{}^{2}}-{2({\bf
r}{\bf v}){\bf r}_{\perp }\over r^{4}}\biggr ).
\label{sf10}
\end{equation}
Using an integration by parts, the conditional moment of the acceleration for a given velocity fluctuation ${\bf V}$ can be expressed as \cite{csire1}:
\begin{equation}
W({\bf V})\langle A^{2}\rangle_{\bf V}=-{1\over \pi^{2}}\int {\partial A\over\partial (\sigma^{2})}({\mb\rho},{\bf 0})e^{-i{\mb\rho}{\bf V}}
d^{2}{\mb\rho}.
\label{sf11}
\end{equation}
We therefore need to Taylor expand the function $A({\mb\rho},{\mb
\sigma})$ for $|{\mb\sigma}|\rightarrow 0$ and carry out the
integration in Eq. (\ref{sf11}). The calculation is relatively tricky
but can be done analytically \cite{csire1}. Substituting the resulting
expression for $\langle A^{2}\rangle_{\bf V}$ in Eq. (\ref{sf3}), we
obtain the following estimate for the speed of fluctuations:
\begin{equation}
T({ V})= {4\sqrt{\pi}V\over n^{3/2}\gamma^{2}\ln N} e^{-{2\pi\over
n{\gamma^{2}}\ln N}V^{2}}\quad (V\ll V_{crit}(N)),    
\label{sf12}
\end{equation}
\begin{equation}
T({V})= {1\over \sqrt{\pi n \ln N}}{1\over V}  \quad (V\gg V_{crit}(N)).    
\label{sf13}
\end{equation}
For weak and large fluctuations, $T(V)$ decreases to zero like $V$ and
$V^{-1}$ respectively.  These asymptotic behaviors have a clear
physical meaning in the nearest neighbor approximation.  When
$r={\gamma/ 2\pi V}$ is small, corresponding to large velocities, it
is highly improbable that another vortex will enter the disk of radius
$r$ before long. By contrast, on a short time scale $T\sim
{r/\overline{v}}\sim (\gamma/\overline{v})V^{-1}$, the vortex will have
left the disk. When $r={\gamma/ 2\pi V}$ is large, corresponding to
small velocities, the probability that the vortex will remain alone in
the disk is low. The characteristic time before another vortex enters
the disk varies like the inverse of the number of vortices expected to
be present in the disk, i.e.  $T\sim {(r/\overline{v})}{1/ n\pi
r^{2}}\sim {(1/ n \gamma
\overline{v})}V$. The demarcation between weak and strong 
fluctuations corresponds to $V\sim \gamma n^{1/2}$, i.e. to the
velocity produced by a vortex distant $n^{-1/2}$ from the point under
consideration. These asymptotic behaviors are also consistent with the
theory of Smoluchowski concerning the persistence of fluctuations (see discussion in \cite{csire1}).

The average duration of a velocity fluctuation is defined by
\begin{equation}
\langle T\rangle =\int_{0}^{+\infty} T(V)W(V)2\pi V dV.
\label{sf14}
\end{equation}
To leading order in $\ln N$, we obtain 
\begin{equation}
\langle T\rangle= {4\over 3}\biggl ({\pi\over 6}\biggr )^{1/2} {1\over
n\gamma\sqrt{\ln N}}.
\label{sf15}
\end{equation}
This formula shows that the typical duration of a velocity
fluctuation scales like 
\begin{equation}
T_{typ}\sim {1\over n\gamma\sqrt{\ln N}}.
\label{sf16}
\end{equation}
This corresponds to the typical time needed by a vortex moving with a
typical velocity $\langle V^{2}\rangle^{1/2}$ (given by Eq. (\ref{mg18})) to
cross the interparticle distance $d\sim n^{-1/2}$, as expected from general physical grounds \cite{weiss}.

\subsection{The diffusion coefficient}
\label{sec_diffusion}

According to the previous discussion, we can characterize the fluctuations of
the velocity of a point vortex by two functions: a
function $W({\bf V})$ which governs the occurrence of the velocity  ${\bf V}$
and a function $T(V)$ which determines the typical time during which the vortex
moves with this velocity. Since the velocity fluctuates on a typical time
$T_{typ}=d/ \sqrt{\langle V^{2}\rangle}$ which is much smaller than the
dynamical time $t_{D}=R/ \sqrt{\langle V^{2}\rangle}$ needed by the
vortex to cross the entire domain, the motion of the vortex will be essentially
stochastic. If we denote by $P({\bf r},t)$ the probability density that the
particle be found in ${\bf r}$ at time $t$, then $P({\bf r},t)$ will satisfy the
diffusion equation 
\begin{equation}
{\partial P\over\partial t}=D\Delta P.
\label{diff1}
\end{equation}
If the particle is at ${\bf r}={\bf r}_{0}$ at time $t=0$, the solution of
Eq. (\ref{diff1}) is clearly 
\begin{equation}
P({\bf r},t|{\bf r}_{0})={1\over 4\pi D t}e^{-{|{\bf r}-{\bf r}_{0}|^{2}\over 4
D t}},
\label{diff2}
\end{equation}
where $D$ is the diffusion coefficient. The mean square displacement
that the particle is expected to suffer during an interval of time
$\Delta t$ large with respect to the fluctuation time $T_{typ}$, is
\begin{equation}
\langle (\Delta {\bf r})^{2}\rangle =4 D \Delta t.
\label{diff3}
\end{equation}
We can obtain another expression for $\langle (\Delta {\bf
r})^{2}\rangle$ in terms of the functions $W({\bf V})$ and $T(V)$
defined in the previous sections.  Indeed, dividing the interval
\begin{equation}
\Delta {\bf r}=\int_{t}^{t+\Delta t}{\bf V}(t') dt',
\label{diff4}
\end{equation}
into a succession of discrete increments in position with amount
$T(V_{i}){\bf V}_{i}$, we readily establish that
\begin{equation}
\langle (\Delta {\bf r})^{2}\rangle = \langle T(V)V^{2}\rangle\Delta t.
\label{diff5}
\end{equation}
Combining Eqs. (\ref{diff3}) and (\ref{diff5}) we obtain an alternative
expression for the diffusion coefficient in the form
\begin{equation}
D={1\over 4}\int T(V)W({\bf V})V^{2} d^{2}{\bf V}.
\label{diff6}
\end{equation}
Substituting for $T(V)$ and $W({\bf V})$ in the foregoing expression, we obtain
to leading order in $\ln N$ \cite{csire1},
\begin{equation}
D={1\over 72}\biggl ({6\over\pi}\biggr )^{1/2}\gamma\sqrt{\ln N}. 
\label{diff7}
\end{equation} 
We should not give too much credit to the numerical factor appearing
in Eq. (\ref{diff7}) since the definition (\ref{sf3}) of $T(V)$
is just an order of magnitude. Note that the scaling form of $D$ is
consistent with the expression
\begin{equation}
D\sim T_{typ}\langle V^{2}\rangle \sim\gamma\sqrt{\ln N}, 
\label{diff8}
\end{equation}
that one would expect on general physical grounds.

\subsection{Application to 2D decaying turbulence}
\label{sec_anomalous}

The previous results have some direct applications to the context of
2D decaying turbulence. The relaxation of 2D decaying turbulence is a
three-stage process: during an initial transient period, the fluid
organizes itself from random fluctuations and a population of coherent
vortices emerges. Then, when two like-sign vortices come into contact
they merge and form a bigger structure. As time goes on, the vortex
number decreases and their average size increases, in a process
reminiscent of a coarsening stage. Finally, when only one dipole is
left, it decays diffusively due to inherent viscosity. Direct
numerical simulations \cite{pomeau} and experiments \cite{hansen}
show that the typical core vorticity $\omega$ remains constant during
the course of the evolution and that the density decreases like $n\sim
t^{-\xi}$ with $\xi\approx 0.7$. As the energy $E\sim \int {\bf
u}^{2}d^{2}{\bf r}\sim n\omega^{2}a^{4}\sim n a^{4}$ is conserved
throughout the evolution, the typical vortex radius increases like
$a\sim t^{\xi/4}$.

In the {\it punctuated Hamiltonian model} \cite{weiss2}, the flow is
approximated by a collection of vortices with constant vorticity
$\pm\omega$ (and circulation $\gamma\sim
\pm\omega a_{i}^{2}$) whose centers follow the Hamiltonian 
dynamics (\ref{pv3})(\ref{pv4}). When two like-sign vortices with
radii $a_{1}$ and $a_{2}$ enter in collision, the resulting vortex
keeps the same vorticity and takes a radius $a$ such that
$a^{4}=a_{1}^{4}+a_{2}^{4}$ which ensures the conservation of energy
(as discussed above). This effective model reproduces the results of
the numerical simulations and experiments, with again $\xi\simeq
0.7$. Since the average distance between vortices, of order $d\sim
t^{\xi/2}$, increases more rapidly than their size, the point vortex
model should provide increasing accuracy. We emphasize, however, that
in the above mentioned studies the density only changes by a factor of
order $4\sim 5$ between the initial time and the final time so that
the scaling exponent is measured on less than one decade.

The punctuated Hamiltonian model has been re-investigated recently by
Sire \& Chavanis \cite{sirec} using a renormalization group procedure
which allows for much longer time simulations that could otherwise be
achieved. It is found that the scaling regime is achieved for very
late times and is characterized by a decay exponent $\xi=1$ (an
effective exponent $\xi\simeq 0.7$ is recovered for shorter times). In
addition, the decay of the total area occupied by the vortices results
in a physical process by which merging occurs principally via
$3$-vortex collisions involving vortices of different sign. A simple
kinetic theory based on an effective $3$-vortex interaction returns
this value $\xi=1$. These theoretical results tend to be confirmed by
recent direct numerical simulations \cite{laval}.

During the decay, the vortices diffuse with
a coefficient $D\sim \gamma$ given by Eq. (\ref{diff7}), where
$\gamma\sim \omega a^{2}$ is their circulation (we ignore here logarithmic
corrections). If the diffusion coefficient were constant, then the
dispersion of the vortices
\begin{equation}
\langle r^{2} \rangle\sim D t,
\label{dec1}
\end{equation}
would increase linearly with time as in ordinary  Brownian motion. However,
since $D$ varies with time according to
\begin{equation}
D\sim \omega a^{2} \sim t^{\xi/2},
\label{dec2}
\end{equation}
we expect \index{Anomalous diffusion} anomalous diffusion, i.e. 
\begin{equation}
\langle r^{2} \rangle\sim t^{\nu},
\label{dec3}
\end{equation}
with $\nu\neq 1$. Substituting Eq. (\ref{dec2}) in Eq. (\ref{dec1}),
we obtain the following relation between $\nu$  and $\xi$:
\begin{equation}
\nu =1+{\xi\over 2}.
\label{dec4}
\end{equation}     
This formula is in perfect agreement with long time numerical
simulations ($\nu=3/2$ for $\xi=1$) \cite{sirec} and experiments
($\nu_{exp}\sim 1.3-1.4$ in the regime where $\xi=0.7$) \cite{hansen}. This
hyperdiffusive behavior can be interpreted in terms of L\'evy flights
\cite{sirec}, with a large time flight distribution $P(\tau)\sim
\tau^{-\mu}$ with $\mu=3-\xi/2$ in agreement with experiments ($\mu_{exp}\sim 2.6\pm 0.2$ for $\xi=0.7$)\cite{hansen}.

\subsection{The spatial correlations in the velocities arising from a random distribution of point vortices}
\label{sec_spatialcorr}

The stochastic approach described in Sec. \ref{sec_marginal} can be
generalized to obtain exact results concerning the spatial
correlations of the velocity fluctuations \cite{csire2}. First of all, the
distribution of the velocity increments between two neighboring points
is given by the 2D Cauchy law \cite{jimenez,min,csire2}:
\begin{equation}
W(\delta {\bf V})={2\over \pi n^{2}\gamma^{2}|\delta {\bf r}|^{2}}\biggl (1+{4|\delta {\bf V}|^{2}\over n^{2}\gamma^{2}|\delta {\bf r}|^{2}}\biggr )^{-3/2},
\label{sc1}
\end{equation}  
which is a particular L\'evy law.  It is also possible to determine an
analytical expression for the conditional moment $\langle \delta {\bf
V}\rangle_{\bf V}$. We find:
\begin{equation}
\langle \delta {\bf V}\rangle_{\bf V}=-{n\gamma\over 2}B\Biggl ({4\pi V^{2}\over n\gamma^{2}\ln N}\biggr )\biggl\lbrace \delta {\bf r}_{\perp}+2{({\bf V}_{\perp}\delta {\bf r})\over V^{2}}{\bf V}\biggr \rbrace \quad (V\ll V_{crit}(N)),
\label{sc1a}
\end{equation} 
\begin{equation}
\langle \delta {\bf V}\rangle_{\bf V}=-{2\pi\over \gamma} V^{2}\biggl\lbrace \delta {\bf r}_{\perp}+2{({\bf V}_{\perp}\delta {\bf r})\over V^{2}}{\bf V}\biggr \rbrace \quad (V\gg V_{crit}(N)),
\label{sc1b}
\end{equation} 
where $B(x)$ denotes the function
\begin{equation}
B(x)={1\over x}(e^{x}-1-x).
\label{sc1c}
\end{equation}

On the other hand, the spatial auto-correlation function of the
velocity between two points separated by an arbitrary distance is
given by \cite{csire2}:
\begin{equation}
\langle {\bf V}_{0}{\bf V}_{1}\rangle={n\gamma^{2}\over 2\pi}\ln\biggl ({R\over r_{1}}\biggr ).
\label{sc2}
\end{equation}  
This leads to an energy spectrum
\begin{equation}
E(k)={n\gamma^{2}\over 4\pi k}(1-J_{0}(kR)),
\label{sc3}
\end{equation} 
which reduces to Novikov's result $E(k)= n\gamma^{2}/4\pi k$ for $k\rightarrow +\infty$ \cite{novikov}. We observe that the velocity correlation function (\ref{sc2}) diverges logarithmically as $r_{1}\rightarrow 0$. This is consistent with the divergence of the variance of the velocity distribution (\ref{mg15})-(\ref{mg17}). Therefore, it is more proper to characterize the spatial correlations of the velocity by the function $\langle V_{1\|}\rangle=\langle {\bf V}_{0}{\bf V}_{1}/|{\bf V}_{0}|\rangle$ which remains finite as $r_{1}\rightarrow 0$. Its evaluation is more complex and leads to the result
\begin{eqnarray}
\langle V_{1\|}\rangle=\langle |{\bf V}_{0}|\rangle-\biggl ({n\gamma^{2}\over \pi\ln N}\biggr )^{1/2}\int_{0}^{+\infty}{J_{1}(z)\over z^{2}}\Gamma_{s^{2}z^{2}}\biggl ({1\over 2}\biggr )dz,
\label{sc4}
\end{eqnarray} 
where 
\begin{eqnarray}
\Gamma_{x}(p+1)=\int_{0}^{x}e^{-t}t^{p}dt,
\label{sc4a}
\end{eqnarray} 
is the incomplete $\Gamma$-function and 
\begin{eqnarray}
\langle |{\bf V}_{0}|\rangle=\biggl ({n\gamma^{2}\over 16}\ln N\biggr )^{1/2}, \qquad s=\biggl ({\pi n\ln N\over 4}\biggr )^{1/2}r_{1}.
\label{sc5}
\end{eqnarray} 
Similarly, the correlation function $K(r_{1})=\langle {\bf V}_{0}{\bf V}_{1}/{V}_{0}^{2}\rangle$ is given by
\begin{eqnarray}
K(s)=1-{4s^{2}\over\ln N}\biggl (1-e^{-1/4s^{2}}+{1\over 4s^{2}}E_{1}\biggl ({1\over 4s^{2}}\biggr )\biggr ),
\label{sc5a}
\end{eqnarray} 
where $E_{1}(z)$ denotes the exponential integral
\begin{eqnarray}
E_{1}(z)=\int_{z}^{+\infty}{e^{-t}\over t}dt.
\label{sc5b}
\end{eqnarray} 
Other results concerning the characterization of the spatial velocity correlations can be found in \cite{csire2}.

\subsection{Statistics of fluctuations of the gravitational field}
\label{sec_gravfluc}

The stochastic approach developed in the preceding sections was
inspired by the famous work of Chandrasekhar \& von Neumann
\cite{c0,cn1,cn2,c3,c4} concerning the fluctuations of the
gravitational field \index{Astrophysics} created by a random
distribution of stars. Let us consider a collection of $N$ stars with
mass $m$ randomly distributed in a sphere of radius $R$ with a uniform
density $n$ in average. The force by unit of mass created at the
center $O$ of the domain is
\begin{equation}
{\bf F}=\sum_{i=1}^{N}{\mb\Phi}_{i},\qquad {\mb\Phi}_{i}=-{Gm \over r_{i}^{3}}{\bf r}_{i}.
\label{gf1}
\end{equation}
As before, the problem consists in determining the distribution of a
sum of random variables.  In the present case, the variance of the
force created by one star diverges algebraically
\begin{equation}
\langle\Phi^{2}\rangle\sim {3\over 4\pi R^{3}}\int\biggl ({Gm\over r^{2}}\biggr )^{2}4\pi r^{2} dr\sim {1\over r},
\label{gf2}
\end{equation} 
so that the distribution of the total force is a L\'evy law. This particular L\'evy law is known as the {\it Holtsmark distribution} since it was first determined by Holtsmark in the context of the electric field created by a gas of simple ions \cite{holtsmark}. In the thermodynamic limit  $N,R\rightarrow +\infty$ with $n={3N\over 4\pi R^{3}}$ finite, the distribution of ${\bf F}$ can be expressed by the Fourier transform
\begin{eqnarray}
W({\bf F})={1\over 2\pi^{2}F}\int_{0}^{+\infty}k\sin (kF)e^{-ak^{3/2}}dk,
\label{gf3}
\end{eqnarray}
with 
\begin{eqnarray}
  a={4\over 15}(2\pi G)^{3/2}nm^{1/2}. 
\label{gf4}
\end{eqnarray}
It has the asymptotic behavior
\begin{equation}
W({\bf F})\sim {1\over 2}G^{3/2}m^{1/2}nF^{-9/2}\qquad (F\rightarrow +\infty),
\label{gf5}
\end{equation}
which can be shown to coincide with the distribution due to the
nearest neighbor. The typical force due to the
nearest neighbor $F_{n.n}\sim Gm/d^{2}\sim Gmn^{2/3}$ is precisely of
the same order as the average value of the force due to all the stars
\begin{equation}
\langle F\rangle=4\Gamma\biggl ({1\over 3}\biggr )\biggl ({8\sqrt{2}\over 15}\biggr )^{2/3}Gmn^{2/3},
\label{gf6}
\end{equation}
determined from Eq. (\ref{gf3}). This shows that only stars close to the star under consideration determine the fluctuations of the gravitational field. In fact, it is possible to show that the ``effective'' force created by a star at  distance $r$ from the star under consideration is \cite{agekyan}:
\begin{equation}
F_{eff}={Gm\over r^{2}}{1\over 1+{r^{2}/\Lambda^{2}}},
\label{gf7}
\end{equation}
where $\Lambda\sim d$ is of the order of the interparticle distance.
For weak separations, one has $F_{eff}\rightarrow Gm/r^{2}$ but for
large separations $r\gg d$, the effect of individual stars compensate
each other and the resulting force is reduced by a factor
$(r/d)^{2}$. This corresponds to a shielding of the interaction, in a
statistical sense, due to collective effects.

Chandrasekhar \& von Neumann have used this stochastic model to
determine the speed of fluctuations $T(F)$ \cite{cn1}, the diffusion
coefficient of stars (a calculation completed by Kandrup \cite{krev})
and the spatial \cite{cn2,c3} and temporal \cite{c4} correlations of
the gravitational field etc... There is a complete parallel with the
results obtained by Chavanis \& Sire \cite{csire1,csire2,eff} for
point vortices and this is another manifestation of the deep formal
analogy between the two systems.

\section{Relaxation of a point vortex in a thermal bath}
\label{sec_bath1}

\subsection{Analogy with Brownian motion}
\label{sec_brownian}

We would like now to develop a kinetic theory of point vortices in
order to describe their relaxation \index{Non-equilibrium phenomena}
towards equilibrium. It has to be noted that an equilibrium state can
be achieved in very different ways so that the kinetic theory of
vortices is not unique and depends on the situation contemplated (see
Sec. \ref{sec_violrelax}). In order to develop an intuition on the
problem, we first propose a naive kinetic theory based on an analogy
with Brownian theory \cite{drift,kin2}. The starting point of this
analogy is to realize that the velocity of a point vortex can be
decomposed in two terms: a smoothly varying function of position and
time $\langle {\bf V}\rangle({\bf r},t)$ and a function $\mb{\cal
V}(t)$ taking into account the ``granularity" of the system. The total
velocity of a point vortex can therefore be written:
\begin{equation}
{\bf V}=\langle {\bf V}\rangle ({\bf r},t)+\mb{\cal V}(t).
\label{bm1}
\end{equation}
The velocity $\langle {\bf V}\rangle ({\bf r},t)$ reflects the influence of the
system as a {\it whole} and is generated by the mean vorticity
$\langle\omega\rangle({\bf r},t)$ according to the Biot \& Savart formula
\begin{equation}
\langle {\bf V}\rangle ({\bf r},t)=-{1\over 2\pi}{\bf z}\times\int  {{\bf
r}'-{\bf r}\over |{\bf r}'-{\bf r}|^{2}}\langle\omega\rangle({\bf
r}',t)d^{2}{\bf r}'.
\label{bm2}
\end{equation} 
The fluctuation $\mb{\cal V}(t)$ arises from the difference between
the exact distribution $\omega_{exact}({\bf r},t)$ of point vortices
given by Eq. (\ref{pv1}) and their ``smoothed-out'' distribution
$\langle\omega\rangle({\bf r},t)=n\gamma$. Indeed, if we consider a
surface element $\sigma$, the number of point vortices actually
present in this area will fluctuate around the mean number
$n\sigma$. These fluctuations will be governed by a Poisson
distribution with variance $n\sigma$. On account of these
fluctuations, the velocity of a vortex will depart from its mean field
value $\langle {\bf V}\rangle$. The velocity fluctuation $\cal V$, of
order ${\gamma/ d}$ (where $d\sim n^{-1/2}$ is the inter-vortex
distance), is much smaller than the average velocity $\langle
V\rangle$, of order $n\gamma R$ (where $R$ is the domain size), but
this term has a cumulative effect which gives rise to a process of
diffusion. It makes sense therefore to introduce a stochastic
description of the vortex motion such as that for colloidal
suspensions in a liquid \cite{kramers,chandrabrownien} or stars in
globular clusters \cite{chandraS}. However, contrary to the ideal
Brownian motion, point vortex systems have relatively long correlation
times so that ${\cal V}(t)$ is {\it not} a white noise. This makes the
study much more complicated than usual and the technical developments of
Sec. \ref{sec_kin} are required. However, in order to gain some
physical insights into the problem, we shall ignore this difficulty
for the moment and describe the system by traditional stochastic
processes.

According to Eq. (\ref{bm1}), we would naively expect that the
evolution of the density probability $P({\bf r},t)$ would be governed
by a diffusion equation of the form
\begin{equation}
{\partial P\over \partial t}+\langle {\bf V}\rangle\nabla P=D\Delta P,
\label{bm3}
\end{equation} 
where $D$ is the diffusion coefficient. This would in fact be the case
for a passive particle having no retroaction on the vortices or when
the distribution of vortices is uniform in average like in
Sec. \ref{sec_fluc}. However, this diffusion equation cannot be valid when the
system is inhomogeneous. Indeed, Eq. (\ref{bm3}) does not
converge towards the Boltzmann distribution (\ref{me5}) when
$t\rightarrow +\infty$. It seems therefore that a term is missing to
act against the diffusion.

This problem is similar to the one encountered by Chandrasekhar in his
stochastic approach of stellar dynamics
\cite{chandrafric}. Chandrasekhar solved the problem by introducing a
{\it dynamical friction} in order to compensate for the effect of
diffusion. The occurence of this frictional force is a manifestation
of the ``fluctuation-dissipation" theorem in statistical mechanics.
In the present context, the dynamical friction is replaced by a {\it
systematic drift} of the vortices \cite{drift}. This drift appears
when the distribution of vorticity is inhomogeneous. It can be
understood in terms of a polarization process like in a phenomenon of
induction.  In Sec. \ref{sec_drift}, we shall derive the drift term
directly from the Liouville equation by using a linear response
theory. For the moment, we introduce this term {\it by hands} and
rewrite the decomposition (\ref{bm1}) in the form
\begin{equation}
{\bf V}=\langle {\bf V}\rangle-\xi\nabla\psi+{\mb{\cal V}}(t),
\label{bm6}
\end{equation}
where $\xi$ is the drift coefficient.  Equation (\ref{bm6}) must be
viewed as a stochastic equation analogous to the Langevin equation in
the ordinary Brownian theory. The corresponding Fokker-Planck equation
can be written \cite{drift}:
\begin{equation}
{\partial P\over\partial t}+\langle {\bf V}\rangle\nabla P=\nabla(D\nabla P+\xi
P\nabla\psi).
\label{bm7}
\end{equation} 
The physical interpretation of each term is straightforward. The left
hand side (which can be written $dP/dt$) is an advection term due to
the smooth mean field velocity $\langle {\bf V}\rangle$.  The right
hand side can be written as the divergence of a current $-\nabla {\bf
J}$ and is the sum of two terms: the first term is a diffusion due to
the erratic motion of the vortices caused by the fluctuations of the
velocity (see Sec. \ref{sec_fluc}) and the second term accounts for
the systematic drift of the vortices due to the inhomogeneity of the
vortex cloud. At equilibrium, the drift precisely balances random
scatterings and the Boltzmann distribution (\ref{me5}) is settled. More
precisely, the condition that the Boltzmann distribution
satisfies Eq. (\ref{bm7}) identically requires that $D$ and $\xi$
be related according to the relation
\begin{equation}
\xi= D\beta\gamma,
\label{bm8}
\end{equation} 
which is a generalization of the Einstein formula to the case of point
vortices. It is remarkable that we can obtain such a general relation
without, at any point, being required to analyze the mechanism of
``collisions''. A more rigorous justification of this relation will be
given in the next sections in which the diffusion coefficient and the
drift term are calculated explicitly. Note that the diffusion current
can be written ${\bf J}=\chi\nabla\alpha$ where $\alpha=\ln\langle
\omega\rangle+\beta\gamma\psi$ is a ``generalized potential'' which is
uniform at equilibrium, see Eq. (\ref{me5}). Therefore, the
Fokker-Planck equation (\ref{bm7}) is consistent with the linear
thermodynamics of Onsager which relates the diffusion currents to the
gradients of generalized potentials. The Fokker-Planck equation
(\ref{bm7}) can also be obtained from a variational formulation (the
so-called Maximum Entropy Production Principle): it can be shown to
maximize the rate of entropy production $\dot S$ under the constraints
brought by the dynamics
\cite{kin2}.

\subsection{Diffusion coefficient of point vortices in an inhomogeneous medium}
\label{sec_diff}

In this section, we determine the value of the diffusion coefficient
$D$ which enters in the Fokker-Planck equation (\ref{bm7}). The
essential difference with the calculation of Sec. \ref{sec_fluc} is
that the vorticity distribution is now {\it inhomogeneous} so that
there exists a shear in the system. This shear considerably modifies
the expression of the diffusion coefficient \cite{drift,kin2}. An
explicit expression for $D$ can be obtained in a ``thermal bath
approximation'' which is valid if we are sufficiently close to
equilibrium. To be specific, we consider a system of $N$ point
vortices at statistical equilibrium with an inverse temperature
$\beta_{eq}$. These ``field vortices'' form the thermal bath. We now
introduce a ``test vortex'' in the system an study its diffusion in
the sea of ``field vortices''.  For simplicity, we restrict ourselves
to simple equilibrium flows which are either unidirectional or
axisymmetric.

In the case of unidirectional flows in the $x$-direction, we define the diffusion coefficient in the $y$ direction by
\begin{equation}
D={\rm lim}_{t\rightarrow +\infty}{\langle (\Delta y)^{2}\rangle\over 2t}\quad {\rm with}\quad  \Delta y=\int_{0}^{t} V_{y}(t')dt',
\label{dc1}
\end{equation} 
where ${\bf V}(t)$ is given by Eq. (\ref{pv2}). It is possible to put the diffusion coefficient in the form of a Kubo formula \cite{kin2}:
\begin{equation}
D=\int_{0}^{+\infty} \langle V_{y}(t)V_{y}(t-\tau)\rangle d\tau,
\label{dc2}
\end{equation} 
where the quantity in bracket is the velocity auto-correlation function at different times. Using Eq. (\ref{pv2}), we have explicitly
\begin{equation}
D=N\int_{0}^{+\infty}d\tau\int d^{2}{\bf r}_{1}V_{y}(1\rightarrow 0,t)V_{y}(1\rightarrow 0,t-\tau)P_{eq}(y_{1}).
\label{dc3}
\end{equation} 
For axisymmetric equilibrium flows, one has similarly 
\begin{equation}
D={\rm lim}_{t\rightarrow +\infty}{\langle (\Delta r)^{2}\rangle\over 2t}\quad {\rm with} \quad \Delta r=\int_{0}^{t} V_{r}(t')dt',
\label{dc4}
\end{equation} 
and 
\begin{equation}
D=N\int_{0}^{+\infty}d\tau\int d^{2}{\bf r}_{1} V_{r(t)}(1\rightarrow 0,t)V_{r(t-\tau)}(1\rightarrow 0,t-\tau)P_{eq}(r_{1}).
\label{dc5}
\end{equation}
The velocity auto-correlation function is calculated in Appendix A
under the assumption that between $t$ and $t-\tau$ the point vortices
follow the equilibrium streamlines, which is valid for sufficiently
strong shears. It is found that the auto-correlation function
decreases like $t^{-2}$ at large times so that the diffusion
coefficient is well-defined. It must be stressed, however, that the
decorrelation is slow so that the fluctuations cannot be described by
a white noise process, as indicated previously \footnote{The problem is even more severe in stellar dynamics where the temporal correlation function of the gravitational force decreases like $t^{-1}$, responsible for logarithmic divergences in the diffusion coefficient \cite{lee}.}. Using the results of
Appendix A, the final expression for the diffusion coefficient in
the presence of a shear can be put in the form
\cite{drift,kin2}:
\begin{equation}
D={\gamma\over 8}{1\over |\Sigma ({\bf r})|}\ln N\langle\omega\rangle_{eq},
\label{dc6}
\end{equation} 
where $|\Sigma ({\bf r})|$ is the local shear of the flow:
$|\Sigma|=-d/dy\langle V\rangle_{eq}(y)$ for unidirectional flows and
$|\Sigma|=r{d/dr}({\langle V\rangle_{eq}(r)/r})$ for axisymmetric
flows. In Sec. \ref{sec_bath}, we shall calculate the diffusion coefficient in
another manner and extend formula (\ref{dc6}) to more general
equilibrium flows with $|\Sigma ({\bf r})|=2\sqrt{-{\rm det}(\Sigma)}$
where $\Sigma$ is the stress tensor.  The expression (\ref{dc6}) for
the diffusion coefficient can be written in the general form (see Sec. \ref{sec_fluc}):
\begin{equation}
D={1\over 4}\tau \langle V^{2}\rangle={\gamma\tau\over 16\pi}\ln N\langle\omega\rangle_{eq},
\label{dc6bis}
\end{equation} 
with a correlation time $\tau=2\pi/|\Sigma({\bf r})|$. Physically, it
corresponds to the time needed by two vortices with relative velocity
$\Sigma d$ (where $d$ is the inter-vortex distance) to be stretched by
the shear on a distance $\sim d$. Of course, when there is no shear,
our approximations break down and the diffusion coefficient is given
by Eq. (\ref{diff7}) which involves a correlation time $\tau\sim
1/\langle\omega\rangle({\bf r})\sqrt{\ln N}$ determined by the
dispersion of the vortices \cite{csire1}. These two results correspond
to a limit of ``strong shear'' and ``weak shear''
respectively. Clearly, a general formula for $D$ should take into
account simultaneously the effect of the shear and the dispersion of
the vortices.

\subsection{Systematic drift experienced by a point vortex in an inhomogeneous  medium: linear response theory}
\label{sec_drift}

In Sec. \ref{sec_brownian}, we have argued on the basis of general
considerations that when the vorticity distribution is inhomogeneous,
a point vortex should experience a systematic drift in addition to its
diffusive motion. In this section, we justify this drift by a linear
response theory, starting directly from the Liouville equation
\cite{drift}. Consider a collection of $N$ point vortices at
statistical equilibrium. In the mean-field limit, the N-particle
distribution function $\mu_{eq}(\lbrace {\bf r}_{k}\rbrace)$ can be
approximated by a product of $N$ one-particle distribution functions
$P_{eq}$, each of which at equilibrium with the same inverse
temperature $\beta_{eq}$:
\begin{equation}
\mu_{eq}(\lbrace{\bf
r}_{k}\rbrace)=\prod_{k=1}^{N}P_{eq}({\bf r}_{k})=\prod_{k=1}^{N}
A e^{ -\beta_{eq} \gamma\psi_{eq}({\bf r}_{k}) }.
\label{drift1}
\end{equation} 
This distribution is stationary, in a statistical sense, since it
corresponds to a maximum entropy state. The introduction of an
additional point vortex, the ``test vortex'', will modify this
equilibrium state. The distribution function of the field vortices
becomes time dependant and can be written in the form
\begin{equation}
{\mu} (\lbrace{{\bf r}}_{k}\rbrace,t)=\mu_{eq} (\lbrace{{\bf
r}}_{k}\rbrace)+{\mu}' (\lbrace {{\bf r}}_{k}\rbrace,t),
\label{drift2}
\end{equation} 
where the perturbation ${\mu}' (\lbrace {{\bf r}}_{k}\rbrace,t)$
reflects the influence of the test vortex on its neighbors, just like
in a polarization process. The $N$-particle distribution function
${\mu} (\lbrace{{\bf r}}_{k}\rbrace,t)$ satisfies the Liouville
equation
\begin{equation}
{\partial {\mu}\over \partial t}+\sum_{i=1}^{N}\biggl
\lbrack \sum_{j\neq i}{{\bf V}}(j\rightarrow i)+{{\bf V}}(0\rightarrow
i)\biggr\rbrack {\partial {\mu}\over \partial {{\bf r}_{i}}}=0.
\label{drift3}
\end{equation} 
Substituting Eq. (\ref{drift2}) in Eq. (\ref{drift3}), we obtain the
evolution equation of the perturbation ${\mu}'$:
\begin{equation}
{\partial {\mu}'\over\partial t}+{\cal L} {\mu}'=\beta_{eq}\gamma\sum_{i=1}^{N}{{\bf V}}_{i}{\partial\psi_{eq}\over\partial{{\bf r}_{i}}}({{\bf r}}_{i})\mu_{eq}(\lbrace {{\bf r}}_{k}\rbrace ),
\label{drift4}
\end{equation} 
where ${\cal L}\equiv \sum_{i=1}^{N} {{\bf V}}_{i}{\partial \over
\partial {{\bf r}}_{i}}$ is a Liouville operator and ${{\bf V}}_{i}=
\sum_{j\neq i}{{\bf V}}(j\rightarrow i)+{{\bf V}}(0\rightarrow i)$
denotes the total velocity of vortex $i$. This equation can be solved
formally with the Greenian
\begin{equation}
G(t,t')\equiv \exp\biggl \lbrace
-\int_{t'}^{t} {\cal L}(\tau)d\tau\biggr \rbrace.
\label{drift5}
\end{equation} 
If $t=0$ is the time at which the test vortex is introduced in the
system, we have ${\mu}'(t=0)=0$. One then finds that
\begin{equation}
{\mu}'(t)=\beta_{eq}\gamma\int_{0}^{t} d\tau G(t,t-\tau)\sum_{i=1}^{N}
{{\bf V}}_{i}{\partial\psi_{eq}\over\partial{{\bf r}_{i}}}({{\bf
r}}_{i})\mu_{eq}(\lbrace{{\bf r}}_{k}\rbrace).
\label{drift6}
\end{equation} 
The average velocity of the test vortex is expressed in term of the distribution function ${\mu}$ of the field vortices by
\begin{equation}
\langle{{\bf V}}\rangle=\int \prod_{k=1}^{N} d^{2}{{\bf r}}_{k}{{\bf V}}{\mu} (\lbrace {{\bf r}}_{k}\rbrace,t),
\label{drift7}
\end{equation} 
where ${{\bf V}}=\sum_{i=1}^{N} {{\bf V}}(i\rightarrow 0)$. Inserting Eqs. (\ref{drift2}) and (\ref{drift6}) into Eq. (\ref{drift7}), one obtains
\begin{eqnarray}
\langle{{\bf V}}\rangle=\int  \prod_{k=1}^{N} d^{2}{\bf r}_{k}{\bf V}\mu_{eq}(\lbrace{\bf r}_{k}\rbrace)+\beta_{eq}\gamma \int  \prod_{k=1}^{N} d^{2}{\bf r}_{k}{\bf{V}} \nonumber\\
\times\int_{0}^{t} d\tau G(t,t-\tau)\sum_{i=1}^{N} {{V}}^{\nu}_{i} {\partial\psi_{eq}\over\partial{ r_{i}^{\nu}}}({\bf r}_{i})\mu_{eq}(\lbrace{\bf r}_{k}\rbrace),
\label{drift8}
\end{eqnarray} 
with summation over repeated greek indices. The two terms arising in this expression have a clear physical meaning. The first term is the mean field velocity $\langle{\bf V}\rangle_{eq}=-{\bf z}\times\nabla\psi_{eq}({\bf r})$ created by the unperturbed distribution function $\mu_{eq} (\lbrace{\bf r}_{k}\rbrace)$. The second term, arising from the perturbation $\mu'$, corresponds to the response of the system to the polarization induced by the test vortex. Because of this back reaction, the test vortex will experience a systematic drift $\langle{\bf V}\rangle_{drift}=\langle{\bf V}\rangle-\langle{\bf V}\rangle_{eq}$. Explicating the action of the Greenian (\ref{drift5}), we obtain
\begin{eqnarray}
\langle{\bf V}\rangle_{drift}=\beta_{eq}\gamma \int  \prod_{k=1}^{N} d^{2}{\bf r}_{k}\sum_{i=1}^{N}{\bf {V}}(i\rightarrow 0,t)\int_{0}^{t} d\tau \sum_{i=1}^{N}\biggl\lbrack \sum_{j\neq i} {V}^{\nu}(j\rightarrow i,t-\tau)\nonumber\\
+{V}^{\nu}(0\rightarrow i,t-\tau)\biggr\rbrack {\partial\psi_{eq}\over\partial{r_{i}^{\nu}}}({\bf r}_{i}(t-\tau))\mu_{eq}(\lbrace{\bf r}_{k}(t-\tau)\rbrace),
\label{drift9}
\end{eqnarray} 
where ${\bf r}_{i}(t-\tau)$ is the position at time $t-\tau$ of the
point vortex $i$ located at ${\bf r}_{i}(t)={\bf r}_{i}$ at time
$t$. This is obtained by solving the Kirchhoff-Hamilton equations of
motion ${d{\bf r}_{i}/ dt}={\bf V}_{i}$ between $t$ and $t-\tau$.

The {exact} expression of the drift (\ref{drift9}) is completely
inextricable in the general case. In order to enlighten its physical
content, we have to make some approximations. We shall consider in the
evaluation of the time integral that the point vortices are advected
by the equilibrium mean-field velocity $\langle{\bf
V}\rangle_{eq}$. This is reasonable in a first approximation because,
when $N\rightarrow \infty$, the typical velocity fluctuations ${\bf
{\cal V}}$, of order ${\gamma/ d}\sim{\gamma}N^{1/2}/R$  are much smaller
than the mean field velocity $\langle {\bf V}\rangle_{eq}$, of order
${N\gamma/ R}$. Of course, this approximation breaks up at scales
smaller than $\delta\sim {R/N}$ when the velocity fluctuations become
comparable to the average velocity. In that case, we cannot ignore the
details of the discrete vortex interactions anymore and a specific
treatment is necessary.  For simplicity, we shall introduce a
small-scale cut-off and replace the exact Greenian $G$ by a smoother
Greenian $\langle G\rangle_{eq}$ constructed with the averaged
Liouville operator $\langle{\cal L}\rangle_{eq}\equiv \sum_{i=1}^{N}
\langle{\bf V}^{i}\rangle_{eq}{\partial \over \partial {{\bf
r}}_{i}}$. In this approximation the correlations involving two
different vortex pairs vanish and we obtain
\begin{eqnarray}
\langle V^{\mu}\rangle_{drift}=\beta_{eq}\gamma \int  \prod_{k=1}^{N} d^{2}{\bf r}_{k}\int_{0}^{t} d\tau \sum_{i=1}^{N}{V}^{\mu}(i\rightarrow 0,t)\nonumber\\
\times {V}^{\nu}(0\rightarrow i,t-\tau) {\partial\psi_{eq}\over\partial r_{i}^{\nu}}({\bf r}_{i}(t-\tau))\prod_{i=k}^{N}P_{eq}({\bf r}_{k}),
\label{drift10}
\end{eqnarray} 
where we have used $P^{eq}({\bf r}_{k}(t-\tau))=P^{eq}({\bf r}_{k}(t))$ since $P_{eq}=f(\psi_{eq})$ is constant along a streamline. Since the vortices are identical, we have equivalently
\begin{eqnarray}
\langle V^{\mu}\rangle_{drift}=N\beta_{eq}\gamma \int d^{2}{\bf r}_{1}\int_{0}^{t} d\tau {V}^{\mu}(1\rightarrow 0,t)\nonumber\\\times {V}^{\nu}(0 \rightarrow 1,t-\tau) {\partial\psi_{eq}\over\partial r_{1}^{\nu}}({\bf r}_{1}(t-\tau))P_{eq}({\bf r}_{1}).
\label{drift11}
\end{eqnarray} 

In the case of a unidirectional equilibrium flow in the $x$ direction, the drift velocity (\ref{drift11}) can be written
\begin{equation}
\langle V_{y}\rangle_{drift}=N\beta_{eq}\gamma \int d^{2}{\bf r}_{1}\int_{0}^{+\infty} d\tau {V}_{y}(1\rightarrow 0,t) {V}_{y}(0 \rightarrow 1,t-\tau) {\partial\psi_{eq}\over\partial y_{1}}(y_{1})P_{eq}(y_{1}),
\label{drift12}
\end{equation} 
where we have used $y_{1}(t-\tau)=y_{1}(t)=y_{1}$ and the time integration has been extended to $+\infty$. Since the space integration diverges as ${\bf r}_{1}\rightarrow {\bf r}_{0}$, we can make a local approximation  $\partial_{y}\psi(y_{1})\simeq\partial_{y}\psi(y)$, neglect the contribution of images in the velocity Kernel and use ${V}_{y}(0 \rightarrow 1)=-{V}_{y}(1 \rightarrow 0)$. The local approximation reflects the strong influence of the nearest neighbor and is only marginally valid as discussed in Sec. \ref{sec_fluc}. The expression of the drift can then be written
\begin{equation}
\langle V_{y}\rangle_{drift}=-\beta_{eq}\gamma D {\partial\psi_{eq}\over\partial y}(y),
\label{drift13}
\end{equation} 
where $D$ is the diffusion coefficient given by formula
(\ref{dc3}). The same type of relation is obtained for an
axisymmetric equilibrium flow. More generaly, one can write the
drift as \cite{drift,kin2}:
\begin{equation}
\langle {\bf V}\rangle_{drift}=-\beta_{eq}\gamma D\nabla\psi_{eq}, 
\label{drift14}
\end{equation} 
where $D$ is given by Eq. (\ref{dc6}). We have thus derived, from a
kinetic theory, an Einstein relation $\xi=D\gamma\beta_{eq}$ for the
point vortex gas. This relation is of great conceptual importance and
could be checked by direct numerical simulations of point vortex
dynamics in the thermal bath approach.

The direction of the drift has also important physical implications.
First, we note that the drift is always normal to the mean field
velocity $\langle {\bf V}\rangle_{eq}=-{\bf
z}\times\nabla\psi_{eq}$. Since $P_{eq}=f(\psi_{eq})$ for an
equilibrium flow, this implies that the drift is directed along the
vorticity gradient. This is clear at first sight if we write the
expression of the drift in the form $\langle {\bf V}\rangle_{drift}=
D\nabla\ln \langle\omega\rangle_{eq}$, using Eqs. (\ref{drift14}) and
(\ref{me5}). Since $D\ge 0$, this formula indicates that the test vortex
always {\it ascends} the vorticity gradient. In fact, we have assumed
in the previous discussion that all the vortices have the same
circulation. It is straightforward to generalize our calculations for
a test vortex with negative circulation evolving in a bath of vortices
with positive circulations. Similar results are obtained except that
now the test vortex {\it descends} the vorticity gradient. 

If we now take into account simultaneously the drift and the diffusion of the test vortex, we can argue that the evolution of the density probability $P({\bf r},t)$ is governed by the Fokker-Planck equation 
\begin{equation}
{\partial P\over\partial t}+\langle {\bf V}\rangle_{eq}\nabla P=\nabla(D(\nabla P+\beta_{eq}\gamma P\nabla\psi_{eq})),
\label{drift15}
\end{equation} 
where we recall that $\psi_{eq}$ is the {\it stationary}
streamfunction generated by the equilibrium vorticity
$\langle\omega\rangle_{eq}$ via the Poisson equation (\ref{pf7}). In
the case of a unidirectional flow, this Fokker-Planck equation can be
transformed into a Schr\"odinger equation (with imaginary time) which
can be solved analytically \cite{scD}.

\subsection{A relaxation equation for point vortices}
\label{sec_polarization}

If we are sufficiently close to equilibrium, we can try to apply our
previous results to {\it all} vortices in the system, eliminating the
somewhat arbitrary distinction between test and field vortices. We
propose to describe the relaxation of a cloud of point vortices
towards statistical equilibrium by the following set of equations
\begin{equation}
{\partial P\over\partial t}+\langle {\bf V}\rangle\nabla P=\nabla(D(\nabla P+\beta\gamma P\nabla\psi)),
\label{pol1}
\end{equation} 
\begin{equation}
\Delta\psi=-N\gamma P,
\label{pol2}
\end{equation} 
consisting of the Fokker-Planck equation (\ref{drift15}) {\it coupled}
to the Poisson equation (\ref{pf7}). This model is expected to be
valid only close to equilibrium so that the inverse temperature
$\beta$ has a clear physical interpretation. In this model, a point
vortex is assumed to undergo a {diffusion} process due to the
fluctuations of the velocity and a {\it systematic drift} $\langle
{\bf V}\rangle_{drift}=-\beta\gamma D\nabla\psi$ due to the
inhomogeneity of the vortex cloud.  At negative temperatures, the
drift is directed inward and the vortices tend to {\it cluster} while
at positive temperatures, the drift is directed outward and the
vortices tend to {\it repell} each other and accumulate on the
boundary. For $\beta=0$, there is no drift and
Eq. (\ref{pol1}) reduces to the pure diffusion equation (\ref{bm3});
in that case, the vorticity distribution is uniform in average. These
results are consistent with the thermodynamical approach of Onsager
\cite{onsager} but the drift provides a physical mechanism to
understand the clustering of point vortices at negative
temperatures. It can also be noted that Eq. (\ref{pol1}) is formally
similar to the Smoluchowski equation describing the relaxation of
colloidal suspensions in an external gravitational field \cite{risken}. In the present context, however, the field $\psi$ is not fixed
but is generated by the distribution of particles itself via the Poisson
equation. The resulting Smoluchowski-Poisson system has been studied
in \cite{crs,scD} for various space dimensions.

The system of equations (\ref{pol1}) and (\ref{pol2}) is also similar
to the model introduced by Chandrasekhar \cite{chandrafric} in his
stochastic description of \index{Astrophysics} stellar dynamics:
\begin{equation}
{\partial f\over\partial t}+{\bf v}{\partial f\over\partial {\bf r}}+\langle {\bf F}\rangle{\partial f\over\partial {\bf v}}={\partial\over\partial {\bf v}}\biggl\lbrace D\biggl ({\partial f\over\partial {\bf v}}+\beta m f {\bf v}\biggr )\biggr\rbrace,
\label{pol3}
\end{equation}
\begin{equation}
\Delta\Phi=4\pi G\int f d^{3}{\bf v}.
\label{pol4}
\end{equation}
In that model, a star undergoes a diffusion process (in velocity
space) due to the fluctuations of the gravitational force and a {\it
dynamical friction} $\langle {\bf F}\rangle_{friction}=-D\beta m {\bf
v}$ resulting from close encounters. Fundamentally, this friction is due to the
inhomogeneity of the velocity distribution. The coefficient of
dynamical friction is given by an Einstein formula $\xi=D\beta m$ in
which the velocity dispersion $1/\beta$ of the stars enters explicitly. 

The morphological similarity of the two models
(\ref{pol1})-(\ref{pol2}) and (\ref{pol3})-(\ref{pol4}) is striking
although the physical content of these equations is, of course, very
different. In this analogy, we see that the systematic drift of the
vortices is the counterpart of the dynamical friction of stars. In
fact, Chandrasekhar's dynamical friction can be derived from a linear
response theory \cite{kandrup} exactly like we have derived the
systematic drift of a vortex. In addition, both terms can be
understood physically as a result of a polarization process (see
discussion in \cite{kin2}). This is another mark of the formal
analogy between point vortices and stars.

\section{Kinetic theory of point vortices}
\label{sec_kin}

The previous relaxation equations are only valid for a test vortex
evolving in a bath of field vortices, or for a cloud of point vortices
close to statistical equilibrium. We would like now to relax this
``thermal bath approximation'' and describe more general situations
which do not explicitly rely on the existence of a well-defined temperature
or equilibrium state. Hence, we would like to develop a \index{Kinetic theory} more complete kinetic
theory of point vortices \cite{kin2}.

\subsection{The Liouville equation}
\label{sec_liouville}

Let us consider a collection of $N+1$ point vortices with identical circulation
$\gamma$.  Let $\mu({\bf r},{\bf r}_{1},...,{\bf r}_{N},t)$ denote
the $N+1$ particle distribution of the system, i.e. $\mu({\bf r},{\bf
r}_{1},...,{\bf r}_{N},t)d^{2}{\bf r}d^{2}{\bf r}_{1}...d^{2}{\bf r}_{N}$
represents the probability that point vortex $0$ be in the cell $({\bf r},{\bf
r}+d{\bf r})$, point vortex $1$ in the cell $({\bf r}_{1},{\bf r}_{1}+d{\bf
r}_{1})$... and point vortex $N$ in the cell $({\bf r}_{N},{\bf r}_{N}+d{\bf
r}_{N})$ at time $t$. The $(N+1)$-particle distribution function $\mu(t)$
satisfies the Liouville equation
\begin{equation}
{\partial\mu\over\partial t}+\sum_{i=0}^{N}{\bf V}_{i}{\partial\mu\over\partial
{\bf r}_{i}}=0,
\label{Liouville1}
\end{equation}
where ${\bf V}_{i}$ is the velocity of vortex $i$ produced by the
other vortices according to Eq. (\ref{pv2}). We also introduce the
one- and $N$-particle distribution functions defined by
\begin{equation}
P({\bf r},t)=\int \mu(\lbrace{\bf r}_{k}\rbrace,t) \prod_{k=1}^{N}d^{2}{\bf
r}_{k},
\label{L2}
\end{equation}
\begin{equation}
\mu_{sys}({\bf r}_{1},...,{\bf r}_{N},t)=\int \mu(\lbrace{\bf
r}_{k}\rbrace,t)d^{2}{\bf r}.
\label{L3}
\end{equation}
We write the distribution function
$\mu$ in the suggestive form
\begin{equation}
\mu({\bf r},{\bf r}_{1},...,{\bf r}_{N},t)=P({\bf r},t)\mu_{sys}({\bf
r}_{1},...,{\bf r}_{N},t)+\mu_{I}({\bf r},{\bf r}_{1},...,{\bf r}_{N},t),
\label{muI}
\end{equation}
where the quantity $\mu_{I}$ reflects the effect of correlations between vortices.

The Liouville equation (\ref{Liouville1}) provides the correct starting point
for the analysis of the dynamics of our vortex system. However, when $N$ is
large, this equation contains much more information than one can interpret.
Consequently, what one would like to do is to describe the system in some
average sense by a one-particle distribution function.

\subsection{The projection operator formalism}
\label{sec_projection}

Our first objective is to derive some {\it exact} kinetic equations
satisfied by $P({\bf r},t)$ and $\mu_{sys}({\bf r}_{1},...,{\bf
r}_{N},t)$. This can be achieved by using the projection operator
formalism developed by Willis \& Picard \cite{wp}. This formalism was
also used by Kandrup \cite{Kandrup} in the context of stellar dynamics
to derive a generalized Landau equation describing the time evolution
of the distribution function of stars in an inhomogeneous medium. We
shall just recall the main steps of the theory. More details can be
found in the original paper of Willis \& Picard \cite{wp}  and in
Kandrup \cite{Kandrup}. To have similar notations, we set $x\equiv
\lbrace{\bf r}\rbrace$ and $y\equiv \lbrace {\bf r}_{1},...,{\bf
r}_{N}\rbrace$. Then, Eq. (\ref{muI}) can be put in the form
\begin{equation}
\mu(x,y,t)=\mu_{R}(x,y,t)+\mu_{I}(x,y,t),
\label{muIbis}
\end{equation}
with
\begin{equation}
\mu_{R}(x,y,t)=f(x,t)g(y,t),
\label{muR}
\end{equation}
where we have written $f(x,t)\equiv P({\bf r},t)$ and $g(y,t)\equiv
\mu_{sys}({\bf r}_{1},...,{\bf r}_{N},t)$. The Liouville equation is also cast
in the form
\begin{equation}
{\partial\mu\over\partial t}=-i L\mu=-i(L_{0}+L_{sys}+L')\mu,
\label{Liouville2}
\end{equation}
where $L_{0}$ and $L_{sys}$ act respectively only on the variables $x$ and $y$,
whereas the interaction Liouvillian $L'$ acts upon both $x$ and $y$ (the complex
number $i$ is here purely formal and has been introduced only to have the same
notations as in \cite{wp,Kandrup}). 

Following Willis \& Picard \cite{wp}, we introduce the time-dependant
projection operator
\begin{equation}
{ P}(x,y,t)=g(y,t)\int dy+f(x,t)\int dx-f(x,t)g(y,t)\int dx\int dy.
\label{Proj}
\end{equation}
We can easily check that
\begin{equation}
{ P}(x,y,t)\mu(x,y,t)=\mu_{R}(x,y,t),
\label{Pmu}
\end{equation}
\begin{equation}
\lbrack 1-{ P}(x,y,t)\rbrack \mu(x,y,t)=\mu_{I}(x,y,t).
\label{Perpmu}
\end{equation}
We also verify that ${ P}$ is a projection in the sense that ${ P}^{2}(t)={
P}(t)$. Applying ${ P}$ and $1-{ P}$ on the Liouville equation
(\ref{Liouville2}), we obtain the coupled equations
\begin{equation}
\partial_{t}\mu_{R}(x,y,t)=-i{ P}L\mu_{R}-iPL\mu_{I},
\label{Sys1}
\end{equation}
and 
\begin{equation}
\partial_{t}\mu_{I}(x,y,t)=-i(1-P)L\mu_{R}-i(1-P)L\mu_{I}.
\label{Sys2}
\end{equation}
These equations  describe the separation between a
``macrodynamics'' and a ``subdynamics''.

Introducing the Greenian
\begin{equation}
{\cal G}(t,t')\equiv \exp \Biggl \lbrace -i\int_{t'}^{t}dt'' \lbrack 
1-{ P}(t'')\rbrack{L}\Biggr \rbrace,
\label{Greenian}
\end{equation}
we can immediately write down a formal solution of Eq. (\ref{Sys2}),
namely
\begin{equation}
\mu_{I}(x,y,t)=-\int_{0}^{t}dt' {\cal G}(t,t')i\lbrack 1-{P}(t')\rbrack { L}\mu_{R}(x,y,t'),
\label{muIexp}
\end{equation}
where we have assumed that, initially, the particles are uncorrelated so that
$\mu_{I}(x,y,0)=0$. Substituting for $\mu_{I}(x,y,t)$ from Eq.
(\ref{muIexp}) in Eq. (\ref{Sys1}), we obtain 
\begin{equation}
\partial_{t}\mu_{R}(x,y,t)=-i{ P}L\mu_{R}-\int_{0}^{t}dt' P(t)L {\cal G}(t,t')
\lbrack 1-P(t')\rbrack { L}\mu_{R}(x,y,t').
\label{sys1}
\end{equation}
The integration over $y$ will yield an equation describing the evolution of $f$.
Using some mathematical properties of the projection operator (\ref{Proj}), the
final result can be put in the nice symmetrical form 
given by Willis \& Picard \cite{wp},
\begin{equation}
\partial_{t}f(x,t)+i L_{0}f+i\langle L'\rangle_{sys}f=-\int_{0}^{t}dt'\int
dy\Delta_{t}L'{\cal G}(t,t')\Delta_{t'}L' g(y,t')f(x,t'),
\label{evolf}
\end{equation}
where the notations stand for
\begin{equation}
\langle L'\rangle_{sys}=\int dy' L'(x,y') g(y',t),
\label{Lprime}
\end{equation}
\begin{equation}
\langle L'\rangle_{0}=\int dx' L'(x',y) f(x',t),
\label{Lprime0}
\end{equation}
\begin{equation}
\Delta_{t}L'=L'-\langle L'\rangle_{sys}-\langle L'\rangle_{0}.
\label{DeltaLprime}
\end{equation}
Similarly, after integrating over $x$ we find the equation satisfied by $g$,
\begin{equation}
\partial_{t}g(y,t)+i L_{sys}g-i\langle L'\rangle_{1}g=-\int_{0}^{t}dt'\int
dx\Delta_{t}L'{\cal G}(t,t')\Delta_{t'}L' g(y,t')f(x,t').
\label{evolg}
\end{equation}

\subsection{Application to the point vortex system}
\label{sec_application}

The previous theory is completely general and we now consider its
application to a system of point vortices \cite{kin2}. Let us first rewrite the
Liouville equation (\ref{Liouville1}) in a form that separates the
contribution of the test vortex from the contribution of the field
vortices:
\begin{eqnarray}
{\partial\mu\over\partial t}+\sum_{i=1}^{N}{\bf V}(i\rightarrow
0){\partial\mu\over\partial {\bf r}}+\sum_{i=1}^{N}{\bf V}(0\rightarrow
i){\partial\mu\over\partial {\bf r}_{i}}\nonumber\\
+\sum_{i=1}^{N}\sum_{j=1,j\neq i}^{N}{\bf V}(j\rightarrow
i){\partial\mu\over\partial {\bf r}_{i}}=0.
\label{Liouvillesep}
\end{eqnarray}
Applying the general theory of Willis \& Picard \cite{wp}, we obtain the
following kinetic equation for the one-particle distribution function
of a vortex system \cite{kin2}:
\begin{eqnarray}
{\partial P\over\partial t}+\langle {\bf V}\rangle{\partial P\over\partial {\bf
r}}={\partial\over\partial r^{\mu}}\int_{0}^{t}d\tau\int\prod_{k=1}^{N}d^{2}{\bf
r}_{k}\sum_{i=1}^{N}\sum_{j=1}^{N}  {\cal V}^{\mu}(i\rightarrow 0)\nonumber\\
\times{\cal G}(t,t-\tau)\biggl ({\cal V}^{\nu}(j\rightarrow
0){\partial\over\partial {r}^{\nu}}+{{\cal V}}^{\nu}(0\rightarrow
j){\partial\over\partial { r}_{j}^{\nu}}\biggr ) P({\bf
r},t-\tau)\mu_{sys}(\lbrace {\bf r}_{k}\rbrace,t-\tau),\nonumber\\
\label{kin2}
\end{eqnarray}
where the Greek indices refer to the components of ${\mb{\cal V}}$ in
a fixed system of coordinates and ${\mb{\cal V}}(i\rightarrow 0)={\bf
V}(i\rightarrow 0)-\langle{\bf V}(i\rightarrow 0)\rangle$ denotes the
velocity fluctuation. We can note that equation (\ref{kin2}) already
shares some analogies with the Fokker-Planck equation of
Sec. \ref{sec_bath1}. Indeed, the first term on the right hand side
corresponds to a diffusion and the second term to a drift. For a
passive particle ${\cal V}^{\nu}(0\rightarrow j)=0$ and the drift
cancels out, as expected.

\subsection{The factorization hypothesis}
\label{sec_factorization}

If the vortices are initially decorrelated then, for sufficiently short times,
they will remain decorrelated. This means that the $(N+1)$-particle distribution
function can be factorized in a product of ($N+1$) one-particle distribution
functions
\begin{equation}
\mu({\bf r},{\bf r}_{1},...,{\bf r}_{N},t)=\prod_{k=0}^{N} P({\bf r}_{k},t).
\label{decorr}
\end{equation}
If we integrate the Liouville equation (\ref{Liouville1}) on the
positions of the $N$ vortices $1,...,N$ and use the factorization
(\ref{decorr}), we directly obtain \cite{kin2}:
\begin{equation}
{\partial P\over\partial t}+\langle {\bf V}\rangle\nabla P=0.
\label{Eulershort}
\end{equation} 
Therefore, for sufficiently short times, the average vorticity
$\langle\omega\rangle$ satisfies the 2D Euler equation (called the
Vlasov equation in other circumstances). However, at later times, the
distribution function $\mu$ differs from the pure product
(\ref{decorr}) and the Euler equation does not provide a good
approximation anymore.  In Sec. \ref{sec_application} we have
determined an exact equation (\ref{kin2}) satisfied by the one-particle
distribution function which is valid at any
time. This equation is not closed, however, since it involves the
$N$-vortex distribution function $\mu_{sys}$. We shall close the
system by assuming that $\mu_{sys}$ can be approximated by a product
of N one-particle distribution functions in the form
\begin{equation}
\mu_{sys}({\bf r}_{1},...,{\bf r}_{N},t)\simeq \prod_{k=1}^{N} P({\bf r}_{k},t).
\label{decorrsys}
\end{equation}

Inserting Eq. (\ref{decorrsys}) in Eq. (\ref{kin2}), we obtain
\begin{eqnarray}
{\partial P\over\partial t}+\langle {\bf V}\rangle{\partial P\over\partial {\bf
r}}={\partial\over\partial r^{\mu}}\int\prod_{k=1}^{N}d^{2}{\bf
r}_{k}\int_{0}^{t}d\tau\sum_{i=1}^{N}\sum_{j=1}^{N}  {\cal V}^{\mu}(i\rightarrow
0)\nonumber\\ 
\times{\cal G}(t,t-\tau)\biggl ({\cal V}^{\nu}(j\rightarrow
0){\partial\over\partial {r}^{\nu}}+{{\cal V}}^{\nu}(0\rightarrow
j){\partial\over\partial { r}_{j}^{\nu}}\biggr ) P({\bf
r},t-\tau)\prod_{k=1}^{N} P({\bf r}_{k},t-\tau).\nonumber\\
\label{genK1}
\end{eqnarray}
If we assume that between
$t$ and $t-\tau$ the trajectories of the particles are determined from the
smooth velocity field created by the vorticity distribution
$\langle\omega\rangle=N\gamma P({\bf r},t)$, the foregoing equation simplifies 
in
\begin{eqnarray}
{\partial P\over\partial t}+\langle {\bf V}\rangle{\partial P\over\partial {\bf
r}}=N{\partial\over\partial r^{\mu}}\int_{0}^{t}d\tau\int d^{2}{\bf
r}_{1}{V}^{\mu}(1\rightarrow 0)_{t}\nonumber\\
 \times\biggl \lbrace
{V}^{\nu}(1\rightarrow 0)P_{1}{\partial P\over\partial {r}^{\nu}}+{{
V}}^{\nu}(0\rightarrow 1)P{\partial P_{1}\over\partial { r}_{1}^{\nu}}\biggr
\rbrace_{t-\tau}, 
\label{new1}
\end{eqnarray}
where $P=P({\bf r},t)$ and $P_{1}=P({\bf r}_{1},t)$.  Eq. (\ref{new1})
is a non Markovian integrodifferential equation since the probability
density $P({\bf r},t)$ in ${\bf r}$ at time $t$ depends on the value
of the whole distribution of probability $P({\bf r}_{1},t-\tau)$ at
earlier times through an integration on ${\bf r}_{1}$ and
$\tau$. Equation (\ref{new1}) is therefore non local in space and
time. It can be shown \cite{kin2} that this kinetic equation
rigorously conserves angular momentum in a circular domain and linear
impulse in a channel (or in an infinite domain). However, under this
form, it has not been possible to prove the conservation of energy and the
H-theorem for the Boltzmann entropy (\ref{me2}). 

\subsection{The case of short decorrelation times}
\label{sec_short}

If we assume that the decorrelation time $\tau$ is short (which does
not need to be the case) and implement a strong Markov approximation,
we obtain
\begin{eqnarray}
{\partial P\over\partial t}+\langle {\bf V}\rangle{\partial P\over\partial {\bf
r}}={N\tau\over 2}{\partial\over\partial r^{\mu}}\int d^{2}{\bf
r}_{1}{V}^{\mu}(1\rightarrow 0)\nonumber\\
\times \biggl \lbrace {V}^{\nu}(1\rightarrow
0)P_{1}{\partial P\over\partial {r}^{\nu}}+{{ V}}^{\nu}(0\rightarrow
1)P{\partial P_{1}\over\partial { r}_{1}^{\nu}}\biggr \rbrace.
\label{new2}
\end{eqnarray}
In the case of an infinite domain  ${\bf V}(0\rightarrow 1)=-{\bf
V}(0\rightarrow 1)$ and we have the further simplification 
\begin{eqnarray}
{\partial P\over\partial t}+\langle {\bf V}\rangle{\partial P\over\partial {\bf
r}}={N\gamma^{2}\over 8\pi^{2}}\tau{\partial\over\partial r^{\mu}}\int d^{2}{\bf
r}_{1} K'^{\mu\nu}({\mb\xi})\biggl ( P_{1}{\partial P\over\partial {r}^{\nu}}-
P{\partial P_{1}\over\partial { r}_{1}^{\nu}}\biggr ),
\label{genK4}
\end{eqnarray}
where 
\begin{eqnarray}
K'^{\mu\nu}({\mb\xi})={\xi_{\perp}^{\mu}\xi_{\perp}^{\nu}\over\xi^{4}}=
{\xi^{2}\delta^{\mu\nu}-\xi^{\mu}\xi^{\nu}\over\xi^{4}}
\label{Kmunu}
\end{eqnarray}
and ${\mb\xi}={\bf r}_{1}-{\bf r}$. To arrive at Eq.  (\ref{Kmunu}) we
have explicitly used the form of the Kernel (\ref{pv2}), and to get
the second equality we have used the fact that we are in two
dimensions.  Note that the symmetrical form of Eq. (\ref{genK4}) is
reminiscent of the Landau equation introduced in \index{Plasmas
physics} plasma physics and
\index{Astrophysics} in stellar dynamics (see, e.g., \cite{Kandrup}):
\begin{eqnarray}
{\partial f\over\partial t}={\partial \over\partial v^{\mu}}\int d^{3}{\bf v}_{1}K^{\mu\nu}\biggl (f_{1}{\partial f\over\partial v^{\nu}}-f{\partial f_{1}\over\partial v_{1}^{\nu}}\biggr ),
\label{landau1}
\end{eqnarray}
with 
\begin{eqnarray}
K^{\mu\nu}=2\pi NG^{2}m^{2}\ln\biggl ({L_{max}\over L_{min}}\biggr ){u^{2}\delta^{\mu\nu}-u^{\mu}u^{\nu}\over u^{3}},
\label{landau2}
\end{eqnarray}
and ${\bf u}={\bf v}-{\bf v_{1}}$, $f=f({\bf v},t)$, $f_{1}=f({\bf
v}_{1},t)$.  In this analogy, the position ${\bf r}$ of the vortices
plays the role of the velocity ${\bf v}$ of the electric charges or
stars and the spatial distribution $P({\bf r},t)$ the role of the
velocity distribution $f({\bf v},t)$. Therefore, we can directly infer
the conservation of linear impulse ${\bf P}_{\perp}=\int\langle
\omega\rangle {\bf r} d^{2}{\bf r}$ and angular momentum $L=\int
\langle\omega\rangle r^{2}d^{2}{\bf r}$ which play respectively the
role of impulse ${\bf P}=\int f {\bf v} d^{3}{\bf v}$ and kinetic
energy $K=\int f {v^{2}\over 2}d^{3}{\bf v}$ in plasma physics. We can
also prove a $H$-theorem for the Boltzmann entropy (\ref{me2}) exactly
like in the case of the Landau equation.  Finally, the solutions of
Eq. (\ref{genK4}) converge towards the Gaussian vortex (the equivalent
of the Maxwellian distribution in plasma physics with ${\bf r}$ in
place of ${\bf v}$):
\begin{eqnarray}
P({\bf r})=A e^{-{1\over 2}\alpha\gamma ({\bf r}-{\bf r}_{0})^{2}}
\label{Gauss}
\end{eqnarray}
which is the maximum entropy state at fixed circulation, angular
momentum and impulse. It is in general {\it different} from the Boltzmann
distribution (\ref{me5}) with the relative streamfunction $\psi'=\psi+{\Omega\over 2}r^{2}-{\bf U}_{\perp}{\bf r}$, except in the particular limit $\beta\rightarrow 0$, $\Omega\rightarrow +\infty$ with fixed $\alpha=\beta\Omega/2$ corresponding to the statistical equilibrium (\ref{ud5}). This clearly indicates that Eq. (\ref{genK4}) does {\it not}  conserve energy.

\subsection{A generalized kinetic equation}
\label{sec_gen}

Now, if we account properly for memory effects in Eq. (\ref{new1}), we
can obtain a generalized kinetic equation which guaranties the
conservation of energy (in addition to the other constraints) and is
therefore more satisfactory. If the distribution of vortices is
axisymmetric, it is possible to calculate the memory function
appearing in Eq. (\ref{new1}) explicitly if we assume that the
correlation time is smaller than the typical time on which the average
vorticity changes appreciably \cite{kin2}. In this approximation, the point
vortices follow, between $t$ and $t-\tau$, circular trajectories with
angular velocity $\Omega(r,t)=\langle V_{\theta}\rangle(r,t)/r$ and
Eq.  (\ref{new1}) simplifies in (see Appendix B): 
\begin{eqnarray}
{\partial P\over\partial t}=-{N\gamma^{2}\over 4 r}{\partial\over\partial
r}\int_{0}^{+\infty}r_{1}dr_{1} \delta(\Omega-\Omega_{1})\ln\biggl \lbrack
1-\biggl ({r_{<}\over r_{>}}\biggr )^{2}\biggr \rbrack\biggl\lbrace {1\over
r}P_{1}{\partial P\over\partial r}- {1\over r_{1}}P{\partial P_{1}\over\partial
r_{1}} \biggr\rbrace, \nonumber\\  
\label{day1}
\end{eqnarray}
where $\Omega=\Omega(r,t)$, $\Omega_{1}=\Omega(r_{1},t)$ and $r_{>}$ (resp.
$r_{<}$) is the biggest (resp. smallest) of $r$ and $r_{1}$. The angular velocity is related to the vorticity by
\begin{eqnarray}
\langle\omega\rangle ={1\over r}{\partial\over\partial r}(\Omega r^{2}).
\label{qw1}
\end{eqnarray}

We can propose an approximation of the general kinetic
equation (\ref{new1}) which encompasses the axisymmetric form
previously derived. Memory effects are not neglected, unlike in
Eq. (\ref{genK4}), but they are simplified in a way which preserves
all the conservation laws of the system (as discussed below). We
propose the generalized kinetic equation \cite{kin2}:
\begin{eqnarray}
{\partial P\over\partial t}+\langle {\bf V}\rangle \nabla P={N\gamma^{2}\over
8}{\partial \over\partial r^{\mu}}\int d^{2}{\bf r}_{1}K^{\mu\nu}\delta ({\mb
\xi}.{\bf v})\biggl (P_{1}{\partial P\over\partial r^{\nu}}-P{\partial
P_{1}\over\partial r_{1}^{\nu}}\biggr ),
\label{sun1}
\end{eqnarray}
with  
\begin{eqnarray}
K^{\mu\nu}({\mb\xi})={\xi_{\perp}^{\mu}\xi_{\perp}^{\nu}
\over\xi^{2}}={\xi^{2}\delta^{\mu\nu}-\xi^{\mu}\xi^{\nu}\over \xi^{2}},
\label{Kmunu2}
\end{eqnarray}
and ${\mb\xi}={\bf r}_{1}-{\bf r}$, ${\bf v}=\langle {\bf V}\rangle
({\bf r}_{1},t)-\langle {\bf V}\rangle ({\bf r},t)$. In the
thermodynamic limit $N\rightarrow +\infty$ and $\gamma\sim 1/N$ (see
Sec. \ref{sec_field}), the kinetic equation (\ref{sun1}) reduces to
the Vlasov equation (\ref{Eulershort}). However, in practice, $N$ is
always finite and the correlations between point vortices must be
taken into account. The ``collision term'' in Eq. (\ref{sun1}) gives
the first order correction $O(1/N)$ to the Vlasov limit.

From Eqs. (\ref{day1}) and (\ref{sun1}), it is clear that the
relaxation of the point vortices is due to a phenomenon of
resonance. Only the points ${\bf r}_{1}$ satisfying the condition
${\mb \xi}\cdot{\bf v}=0$ with ${\bf r}_{1}\neq {\bf r}$ contribute to
the diffusion current in ${\bf r}$. In the axisymmetric case, this
condition of resonance reduces to $\Omega(r_{1})=\Omega(r)$, which
supposes that the angular velocity profile is non monotonic. The
occurence of a $\delta$-function in Eqs. (\ref{day1}) and (\ref{sun1})
is the main difference with the Landau equation (\ref{landau1}). In
the present context, it ensures the conservation of the ``potential''
energy of the vortices $E={1\over 2}\int\omega\psi d^{2}{\bf r}$ which has
no counterpart in Landau's theory applying to spatially uniform
plasmas. It can also be noted that, contrary to the Landau equation,
the kinetic equations (\ref{genK4}) and (\ref{sun1}) do {\it not}
suffer the well-known logarithmic divergence appearing in the context
of Coulombian plasmas and stellar systems (see, e.g.,
\cite{Kandrup}). This is due essentially to the lower dimension of
space ($D=2$ instead of $D=3$) and to the different nature of the
interactions.

\subsection{Conservation laws and H-theorem}
\label{sec_lawsH}

We now derive the conservation laws and the H-theorem satisfied by Eq.
(\ref{sun1}). The conservation of the circulation is straightforward
since the right hand side of Eq.  (\ref{sun1}) can be written as the
divergence of a current. To prove the conservation of angular
momentum, we take the time derivative of $L=N\gamma\int
Pr^{2}d^{2}{\bf r}$, substitute for Eq. (\ref{sun1}), permut the dummy
variables ${\bf r}$ and ${\bf r}_{1}$ and add the resulting
expressions. This yields
\begin{eqnarray}
\dot L={N^{2}\gamma^{3}\over 8}\int d^{2}{\bf r}d^{2}{\bf
r}_{1}K^{\mu\nu}\xi^{\mu}\delta ({\mb\xi}.{\bf v})\biggl  (P_{1}{\partial
P\over\partial r^{\nu}}-P{\partial P_{1}\over\partial r_{1}^{\nu}}\biggr ).
\label{mon3}
\end{eqnarray}
From Eq. (\ref{Kmunu2}), we immediately verify that
\begin{eqnarray}
K^{\mu\nu}\xi^{\mu}=0,
\label{mont3}
\end{eqnarray}
which proves the conservation of angular momentum. We can prove the
conservation of linear impulse in a similar manner \cite{kin2}. For the
conservation of energy, we start from Eq. (\ref{mf12}) and follow
the same procedure. This yields
\begin{eqnarray}
\dot E={N^{2}\gamma^{3}\over 16}\int d^{2}{\bf r}d^{2}{\bf
r}_{1}K^{\mu\nu}v_{\perp}^{\mu}\delta ({\mb\xi}.{\bf v})\biggl  (P_{1}{\partial
P\over\partial r^{\nu}}-P{\partial P_{1}\over\partial r_{1}^{\nu}}\biggr ).
\label{mon7}
\end{eqnarray}
Considering the form of the tensor (\ref{Kmunu2}), we have 
\begin{eqnarray}
K^{\mu\nu}v_{\perp}^{\mu}={\xi_{\perp}^{\nu}\over \xi^{2}}({\mb\xi}.{\bf v}).
\label{mon8}
\end{eqnarray}
When substituted in Eq. (\ref{mon7}), we see that the occurence of the
$\delta$-function in the kinetic equation implies $\dot E=0$. Finally,
for the rate of entropy production we have, according to Eqs.
(\ref{me2}) and (\ref{sun1}):
\begin{eqnarray}
\dot S={N^{2}\gamma^{2}\over 8}\int d^{2}{\bf r}d^{2}{\bf r}_{1}{1\over P
P_{1}}P_{1}{\partial P\over\partial r^{\mu}} K^{\mu\nu}\delta ({\mb\xi}.{\bf
v})\biggl  (P_{1}{\partial P\over\partial r^{\nu}}-P{\partial P_{1}\over\partial
r_{1}^{\nu}}\biggr ).
\label{mon9}
\end{eqnarray}
Permutting the dummy variables ${\bf r}$ and ${\bf r}_{1}$ and adding the
resulting expression to Eq. (\ref{mon9}), we obtain
\begin{eqnarray}
\dot S={N^{2}\gamma^{2}\over 16}\int d^{2}{\bf r}d^{2}{\bf r}_{1}{1\over P
P_{1}}\delta ({\mb\xi}.{\bf v}) \biggl  (P_{1}{\partial P\over\partial
r^{\mu}}-P{\partial P_{1}\over\partial r_{1}^{\mu}}\biggr )  K^{\mu\nu}\biggl
(P_{1}{\partial P\over\partial r^{\nu}}-P{\partial P_{1}\over\partial
r_{1}^{\nu}}\biggr ).\nonumber\\
\label{mon10}
\end{eqnarray}
Now, for any vector, $A^{\mu}K^{\mu\nu}A^{\nu}=({\bf
A}\cdot {\mb\xi}_{\perp})^{2}/\xi^{2}\ge 0$. This proves a H-theorem ($\dot
S\ge 0$) for the kinetic equation (\ref{sun1}). It should be
emphasized that the conservation laws and the H-theorem result
essentially from the {\it symmetry} of the kinetic equation. This is
satisfying from a physical point of view.

It is also easy to show that the Boltzmann distribution
\begin{eqnarray}
P=Ae^{-\beta\gamma (\psi+{\Omega\over 2}r^{2}-{\bf U}_{\perp}{\bf r})},
\label{mon11}
\end{eqnarray}
is a stationary solution of Eq. (\ref{sun1}). Noting that
\begin{eqnarray}
{\partial P\over\partial r^{\nu}}=-\beta\gamma\biggl ({\partial\psi\over\partial
r^{\nu}}+\Omega r^{\nu}-U_{\perp}^{\nu}\biggr )P,
\label{mon12}
\end{eqnarray}
we have successively
\begin{eqnarray}
K^{\mu\nu}\biggl (P_{1}{\partial P\over\partial r^{\nu}}-P{\partial
P_{1}\over\partial r_{1}^{\nu}}\biggr )=\beta\gamma
PP_{1}K^{\mu\nu}(v_{\perp}^{\nu}+\Omega\xi^{\nu})=\beta\gamma
PP_{1}{\xi_{\perp}^{\mu}\over \xi^{2}}({\mb\xi}.{\bf v}),
\label{mon13}
\end{eqnarray}
where we have used Eqs. (\ref{mont3}) and (\ref{mon8}). When
Eq. (\ref{mon13}) is substituted in Eq. (\ref{sun1}), we find that the
right hand side cancels out due to the $\delta$-function. The
advective term is also zero since $P=f(\psi')$. Therefore, the
Boltzmann distribution (\ref{mon11}) is a stationary solution of Eq.
(\ref{sun1}).  Note, however, that this is not the only solution,
unlike for ordinary kinetic equations. Any stationary solution of the
Euler equation satisfying in addition ${\mb\xi}.{\bf v}\neq 0$ for any
couple of points ${\bf r}$, ${\bf r}_{1}$ (with ${\bf r}\neq {\bf
r}_{1}$) is a solution of Eq. (\ref{sun1}). Physically, this implies
that the system needs sufficiently strong resonances to relax towards
the maximum entropy state. If this condition is not realized, the
system can remain frozen in a sort of ``metastable'' equilibrium
state. Further evolution of the system will require non trivial
correlations between point vortices which are not taken into account
in the present theory.

\subsection{The thermal bath approximation}
\label{sec_bath}

A direct connexion between the generalized kinetic equation
(\ref{sun1}) and the Fokker-Planck equation of Sec. \ref{sec_bath1}
can be found.  Introducing a diffusion tensor
\begin{equation}
D^{\mu\nu}={N\gamma^{2}\over 8} \int d^{2}{\bf r}_{1}
K^{\mu\nu}\delta({\mb\xi}.{\bf v})P_{1}, 
\label{Dmunu}
\end{equation}   
and a drift term
\begin{equation}
\eta^{\mu}=-{N\gamma^{2}\over 8} \int d^{2}{\bf r}_{1}
K^{\mu\nu}\delta({\mb\xi}.{\bf v}){\partial P_{1}\over\partial r_{1}^{\nu}},  
\label{etamu}
\end{equation}   
we can rewrite Eq. (\ref{sun1}) in the more illuminating form
\begin{equation}
{\partial P\over\partial t}+\langle {\bf V}\rangle \nabla
P={\partial\over\partial r^{\mu}}\biggl \lbrack D^{\mu\nu}{\partial
P\over\partial r^{\nu}}+P\eta^{\mu}\biggr\rbrack,
\label{FPgenee}
\end{equation}
similar to a general Fokker-Planck equation. Note, however, that
Eq. (\ref{FPgenee}) is an integrodifferential equation since the
density probability $P({\bf r},t)$ in ${\bf r}$ at time $t$ depends on
the value of the whole distribution of probability $P({\bf r}_{1},t)$
at the same time by an integration over ${\bf r}_{1}$. By contrast,
the Fokker-Planck equation (\ref{bm7}) is a differential equation. The
usual way to transform an integrodifferential equation into a
differential equation is to make a guess for the function $P({\bf
r}_{1})$ appearing under the integral sign and refine the guess by
successive iterations. In practice we simply make one sensible guess.
Therefore, if we are close to equilibrium, it seems natural to replace
the function $P_{1}$ appearing in the integrals by the Boltzmann
distribution
\begin{equation}
P({\bf r}_{1})=Ae^{-\beta\gamma\psi({\bf r}_{1})}.
\label{eq}
\end{equation} 
This corresponds to the ``thermal bath approximation'' of
Sec. \ref{sec_bath1}: the vortices have not yet relaxed completely,
but when we focus on the relaxation of a given point vortex (described
by $P$) we can consider, in a first approximation, that the rest of
the system (described by $P_{1}$) is at equilibrium. Within this
approximation, the diffusion coefficient and the drift simplify in
\begin{equation}
\eta^{\mu}=\beta\gamma D^{\mu\nu}{\partial\psi\over\partial r^{\nu}},
\label{driftbis}
\end{equation} 
\begin{equation}
D^{\mu\nu}={N\gamma^{2}\over 8}P({\bf r},t)\int K^{\mu\nu}\delta({\mb\xi}.{\bf
v}) d^{2}{\mb\xi},
\label{diffusion}
\end{equation}
where we have made the local approximation. If we assume
that the correlation time is short, i.e.  if we replace
$\xi^{2}\delta({\mb\xi}.{\bf v})$ by $\tau/\pi^{2}$ (compare Eqs.
(\ref{sun1}) and (\ref{genK4})), we obtain
\begin{equation}
{\mb\eta}=\beta\gamma D\nabla\psi,
\label{tu1}
\end{equation} 
\begin{equation}
D={\gamma\tau\over 16\pi}\ln N \langle\omega\rangle,
\label{tu2}
\end{equation}
and Eq. (\ref{FPgenee}) reduces to the Fokker-Planck equation
\begin{equation}
{\partial P\over \partial t}+\langle {\bf V}\rangle \nabla P=\nabla(D(\nabla
P+\beta\gamma P\nabla\psi)).
\label{rs}
\end{equation} 

This approximation is, however, not very satisfactory since the
decorrelation time $\tau$ appears as a free parameter. In fact, the
decorrelation time can be determined self-consistently from the above
formulae by evaluating properly the $\delta$-function in
Eq. (\ref{diffusion}). Expanding the velocity difference ${\bf
v}=\langle {\bf V}_{1}\rangle -\langle {\bf V}\rangle$ in a Taylor
series in ${\mb\xi}={\bf r}_{1}-{\bf r}$, we obtain to first order in
the expansion
\begin{equation}
{\mb\xi}.{\bf v}=\Sigma^{\mu\nu}\xi^{\mu}\xi^{\nu},
\label{sat1}
\end{equation}
where 
\begin{equation}
\Sigma^{\mu\nu}={1\over 2}\biggl ({\partial\langle V\rangle^{\mu}\over\partial
r^{\nu}}+{\partial\langle V\rangle^{\nu}\over\partial r^{\mu}}\biggr ),
\label{sat2}
\end{equation}
is the stress tensor. It has the property of symmetry 
$\Sigma^{\mu\nu}=\Sigma^{\nu\mu}$. Since the flow is divergenceless, we also
have $\Sigma^{xx}+\Sigma^{yy}=0$.  In terms of the
stress tensor (\ref{sat2}), the diffusion tensor (\ref{diffusion}) can be
rewritten 
\begin{equation}
D^{\mu\nu}={NP\gamma^{2}\over 8}\int
{\xi^{2}\delta^{\mu\nu}-\xi^{\mu}\xi^{\nu}\over
\xi^{2}}\delta(\Sigma^{\mu\nu}\xi^{\mu}\xi^{\nu})  d^{2}{\mb\xi}.
\label{sat3}
\end{equation}
This integral can be performed easily by working in a basis $(\xi'_{1},\xi'_{2})$ in which the tensor
$\Sigma^{\mu\nu}$ is anti-diagonal \cite{kin2}. Then,
\begin{equation}
D^{\mu\nu}={NP\gamma^{2}\over 8}\int
{\xi'^{2}\delta^{\mu\nu}-\xi'^{\mu}\xi'^{\nu}\over \xi'^{2}}\delta
(|\Sigma({\bf r})|\xi'_{1}\xi'_{2})\  d\xi'_{1}d\xi'_{2},
\label{sat4}
\end{equation}
where we have set $|\Sigma({\bf r})|=2\sqrt{-{\rm det}(\Sigma)}$. Clearly,
this quantity is invariant by a change of reference frame and it
measures the local shear of the flow. It is easy to check that the
diffusion is isotropic and that $D^{\mu\nu}=D\delta^{\mu\nu}$ with
\begin{equation}
D={NP\gamma^{2}\over 8}{1\over |\Sigma({\bf r})| }\int {\xi_{2}^{'2}\over
\xi_{1}^{'2}+\xi_{2}^{'2}}\delta (\xi'_{1}\xi'_{2})\ d\xi'_{1}d\xi'_{2}.
\label{sat5}
\end{equation}
Setting $\xi_{1}'=\xi\cos\theta$ and $\xi'_{2}=\xi\sin\theta$ where
$\xi=\xi'=|{\bf r}_{1}-{\bf r}|$, we obtain
\begin{equation}
D={NP\gamma^{2}\over 8}{1\over |\Sigma({\bf r})| }
\int_{0}^{+\infty}\xi d\xi\int_{0}^{2\pi}d\theta \sin^{2}\theta\delta
(\xi^{2}\cos\theta\sin\theta),
\label{sat6}
\end{equation}
or, equivalently,
\begin{equation}
D={NP\gamma^{2}\over 4}{1\over |\Sigma({\bf r})|   }
\int_{0}^{+\infty}{d\xi\over\xi} \int_{0}^{\pi}d\theta \sin\theta\delta
(\cos\theta).
\label{sat7}
\end{equation}
As explained previously, we regularize the logarithmic
divergence by introducing appropriate cut-offs at small and large scales. With
the change of variables $t=\cos\theta$, we finally obtain 
\begin{equation}
D={NP\gamma^{2}\over 8}{1\over |\Sigma({\bf r})|   } \ln N
\int_{-1}^{+1}dt \ \delta (t)={NP\gamma^{2}\over 8}{1\over |\Sigma({\bf r})|  }
\ln N,
\label{sat8}
\end{equation}
which establishes Eq. (\ref{dc6}) in the general case.

\subsection{The collisional relaxation time}
\label{sec_collrelax}

We can deduce from this kinetic theory the ``collisional'' relaxation
time of the point vortex gas. Considering the Fokker-Planck equation
(\ref{rs}), it is easy to check that, for $t\rightarrow +\infty$, the
distribution function $P({\bf r},t)$ will converge towards the
Boltzmann distribution (\ref{me5}). The relaxation time corresponds
typically to the time needed by a vortex to diffuse over a distance
$R$, the system size. Therefore, $t_{relax}\sim R^{2}/D$, with $D\sim
\gamma\ln N$ according to Eq. (\ref{sat8}). Using $\Gamma=N\gamma$ and
introducing the dynamical time $t_{D}\sim
\langle\omega\rangle^{-1}\sim R^{2}/\Gamma$, we obtain the estimate
\begin{equation}
t_{relax}\sim {N\over \ln N}t_{D},
\label{collrelax1}
\end{equation}
as in the case of \index{Astrophysics} collisional stellar systems
\cite{bt}. Since the statistical description is expected to yield
relevant results for large $N$, we conclude that the ``collisional''
relaxation of point vortices towards the Boltzmann distribution
(\ref{me5}) is a {\it very slow process}. It can certainly not account
for most of the numerical simulations and experiments of 2D turbulence
and point vortex dynamics in which an equilibrium state is established
{\it extremely rapidly}. This implies that a more violent relaxation
mechanism must be at work in the system (see
Sec. \ref{sec_violrelax}). This is a remark of crucial importance
because it means that all the results established previously,
including the Boltzmann distribution (\ref{me5}), must be revised.

\section{Violent relaxation of 2D vortices and stellar systems}
\label{sec_violrelax}

\subsection{The Euler and the Vlasov equations}
\label{sec_EV}

In the preceding sections, we have focused our attention to the point
vortex model as an idealization of more realistic flows which
necessarily involve a continuous vorticity distribution. This
approximation is interesting in a first approach because it leads to a
system of $N$ particles in interaction (like electric charges or
stars) for which the methods of statistical mechanics are directly
applicable. In addition, this model keeps the specificity of
two-dimensional vorticity flows such as long-range interactions
between vortices and structure formation. However, there are many
different ways to approximate a continuous vorticity field by a cloud
of point vortices and different approximations can lead to different
statistical equilibrium states (this difficulty was underlined by
Onsager \cite{onsager}). Therefore, if we want to apply the results of
statistical mechanics to realistic situations (e.g., geophysical
flows) it is necessary to go beyond the point vortex model and
develop a statistical mechanics for continuous vorticity fields.

For flows of geophysical or astrophysical  interest, the Reynolds numbers are so high that the molecular viscosity is not expected to play a crucial role in the dynamics. Therefore, these flows are described in the simplest model by the Euler-Poisson system
\begin{equation}
{\partial\omega\over\partial t}+{\bf u}\nabla\omega=0,
\label{EV1}
\end{equation}
\begin{equation}
\omega=-\Delta\psi.
\label{EV2}
\end{equation}
It can be recalled that these equations also model the early dynamics
of a cloud of point vortices before correlations between vortices have
developed (in that case, $\omega$ is proportional to the one-body
distribution function $P({\bf r},t)$, see Eq. (\ref{Eulershort})). As
discussed in Sec. \ref{sec_collrelax}, this is the regime of physical
interest since the ``collisional'' relaxation of point vortices is in
general very slow. 

Similarly, for a majority of stellar systems, including the
\index{Astrophysics} important class of elliptical galaxies, the
relaxation time by two-body encounters is $\sim10^{12}$ times larger
than the age of the universe. Therefore, the dynamics of stars in a
galaxy is essentially {\it collisionless} and appropriately described
by the Vlasov-Poisson system
\begin{equation}
{\partial f\over\partial t}+{\bf v}{\partial f\over\partial {\bf r}}+{\bf F}{\partial f\over\partial {\bf v}}=0,
\label{EV3}
\end{equation}
\begin{equation}
\Delta\Phi=4\pi G\int f d^{3}{\bf v}.
\label{EV4}
\end{equation}

The morphological similarity of the Euler-Poisson and Vlasov-Poisson
systems is another manifestation of the close analogy between 2D
vortices and stellar systems. If we make the correspondance between
the vorticity and the distribution function $(\omega\leftrightarrow
f)$ and between the stream function and the gravitational potential
$(\psi\leftrightarrow \Phi)$, these two equations describe the
advection of a density by an incompressible flow with which it
interacts via a Poisson equation. Then, the density is not advected
passively by the flow but is coupled to its motion. This coupling is
responsible for violent fluctuations of the stream function or
gravitational potential. These fluctuations will mix the vorticity or
the phase elements at small scales and induce a self-organization and
the appearance of structures at larger scales (see
Fig. \ref{chava-fig9}). This \index{Violent relaxation} violently
changing potential provides a mechanism analogous to a relaxation in a
gas, but the specificity of this relaxation is that it is {\it
collisionless} and due to the long-range nature of the
interactions. It is now clear that \index{Chaotic mixing} this
``chaotic mixing'' is the driving source of relaxation in
two-dimensional turbulence and stellar dynamics. The kinetic theory
presented in Sec. \ref{sec_kin} for point vortices (and the one
developed by Chandrasekhar for stars) is only valid in situations in
which this chaotic mixing is prevented or has died away.

\begin{figure}
\centering
\includegraphics[width=0.7\textwidth]{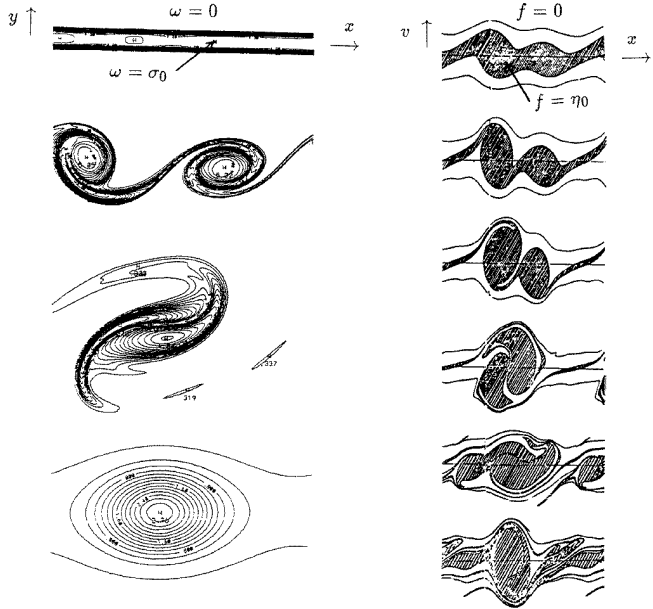}
\caption[]{Violent relaxation of two-dimensional vorticity flows and self-gravitating systems. The left panel corresponds to the nonlinear developement of the Kelvin-Helmholtz instability. The equilibrium state is a large-scale vortex which is well-described by statistical mechanics \cite{staquet}. The right panel corresponds to a simulation of the Vlasov-Poisson system showing a mixing process and the formation of a coherent structure in phase space with a Fermi-Dirac distribution \cite{bertrand}. This simulation is restricted to a one-dimensional system but the process remains the same in higher dimensions.   }
\label{chava-fig9}
\end{figure}

\subsection{The statistical equilibrium}
\label{sec_se}

During the mixing process, the Vlasov-Poisson and the Euler-Poisson
systems generate intermingled filaments at smaller and smaller
scales. Therefore, a {\it deterministic} description of the flow would
require a rapidly increasing amount of information as time goes
on. For that reason, it is appropriate to undertake a {\it
probabilistic} description in order to smooth out the small scales and
concentrate on the locally averaged quantities. This statistical
approach, called the theory of ``violent relaxation'', was introduced
by Lynden-Bell \cite{lb} in 1967 for collisionless stellar systems and
rediscovered independantly by Kuzmin \cite{kuzmin}, Miller
\cite{miller,weichman} and Robert \& Sommeria \cite{rs1} for two-dimensional
vorticity flows. The analogy between these two statistical mechanics
(including the relaxation towards equilibrium) was discussed in detail
by Chavanis \cite{cthese,csr,cfloride,japon}. On the other hand, a
rigourous justification of this statistical approach has been given by
Robert \cite{rob} by using the concept of Young measures and large
deviations.  This theory improves upon previous works based on a
spectral representation of the flow \cite{kraichnan}, which do not
respect all the conservation laws of the invisicd dynamics. In the
following, we present the statistical mechanics of violent relaxation
for two-dimensional vorticity flows.  A discussion of the statistical
mechanics of violent relaxation in stellar systems, closely following
the presentation of this paper, can be found in
\cite{dubrovnik}.

In the statistical approach, the exact knowledge of the ``fine-grained'' or microscopic vorticity field is replaced by the probability density $\rho({\bf r},\sigma)$ of finding the vorticity level $\sigma$ in ${\bf r}$. The normalization condition yields at each point 
\begin{equation}
\int\rho({\bf r},\sigma)d\sigma =1,
\label{se1}
\end{equation} 
and the locally averaged (coarse-grained) vorticity is expressed in terms of the density probability in the form
\begin{equation}
\overline{\omega}=\int\rho({\bf r},\sigma)\sigma d\sigma.
\label{se2}
\end{equation} 
More generally, the moments of the vorticity are defined by
\begin{equation}
\overline{\omega^{n}}=\int \rho({\bf r},\sigma)\sigma^{n}d\sigma.
\label{se2bis}
\end{equation} 
During the evolution, the energy
\begin{equation}
E={1\over 2}\int \overline{\omega}\psi d^{2}{\bf r},
\label{se3}
\end{equation} 
is conserved as well as the total area of each level of vorticity
\begin{equation}
\gamma(\sigma)=\int\rho({\bf r},\sigma)d^{2}{\bf r}.
\label{se4}
\end{equation} 
These last constraints are equivalent to the conservation of the
Casimir integrals $C_{h}=\int h({\omega}) d^{2}{\bf r}$ for any
continuous function $h$. Such integrals are conserved by the Euler
equation because the fluid particles keep their vorticity (on account
of the transport equation $d\omega/dt=0$) and their surface (on
account of the incompressibility of the flow).

After a complex evolution the system is expected to be in the most
probable, i.e. most mixed state, consistent with all the constraints
imposed by the dynamics. We define the mixing entropy as the logarithm
of the number of microscopic configurations associated with the same
macroscopic state (characterized by the probability density $\rho({\bf
r},\sigma)$). We divide the macrocells $({\bf r},{\bf r}+d{\bf r})$
into $\nu$ microcells and denote by $n_{ij}$ the number of microcells
occupied by the vorticity level $\sigma_{j}$ in the $i$-th
macrocell. A simple combinatorial argument indicates that the number
of microstates associated with the macrostate $\lbrace n_{ij}\rbrace$
is
\begin{equation}
W(\lbrace n_{ij}\rbrace)=\prod_{i}N_{j}!\prod_{i}{\nu!\over n_{ij}!},
\label{se5}
\end{equation} 
where $N_{j}=\sum_{i}n_{ij}$ is the total number of microcells
occupied by $\sigma_{j}$. We have to add the normalization condition
$\sum_{j}n_{ij}=\nu$, equivalent to Eq. (\ref{se1}), which prevents
overlapping of different vorticity levels. This constraint plays a
role similar to the Pauli exclusion principle in quantum
mechanics. Morphologically, the statistics (\ref{se5}) corresponds to
a $4^{\rm th}$ type of statistics since the particles are
distinguishable but subject to an exclusion principle \cite{lb}. There
is no such exclusion for point vortices (in the collisional regime)
since they are free a priori to approach each other without
limitation.

Taking the logarithm of $W$ and passing to the continuum limit with the aid 
of the Stirling formula, we get 
\begin{equation}
S=-\int \rho({\bf r},\sigma)\ln\rho({\bf r},\sigma)d^{2}{\bf r}d\sigma.
\label{se6}
\end{equation} 
The most probable macroscopic state is obtained by maximizing the
mixing entropy (\ref{se6}) with fixed energy (\ref{se3}), global
vorticity distribution (\ref{se4}) and local normalization
(\ref{se1}). This problem is treated by introducing Lagrange
multipliers, so that the first variations satisfy
\begin{equation}
\delta S-\beta\delta E-\int\alpha(\sigma)\delta\gamma(\sigma)d\sigma-\int\zeta({\bf r})\delta\biggl (\int \rho({\bf r},\sigma)d\sigma\biggr )d^{2}{\bf r}=0.
\label{se7}
\end{equation} 
The resulting optimal probability is a Gibbs state which can be expressed as
\begin{equation}
\rho({\bf r},\sigma)={1\over Z(\psi)}g(\sigma)e^{-\beta\sigma\psi},
\label{se8}
\end{equation} 
where $Z(\psi)\equiv e^{\zeta({\bf r})+1}$ and $g(\sigma)\equiv
e^{-\alpha(\sigma)}$.  The normalization condition (\ref{se1}) leads to a
value of the partition function $Z$ of the form
\begin{equation}
Z=\int g(\sigma)e^{-\beta\sigma\psi}d\sigma,
\label{se9}
\end{equation} 
and the locally averaged vorticity (\ref{se2}) is expressed as a function of $\psi$ according to 
\begin{equation}
\overline{\omega}={\int g(\sigma)\sigma e^{-\beta\sigma\psi}d\sigma\over \int g(\sigma)e^{-\beta\sigma\psi}d\sigma}=f(\psi).
\label{se10}
\end{equation} 
This expression can be rewritten
\begin{equation}
\overline{\omega}=-{1\over\beta}{d\ln Z\over d\psi}.
\label{se11}
\end{equation} 
Differentiating Eq. (\ref{se10}) with respect to $\psi$, we check that the variance of the vorticity can be written \cite{shallow}
\begin{equation}
\omega_{2}\equiv\overline{\omega^{2}}-\overline{\omega}^{2}=-{1\over\beta}f'(\psi),
\label{se12}
\end{equation} 
or, alternatively,
\begin{equation}
\omega_{2}={1\over\beta^{2}}{d^{2}\ln Z\over d\psi^{2}}.
\label{se13}
\end{equation} 
Therefore, the slope of the function $\overline{\omega}=f(\psi)$ is
directly related to the variance of the vorticity distribution. Since
$\omega_{2}>0$, we find that the function $\overline{\omega}=f(\psi)$
is monotonic; it is decreasing for $\beta>0$ and increasing for
$\beta<0$ (it is constant for $\beta=0$). Another proof of this result
is given in \cite{rs1}.

Two particular cases are worth considering. If the local distribution
of vorticity is Gaussian, then the $\omega-\psi$ relationship
(\ref{se10}) is linear \cite{miller,weichman}. On the other hand, in
the case of a single level of vorticity $\sigma_{0}$ (in addition to
the level $\sigma=0$), the coarse-grained vorticity $\omega=\rho({\bf
r},\sigma_{0})\sigma_{0}$ takes explicitly the form
\begin{equation}
\overline{\omega}={\sigma_{0}\over 1+\lambda e^{\beta\sigma_{0}\psi}}.
\label{se14}
\end{equation} 
This is formally similar to the Fermi-Dirac statistics. Here, the exclusion principle $\overline{\omega}\le\sigma_{0}$ is due to the Liouville theorem (i.e., the conservation of the fine-grained vorticity) not to quantum mechanics. Because of the averaging procedure, the coarse-grained vorticity can only {\it decrease} by internal mixing, as irrotational flow is incorporated into the patch $\sigma_{0}$, and this results in an ``effective'' exclusion principle. In the dilute limit $\overline{\omega}\ll\sigma_{0}$, Eq. (\ref{se14}) reduces to the Boltzmann distribution $\overline{\omega}=A e^{-\beta\sigma_{0}\psi}$ as in the point vortex model \cite{jm,pl}.

Similar results are obtained in the  context of collisionless stellar system \index{Astrophysics} described by the Vlasov-Poisson system \cite{lb,dubrovnik}. The equivalent of Eq. (\ref{se14}) is the Lynden-Bell distribution function 
\begin{equation}
\overline{f}={\eta_{0}\over 1+\lambda e^{\beta\eta_{0}({v^{2}\over 2}+\Phi)}},
\label{se15}
\end{equation} 
which formally coincides with the distribution function of the self-gravitating Fermi gas. In the non-degenerate limit $\overline{f}\ll\eta_{0}$, it reduces to the Maxwell-Boltzmann distribution $\overline{f}=Ae^{-\beta\eta_{0}({v^{2}\over 2}+\Phi)}$. Therefore, the theory of violent relaxation naturally explains the observed isothermal cores of elliptical galaxies without recourse to collisions \cite{lb,hjorth}.

\subsection{The Maximum Entropy Production Principle}
\label{sec_MEPP}

Basically, a two-dimensional incompressible turbulent flow at high
Reynolds numbers is described by the Euler-Poisson system. In
principle, these equations determine completely the evolution of the
vorticity field $\omega({\bf r},t)$. However, in practice, we are not
interested by the finely striated structure of the flow but only by
its smoothed-out structure. Indeed, the observations and the numerical
simulations are always realized with a finite resolution. Moreover,
the coarse-grained vorticity $\overline{\omega}$ is expected to
converge towards an equilibrium state, the Gibbs state (\ref{se8}),
contrary to the exact vorticity field $\omega$ which develops smaller
and smaller scales. There is also a technical difficulty to simulate
an inviscid dynamics due precisely to the developement of this
small-scale motion. Contour dynamics methods need to introduce a
``surgery'' and spectral codes a ``viscosity'' in order to prevent
numerical instabilities. However, this artificial viscosity breaks the
conservation laws of the Euler equations. What we would like to obtain
is a set of \index{Relaxation} relaxation equations which smooth out
the small scales while conserving all the constraints of the Euler
equation. Such a parametrization has been proposed by Robert \&
Sommeria \cite{rs} in terms of a phenomenological Maximum Entropy
Production Principle (MEPP).

Let us decompose the vorticity $\omega$ and velocity ${\bf u}$ into
a mean and a fluctuating part, namely $\omega=\overline{\omega
}+\tilde{\omega }$,
${\bf{u}}=\overline{\bf{u}}+\tilde{\bf{u}}$. Taking the local average
of the Euler equation (\ref{EV1}), we get
\begin{equation}
\label{mepp1}
{\partial \overline{\omega} \over \partial t}+\nabla(\overline{\omega} 
\ \overline{\bf  u})=-\nabla {\bf  J}_{\omega},
\end{equation}
where the current $\bf{J}_{\omega }=\overline{\tilde{\omega
}\tilde{\mathbf{u}}}$ represents the correlations of the fine-grained
fluctuations. Eq. (\ref{mepp1}) can be viewed as a local conservation
law for the circulation $\Gamma=\int
\overline{\omega}d^{2}{\bf{r}}$. To apply the MEPP, we need to
consider not only the locally averaged vorticity field
$\overline{\omega}$ but the whole probability distribution $\rho
({\bf{r}},{\sigma },t)$ now evolving with time $t$.  The conservation
of the global vorticity distribution $\gamma (\sigma )=\int \rho
d^{2}{\bf{r}}$ can be written in the local form
\begin{equation}
\label{mepp2}
{\partial \rho\over \partial t}+\nabla(\rho{\bf  u})=-\nabla {\bf  J},
\end{equation}
where $\bf{J}({\bf{r}},{\sigma },t)$ is the (unknown) current
associated with the vorticity level $\sigma$. Integrating
Eq. (\ref{mepp2}) over all the vorticity levels $\sigma$, using
Eq. (\ref{se1}), and comparing with Eq. (\ref{pf3}), we find the constraint
\begin{equation}
\label{mepp3}
\int {\bf{J}}({\bf{r}},\sigma ,t)d\sigma ={\bf 0}.
\end{equation}
Multiplying Eq. (\ref{mepp2}) by $\sigma$, integrating over all the vorticity levels, using Eq. (\ref{se2}), and comparing with Eq. (\ref{mepp1}), we get
$\int {\bf{J}}({\bf{r}},\sigma ,t)\sigma d\sigma ={\bf{J}}_{\omega }$.

We can express the time variation of energy in terms of ${\bf J}$,
using Eqs. (\ref{se3}) and (\ref{mepp1}), leading to the energy
conservation constraint
\begin{equation}
\label{mepp4}
\dot{E}=\int {\bf{J}}_{\omega}\nabla\psi \ d^{2}{\bf r}=0.
\end{equation}
Using Eqs. (\ref{se6}) and (\ref{mepp2}), we similarly express the
rate of entropy production as
\begin{equation}
\label{mepp5}
\dot{S}=-\int {\bf{J}} \nabla (\ln \rho )d^{2}{\bf{r}}d\sigma .
\end{equation}

The Maximum Entropy Production Principle (MEPP) consists in choosing
the current ${\bf{J}}$ which maximizes the rate of entropy production
$\dot{S}$ respecting the constraints $\dot{E}=0$, $\int {\bf
J}d\sigma={\bf 0}$ and $\int {J^{2}\over 2\rho }d\sigma \leq
C({\mathbf{r}},t)$.  The last constraint expresses a bound (unknown)
on the value of the diffusion current. Convexity arguments justify
that this bound is always reached so that the inequality can be
replaced by an equality. The corresponding condition on first
variations can be written at each time $t$:
\begin{eqnarray}
\label{mepp6}
\delta \dot{S}-\beta (t)\delta \dot{E}-\int {\mb \zeta} ({\mathbf{r}},t)\delta \biggl (\int {\mathbf{J}}d\sigma \biggr )d^{2}{\mathbf{r}}-\int D^{-1}({\mathbf{r}},t)\delta \biggl (\int {J^{2}\over 2\rho} d\sigma\biggr ) d^{2}{\mathbf{r}}=0,\nonumber\\
\end{eqnarray}
and leads to a current of the form 
\begin{equation}
\label{mepp7}
{\mathbf{J}}=-D({\bf{r}},t)\biggl \lbrack \nabla \rho +\beta (t)\rho (\sigma -\overline{\omega})\nabla\psi\biggr \rbrack .
\end{equation}
The Lagrange multiplier ${\mb\zeta}$ has been eliminated, using the
condition (\ref{mepp3}) of local normalization conservation. The
conservation of energy (\ref{mepp4}) at any time determines the
evolution of the Lagrange multiplier $\beta (t)$ according to
\begin{equation}
\label{mepp8}
\beta (t)=-{\int D\nabla \overline{\omega}\nabla\psi d^{2}{\bf{r}}\over \int D\omega_{2}(\nabla\psi)^{2}d^{2}{\bf{r}}}.
\end{equation}

The entropy production (\ref{mepp5}) can be written
\begin{equation}
\label{mepp9}
\dot S=-\int {\bf J\over\rho}(\nabla\rho+\beta\rho(\sigma-\overline{\omega})\nabla\psi)d^{2}{\bf r}d\sigma+\beta\int {\bf J}(\sigma-\overline{\omega})\nabla\psi d^{2}{\bf r}d\sigma.
\end{equation}
Using Eqs. (\ref{mepp3}) and (\ref{mepp4}), the second integral is seen to
cancel out. Inserting Eq. (\ref{mepp7}) in the first integral, we
find
\begin{equation}
\label{mepp10}
\dot{S}=\int {J^{2}\over D\rho }d^{2}{\bf{r}}d\sigma,
\end{equation}
which is positive provided that $D\geq 0$. A stationary solution
$\dot{S}=0$ is such that ${\bf{J}}={\bf 0}$ yielding, together
with Eq. (\ref{mepp7}),
\begin{equation}
\label{mepp11}
\nabla (\ln \rho )+\beta (\sigma -\overline{\omega})\nabla \psi ={\bf 0}.
\end{equation}
For any reference vorticity level $\sigma_{0}$, it writes 
\begin{equation}
\label{mepp12}
\nabla (\ln \rho _{0})+\beta (\sigma _{0}-\overline{\omega})\nabla \psi ={\bf 0}.
\end{equation}
Substracting Eqs. (\ref{mepp11}) and (\ref{mepp12}), we obtain $\nabla
\ln ({\rho /\rho _{0}})+\beta (\sigma -\sigma _{0})\nabla \psi ={\bf
0}$, which is immediately integrated into
\begin{equation}
\label{mepp13}
\rho ({\bf{r}},\sigma )={1\over Z({\bf{r}})}g(\sigma )e^{-\beta \sigma \psi },
\end{equation}
where $Z^{-1}({\bf{r}})\equiv \rho _{0}({\bf{r}})e^{\beta
\sigma _{0}\psi ({\bf{r}})}$ and $g(\sigma )\equiv e^{A(\sigma
)}$, $A(\sigma )$ being a constant of integration. Therefore, entropy
increases until the Gibbs state (\ref{se8}) is reached, with $\beta
=\lim _{t\rightarrow \infty }\beta (t)$. Furthermore, we can show that
a stationary solution of these relaxation equations is linearly stable
if, and only if, it is an entropy {\it maximum} (in
preparation). Therefore, this numerical algorithm selects the maxima
(and not the minima or the saddle points) among all critical points of
entropy. When several entropy maxima subsist for the same values of
the constraints, the choice of equilibrium depends on a complicated
notion of ``basin of attraction'' and not simply whether the solution
is a local or a global entropy maximum (see Ref. \cite{crs} in a
related context).

The relaxation equations (\ref{mepp2}), (\ref{mepp7}) and (\ref{mepp8})
can be simplified in the single level approximation. In that case,
Eq. (\ref{mepp1}) is explicitly given by
\begin{equation}
\label{mepp14}
{\partial\overline{\omega}\over\partial t}+\overline{{\bf u}}\nabla\overline{\omega}=\nabla\biggl\lbrack D\biggl (\nabla\overline{\omega}+\beta(t)\overline{\omega}(\sigma_{0}-\overline{\omega})\nabla\psi\biggr )\biggr\rbrack.
\end{equation}
In the dilute limit $\overline{\omega}\ll\sigma_{0}$, it takes a form
similar to the Fokker-Planck equation (\ref{pol1}) obtained for point
vortices. These equations both involve a diffusion and a drift, but
these terms have a different physical interpretation in each case.

It is also instructive to apply \index{Astrophysics} this thermodynamical approach to the
Vlasov-Poisson system. In the single level approximation, it leads to
the following equation for the coarse-grained distribution function
\cite{csr,cg}:
\begin{equation}
\label{mepp15}
{\partial \overline{f}\over \partial t}+{\bf
v}{\partial\overline{f}\over\partial{\bf r}}+{\bf
F}{\partial\overline{f}\over\partial{\bf
v}}={\partial\over\partial{\bf v}}\biggl\lbrack D \biggl
({\partial\overline{f}\over\partial{\bf v }}+\beta(t)\overline{f}
(\eta_{0}-\overline{f}){\bf v}\biggr )\biggr\rbrack,
\end{equation}
which is a generalized form of the familiar Fokker-Planck equation
(\ref{pol3}) recovered in the non degenerate limit
$\overline{f}\ll\eta_{0}$. We can check that Eq. (\ref{mepp15}) 
returns the Lynden-Bell distribution function (\ref{se15}) at
equilibrium.

The diffusion coefficient $D$ is not determined by the MEPP as it
depends on the unknown bound $C$ on the current. For the purpose of
reaching the Gibbs state (\ref{se8}), the diffusion coefficient can
simply be chosen arbitrarily (but with a positive value in order to
ensure entropy increase). However, the precise form of the diffusion
coefficient is important in order to determine the relaxation time and
take into account kinetic confinement and incomplete relaxation (see
below). In the context of 2D turbulence, Robert \& Rosier \cite{rr}
have proposed a simple evaluation of $D$ by using an analogy with the
diffusion of a passive scalar $\omega$ subjected to a turbulent
velocity field ${\bf u}=\overline{{\bf u}}+\tilde{\bf u}$ (see also
\cite{csr}, Appendix B). In that case, the mean value
$\overline{\omega}$ satisfies a convection-diffusion equation
\begin{equation}
\label{mepp16}
{\partial\overline{\omega}\over\partial t}+({\bf u}\nabla)\overline{\omega}=\nabla (D\nabla\overline{\omega}),
\end{equation}
with a diffusion coefficient given by
\begin{equation}
\label{mepp17}
D={1\over 4}\tau\overline{{\tilde {\bf u}}^{2}}({\bf r},t),
\end{equation}
where $\tau$ is the decorrelation time of the velocity
fluctuations. Equation (\ref{mepp14}) reduces to Eq. (\ref{mepp17})
when $\beta=0$, i.e. when the energy constraint is not active. In the
Euler-Poisson system, the velocity is produced by the vorticity itself
via the Biot \& Savart formula
\begin{equation}
\label{mepp18}
{\bf u}({\bf r},t)=\int\omega({\bf r}',t){\bf V}({\bf r}'\rightarrow {\bf r})d^{2}{\bf r}',
\end{equation}
where ${\bf V}({\bf r}'\rightarrow {\bf r})=-(1/2\pi){\bf z}\times ({\bf r}'-{\bf r})/|{\bf r}'-{\bf r}|^{2}$. Inserting Eq. (\ref{mepp18}) in Eq. (\ref{mepp17}), we obtain 
\begin{equation}
\label{mepp19}
D={\tau\over 4}\int \overline{\tilde\omega({\bf r}',t)\tilde\omega({\bf r}'',t)}{\bf V}({\bf r}'\rightarrow {\bf r}){\bf V}({\bf r}''\rightarrow {\bf r})d^{2}{\bf r}'d^{2}{\bf r}''.
\end{equation}
If we neglect the spatial correlations of the vorticity fluctuations on scales larger than $\epsilon$, the resolution scale, we have 
\begin{equation}
\label{mepp20}
\overline{\tilde\omega({\bf r}',t)\tilde\omega({\bf r}'',t)}=\epsilon^{2}\overline{\tilde\omega^{2}}({\bf r}',t)\delta ({\bf r}'-{\bf r}''),
\end{equation}
and Eq. (\ref{mepp19}) reduces to
\begin{equation}
\label{mepp21}
D={\tau\over 4}\epsilon^{2}\int \overline{\tilde\omega^{2}}({\bf r}',t){\bf V}^{2}({\bf r}'\rightarrow {\bf r})d^{2}{\bf r}'.
\end{equation}
Making a local approximation and introducing an upper cut-off $a$, we obtain
\begin{equation}
\label{mepp22}
D={\tau\over 4}\epsilon^{2}\overline{\tilde\omega^{2}}({\bf r},t)\int_{\epsilon}^{a} \biggl ({1\over 2\pi \xi}\biggr )^{2}2\pi \xi d\xi,
\end{equation}
with $\overline{\tilde\omega^{2}}=\overline{\omega^{2}}-\overline{\omega}^{2}$.
This leads to the following expression for the diffusion coefficient
\begin{equation}
\label{mepp23}
D=(\overline{\omega^{2}}-\overline{\omega}^{2}){\tau\epsilon^{2}\over 8\pi}\ln\biggl 
({a\over\epsilon}\biggr ). 
\end{equation}
We note that the diffusion coefficient vanishes in regions where there
is no fluctuation of the vorticity at small scales,
i.e. $\omega_{2}=0$. At the contact with the unmixed flow, the
diffusion current also vanishes resulting in a confinement of the
vorticity. This leads to the \index{Incomplete relaxation} concept of
{\it incomplete relaxation}: in a large-scale turbulent flow, the
Gibbs state (\ref{se8}) is satisfied only in restricted regions of
space where mixing is sufficiently efficient to justify an ergodic
hypothesis. Outside these domains, the relaxation is slowed down or
even stopped. This kinetic confinement is illustrated by the numerical
simulations of Robert \& Rosier \cite{rr} and further discussed by
Chavanis \& Sommeria
\cite{csfluide}. In this viewpoint, the vortices of two-dimensional
turbulence can be considered as restricted equilibrium states or {\it
maximum entropy bubbles} \cite{csfluide} separated from each other by
an almost irrotational background. A similar kinetic confinement can
be advocated in the case of stellar systems \cite{csr,dubrovnik} to
account for incomplete relaxation \cite{lb} and solve the infinite
mass problem.

\subsection{Recent developements}
\label{sec_del}

An interesting problem in fluid dynamics is to obtain a classification
of the ``zoology'' of vortices (monopoles, translating and rotating
dipoles, tripoles...) met in two-dimensional flows. The statistical
mechanics approach presented in Sec. \ref{sec_se} provides a general
framework to tackle this problem as it selects the most probable
structures among all possible solutions of the 2D Euler
equations. However, the prediction is not straightforward because, in
the general case, we have to take into account an infinite set of
constraints, namely the conservation of all the Casimirs $C_{h}$, in
addition to energy $E$, angular momentum $L$ and impulse $P$. A
numerical algorithm has been developed by Turkington \& Whitaker
\cite{tw} to solve this problem and several calculations have been
performed in rectangular or circular domains for a finite number of
vorticity levels and for particular values of the integral
constraints. However, many structures are found and it is difficult to
have a clear picture of the bifurcation diagram in parameter
space. For that reason, Chavanis \& Sommeria \cite{csfluide0,csfluide} have
considered a particular limit of the statistical theory, the so-called
``strong mixing limit'', in which the study of these bifurcations can
be performed analytically. This limit corresponds to
$\beta\sigma\psi\ll 1$ so that the equations of the problem can be
expanded in terms of this small parameter (this is like the
Debye-H\"uckel approximation in plasma physics). To zeroth order in
the expansion, the density probability of each level is uniform which
corresponds to a completely mixed state. To first order, the
relationship between vorticity and streamfunction is linear and this
can justify an inviscid minimum enstrophy principle (for the
coarse-grained enstrophy
$\Gamma_{2}^{c.g.}=\int\overline{\omega}^{2}d^{2}{\bf r}$)
\cite{csfluide0}. In that case, the equilibrium flow only depends on
$E$, $L$, $P$ and the first moment $\Gamma$ of the vorticity (the
fine-grained enstrophy
$\Gamma_{2}^{f.g.}=\int\overline{\omega^{2}}d^{2}{\bf r}$ serves as a
normalization factor). This particular limit of the statistical theory
already exhibits a rich bifurcation diagram and often provides a good
approximation of more general situations (in particular for weakly
energetic flows). It is also possible, in principle, to go to higher
orders in the expansion in which case more and more vorticity moments
$\Gamma_{3},\Gamma_{4},...$ are necessary to describe the structure of
equilibrium. This limit makes therefore a hierarchy between the
constraints as it shows that, in many situations, only the lowest
moments of the vorticity are important to characterize the equilibrium
state.  Because of this hierarchy, we can make some relevant
predictions without the complete knowledge of the initial condition.

\begin{figure}
\centering
\includegraphics[angle=0,width=0.5\textwidth]{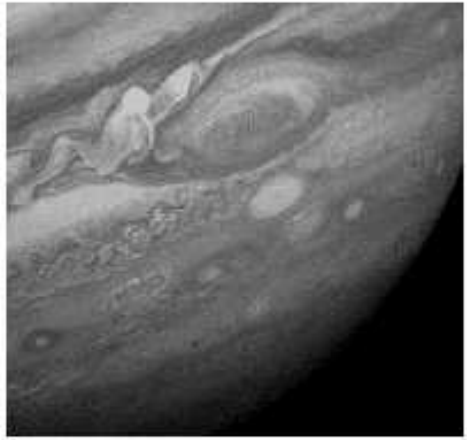}
\caption[]{Jupiter's great red spot}
\label{redspots}
\end{figure}

The statistical mechanics approach can be extended immediately to the
quasi-geostrophic (QG) equations \cite{michel} by simply replacing the
vorticity by the potential vorticity (PV). The formalism has also been
generalized by Chavanis \& Sommeria \cite{shallow} to the
shallow-water (SW) equations.  These equations are more relevant to
describe geophysical flows than the 2D Euler equations
\cite{pedlosky}. In the QG approximation, Bouchet \& Sommeria
\cite{bouchet1} have explained the formation of jets and vortices in
planetary atmospheres in terms of statistical mechanics, as initiated
in \cite{nore}. In particular, the annular jet structure of Jupiter's
Great Red Spot (see Figs. \ref{redspots}-\ref{freddy}) is reproduced
and explained as the coexistence of two thermodynamical phases in
contact (a picture which is rigorously valid in a small Rossby radius
expansion).  These results can be extended to the more general case of
shallow water equations \cite{bouchet2}. In this geophysical context,
the deformation of the fluid surface tends to reduce the range of
interaction between vortices. This situation is comparable to what
happens in a neutral plasma: the interaction between vortices is
shielded on a distance of the order of the Rossby length, the
analogous of the Debye length in plasma physics.

\begin{figure}
\centering
\includegraphics[angle=0,width=0.7\textwidth]{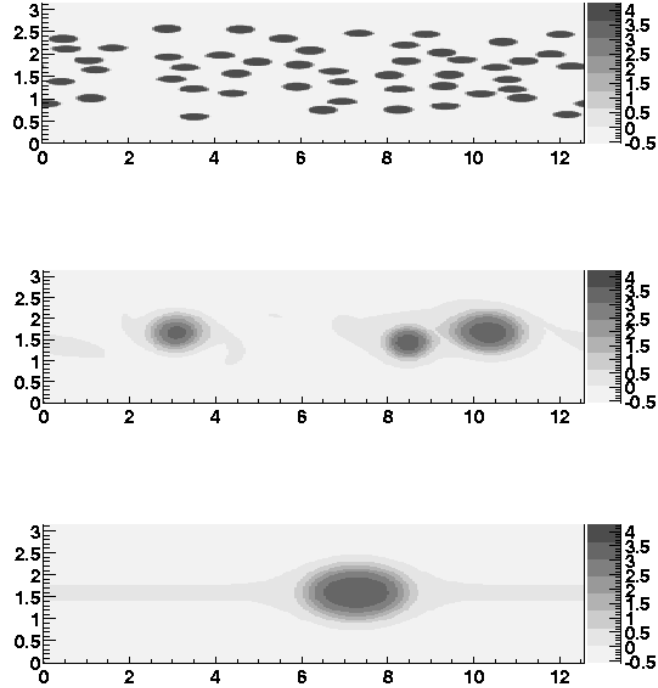}
\caption[]{Relaxation towards statistical equilibrium in a QG model of Jupiter's great red spot (from Ref. \cite{bouchet0}). Three successive potential vorticity (PV) fields are represented as grey levels. The initial condition (top) is made of small PV patches. These patches organize into vortices (middle) that eventually merge into a single one (bottom). This sequence is obtained with the relaxation equations described in Sec. \ref{sec_MEPP}, applied to the Q.G. situation: entropy increases with time while energy is exactly conserved. At equilibrium, the vortex is an oval spot of quasiuniform PV surrounded by strong gradients, corresponding to an annular jet.    }
\label{freddy}
\end{figure}

The relaxation equations presented in Sec. \ref{sec_MEPP} provide a
convenient parameterization of sub-grid scale eddies which drives the
system toward statistical equilibrium by a continuous time
evolution. Such relaxation equations can be used both as a realistic
coarse resolution model of the turbulent evolution, and as a method of
determination of the statistical equilibrium resulting from this
evolution. However, these equations do not preserve the invariance
properties of the Euler equations by a change of reference frame. In
addition, the conservation of energy is enforced by a formal Lagrange
multiplier $\beta(t)$ which is {\it uniform} in space. This may be a
limitation to describe large-scale turbulent flows which organize
locally in several types of structures with a different temperature. A
generalization of the MEPP has been proposed by Chavanis \& Sommeria
\cite{csthermo} as an attempt to solve these difficulties. More general,
but also more complex, relaxation equations are obtained. They involve
a space dependant temperature which tends to be uniform in each vortex
but with an a priori different value from one vortex to the other. A
simplified version of these relaxation equations has been solved
numerically by Kazantzev {\it et al.} \cite{kaz} in an oceanic
context.

Despite its practical interest, the main drawback of the MEPP is its {\it ad
hoc} nature. In \cite{kin1}, we have attempted to justify the
relaxation equation (\ref{mepp14}) from first principles starting
directly from the 2D Euler equation. A systematic derivation can be
obtained in the so-called quasilinear approximation. This
approximation is well-known in plasma physics and stellar dynamics
\cite{kp,sl,dubrovnik} for the Vlasov-Poisson system and we tried to extend it
to the Euler-Poisson system. Substracting Eq. (\ref{EV1}) and
(\ref{mepp1}) and neglecting the non linear terms in the equation for
$\tilde\omega$, the basic equations of the quasilinear theory are
\begin{equation}
\label{q1}
{\partial\overline{\omega}\over\partial t}+\overline{\bf u}\nabla \overline{\omega}=-\nabla\overline{\tilde\omega\tilde {\bf u}},
\end{equation}
\begin{equation}
\label{q2}
{\partial\tilde{\omega}\over\partial t}+\overline{\bf u}\nabla \tilde{\omega}=-\tilde{\bf u}\nabla\overline{\omega}.
\end{equation}
It is possible to solve Eq. (\ref{q2}) formally with the aid of Green's
functions and substitute the resulting expression for
$\tilde\omega({\bf r},t)$ back into Eq. (\ref{q1}). Implementing a closure
relation of the form (\ref{mepp20}), a kinetic equation can be obtained for
$\overline{\omega}$. In the single level approximation, it reads \cite{kin1}:
\begin{eqnarray}
\label{q3}
{\partial\overline{\omega}\over\partial t}+\overline{\bf u}\nabla \overline{\omega}={\epsilon^{2}}{\partial\over\partial r^{\mu}}\int_{0}^{t}ds\int d^{2}{\bf r}'V^{\mu}({\bf r}'\rightarrow {\bf r})_{t}\nonumber\\
\times\biggl\lbrace V^{\nu}({\bf r}'\rightarrow {\bf r})\overline{\omega}'(\sigma_{0}-\overline{\omega}'){\partial\overline{\omega}\over\partial r^{\nu}}+ V^{\nu}({\bf r}\rightarrow {\bf r}')\overline{\omega}(\sigma_{0}-\overline{\omega}){\partial\overline{\omega}'\over\partial r'^{\nu}}\biggr\rbrace_{t-s}.
\end{eqnarray}

Assuming in addition that the decorrelation time $\tau$ is short, as
in the stochastic model of Robert \& Rosier \cite{rr}, we are led to
the following equation
\begin{equation}
\label{q4}
{\partial\overline{\omega}\over\partial t}+\overline{\bf u}\nabla \overline{\omega}={\epsilon^{2}\tau\over 8\pi^{2}}{\partial\over\partial r^{\mu}}\int d^{2}{\bf r}'K^{\mu\nu}({\bf r}'-{\bf r})\biggl\lbrace \overline{\omega}'(\sigma_{0}-\overline{\omega}'){\partial\overline{\omega}\over\partial r^{\nu}}-\overline{\omega}(\sigma_{0}-\overline{\omega}){\partial\overline{\omega}'\over\partial r'^{\nu}}\biggr\rbrace,
\end{equation}
\begin{equation}
\label{q5}
K^{\mu\nu}({\bf r}'-{\bf r})={\xi^{2}\delta^{\mu\nu}-\xi^{\mu}\xi^{\nu}\over \xi^{4}},
\end{equation}
with ${\mb\xi}={\bf r}'-{\bf r}$. This equation includes a diffusion
and a drift, as in the MEPP, but these terms are obtained here
directly from a local average of the Euler equation. On the other
hand, the conservation of angular momentum results from the symmetry
of the diffusion current instead of an ad hoc Lagrange
multiplier. This symmetric structure respects in addition the
invariance properties of the Euler equation. Finally, a $H$-theorem
for the Fermi-Dirac entropy
\begin{equation}
\label{q6}
S=-\int \biggl \lbrace {\overline{\omega}\over\sigma_{0}}\ln {\overline{\omega}\over\sigma_{0}}+\biggl (1-{\overline{\omega}\over\sigma_{0}}\biggr )\ln \biggl (1-{\overline{\omega}\over\sigma_{0}}\biggr )\biggr\rbrace d^{2}{\bf r},
\end{equation}
can be {\it derived} from this kinetic equation instead of being {\it
postulated} as in the MEPP. Our approach provides therefore an
alternative, dynamical, justification of the mixing entropy 
introduced by Miller \cite{miller} and Robert \& Sommeria \cite{rs} at
statistical equilibrium.

Equations (\ref{q4}) and (\ref{q5}) are similar to the kinetic
equations (\ref{genK4}) and (\ref{Kmunu}) obtained for point vortices.
There are, however, two important differences: (i) the drift and the
diffusion involve the product $\overline{\omega}\times
(\sigma_{0}-\overline{\omega})$ instead of
$\langle\omega\rangle$. This nonlinearity ensures that the constraint
$\overline{\omega}\le \sigma_{0}$ is satisfied at any time. (ii) The
diffusion coefficient is proportional to the circulation
$\sigma_{0}\epsilon^{2}$ of a completely filled macrocell, instead of
the circulation $\gamma$ of a point vortex. In general,
$\sigma_{0}\epsilon^{2}\gg\gamma$ so that the relaxation by
collisionless mixing is much more rapid than the collisional
relaxation. From the above theory, we find that the time scale of the
violent relaxation is of order $t_{D}$, the dynamical time, whereas the
collisional relaxation of point vortices operates on a time scale $\sim
(N/\ln N)t_{D}$.

If we are close to equilibrium, we can implement a ``thermal bath
approximation'' and replace the vorticity $\omega'\equiv \omega({\bf
r}',t)$ by its equilibrium form (\ref{se14})\cite{kin1}. Then, Eq. (\ref{q4})
reduces to the equation (\ref{mepp14}) derived from the MEPP and the
diffusion coefficient coincides with the estimate (\ref{mepp23})
based on the passive scalar model 
($\omega_{2}=\overline{\omega}(\sigma_{0}-\overline{\omega})$ in the
single level approximation).  All these results are
satisfactory. However, Eq. (\ref{q4}) does {\it not} conserve energy
and this marks a flaw in the previous description. It is probable that
this constraint is broken by the small correlation time hypothesis,
but we do not see how to simplify Eq. (\ref{q3}) further without this
assumption. We could try to satisfy the energy constraint by
introducing a $\delta$-function term in Eq. (\ref{q4}) as in
Eq. (\ref{sun1}) but it is not clear how one can justify this procedure in
the present context. The conservation of energy (which is intimately
related to the form of the drift term) is the most serious problem
that we have encountered in trying to develop a kinetic theory of 2D
turbulence. Up to date, it has not been answered satisfactory. Further
scrutinity should be given to the non markovian kinetic equation (\ref{q3})
which may conserve energy although we were not able to prove it. 

It has to be emphasized that the quasilinear theory is {\it not} a
theory of violent relaxation in the usual sense since it only applies
to the late quiescent stages of the relaxation (``gentle
relaxation''), when the fluctuations have weaken and a linearization
procedure can be implemented. In order to give a more relevant
description of violent relaxation, it will be important in future
works to make a link between statistical mechanics and
\index{Self-consistent chaos} chaotic dynamics, in particular in the
point vortex model. This may give a new estimate of the diffusion
coefficient and of the decorrelation time $\tau$, which in this
context could be related to a Lyapunov exponent. We feel that this
track is an important one to make progress in the understanding of 2D
turbulence and point vortex dynamics.

In the quasilinear theory, it is implicitly
assumed that the decomposition $\omega=\overline{\omega}+\tilde\omega$
is obvious and that $\overline{\omega}$ should be regarded as a
statistical average of $\omega({\bf r},t,\zeta)$ over different
realizations $\zeta$ of the flow. This implies in particular that
$\overline{\overline{\omega}}=\overline{\omega}$. However, a different
approach is considered by Laval {\it et al.} \cite{dub} and Bouchet
\cite{bouchet0} who define $\overline{\omega}$ as the convolution of
$\omega({\bf r},t)$ with a Gaussian window of size $\epsilon$ (or by a
truncation in Fourier space). In that case
$\overline{\overline{\omega}}\neq\overline{\omega}$ and new terms
arise in Eq. (\ref{q1}), in particular a term ${\bf
J}_{d}=\overline{\omega\overline{\bf u}}-\overline{\omega}\
\overline{\bf u}$ which dominates over the others
\cite{bouchet0}. When $\epsilon\rightarrow 0$, this term becomes
equivalent to an anisotropic diffusion $\epsilon^{2}{\bf
\Sigma}'\nabla\overline{\omega}$ with a diffusion
coefficient (or turbulent viscosity) related to the stress tensor
$\Sigma'_{ij}=\partial_{i}\overline{u}_{j}$. This term alone conserves
energy but there exist directions in which the viscosity is negative
leading to instabilities. In order to circumvent this difficulty,
Bouchet \cite{bouchet0} proposes to project the diffusion current on
directions in which the viscosity is positive and to introduce a drift
term, as in the MEPP, in order to recover the conservation of energy
lost by this procedure. This leads to an operational subgridscale
model of 2D turbulence which appears to be more efficient than other
parametrizations. As for the MEPP, its drawback is its {\it ad hoc}
nature but it is not clear at present if it will be possible one day
to do much better and derive a parametrization of 2D turbulence from
first principles as attempted in the quasilinear theory.

\subsection{The limits $t\rightarrow +\infty$ and $N\rightarrow +\infty$}
\label{sec_tsallis}

We have indicated previously that a system of point vortices or point
mass stars \index{Astrophysics} can achieve two successive equilibrium
states. On a short time scale, the correlations between particles have
not yet developed and the system is described by the Vlasov (or Euler)
equation. In this regime, the dynamics is collisionless. Yet, for
systems with long-range interactions, the collective nature of the
evolution is responsible for an effective relaxation process, called
violent relaxation, which leads to a metaequilibrium state
(\ref{se14})(\ref{se15}) on a very short time scale. On a longer time
scale, the fluctuations of the potential have died away and the
developement of correlations between stars or between point vortices
leads to another, slower, relaxation process. This corresponds to the
``collisional'' regime. This second process is more standard and leads
to a true equilibrium state (\ref{me5})(\ref{gn5}). In the case of
continuous vorticity fields (instead of point vortices), the second
stage is replaced by a viscous decay of the vorticity due to inherent
viscosity.

We can discuss these two successive equilibrium states in a slightly
different manner. Let us consider a fixed interval of time and let the
number of particles $N\rightarrow +\infty$. In that case, the system
is rigorously described by the Vlasov (or Euler) equation, and a
metaequilibrium state is achieved on a timescale independant on
$N$. Alternatively, if we fix $N$ and let $t\rightarrow +\infty$, the
system will relax to the true equilibrium state resulting from a
collisional evolution. These two equilibrium states are of course
physically distinct. This implies that the order of the limits
$N\rightarrow +\infty$ and $t\rightarrow +\infty$ is not
interchangeable.

In the process of violent relaxation, the statistical mechanics is not
as firmly established as in the collisional regime, although it is
often the process of most interest. Indeed, the mixing required for
the validity of the ergodic hypothesis is fed by the fluctuations
of the potential. As these fluctuations decay as we approach
equilibrium (by definition!), the mixing becomes less and less
efficient and this can lead to an {\it incomplete relaxation}.  Since
the Boltzmann-Gibbs entropy does not always give a good description of
the equilibrium state, it has been proposed sometimes to use a wider
class of functionals to describe the process of violent relaxation
\cite{henon}. Among them, the $q$-entropies introduced by Tsallis \cite{Tsallis} have
been shown to give in some cases a good fit of the equilibrium
state. However, since the value of $q$ needs to be adjusted in each
case, it is not clear whether this agreement is the signal of a
generalized thermodynamics
\index{Non extensive thermodynamics} or just a coincidence \footnote{In 2D turbulence, Boghosian
\cite{boghosian} justifies a form of minimum enstrophy principle from
Tsallis thermodynamics in order to interpret the experimenal results
of Huang \& Driscoll \cite{huang} in a magnetized plasma. This is because
the enstrophy $\Gamma_{2}=\int\omega^{2}d^{2}{\bf r}$ is a particular
$q$-entropy. However, this is essentially coincidental and the minimum
enstrophy principle can lead to inconsistencies as discussed in
\cite{brands}.  In astrophysics, Tsallis entropies lead to pure
polytropes \cite{plastino}. These distribution functions are known for
a long time but they do not give a particularly good description of
elliptical galaxies or other stellar systems. Therefore, the relevance
of Tsallis generalized thermodynamics in 2D turbulence and stellar
dynamics remains questionable \cite{brands,poly}.}. 
It is clear that the Boltzmann entropy does not always give a perfect
description of the equilibrium state but there is no convincing
reason, up to date, why the system would select another ``universal''
form of entropy in the context of violent relaxation \cite{grand}. In
any case, the formalism developed by Tsallis and co-workers is nice to
generalize to a wider class of functionals the results obtained with
the Boltzmann entropy. Since this generalization is often analytically
tractable (as it leads to power laws), this may explain the interest
and the attractive nature of this approach.

\section{Conclusion}
\label{sec_conclusion}

The statistical mechanics of 2D vortices and \index{Astrophysics}
stellar systems appear to be remarkably similar despite the different
nature of these systems. We have tried to develop this analogy in
different directions. First of all, the $N$-star and $N$-vortex
problems both involve an unshielded long-range potential generated by
the density of particles themselves. At statistical equilibrium, these
systems are described by a Boltzmann-Poisson equation whose solutions
characterize organized states (at negative temperatures for
vortices). In the case of stars, the relaxation towards equilibrium
can be viewed as a competition between a diffusion and a friction. We
have proposed to describe the relaxation of point vortices similarly
in terms of a diffusion and a drift. The diffusion is due to the
fluctuations of the force experienced by a star or to the fluctuations
of the velocity field moving a vortex. The statistics of these
fluctuations can be studied by similar mathematical methods. The
fluctuations of the gravitational field are described by a particular
L\'evy law, called the Holtzmark distribution, and the fluctuations of
the velocity of vortices are described by a marginal Gaussian
distribution, intermediate between Gaussian and L\'evy laws. The
friction experienced by a star is due fundamentally to the
inhomogeneity of the velocity distribution of the stellar
cloud. Analogously, the drift experienced by a vortex results from the
spatial inhomogeneity of the vortex cloud. The friction and the drift
can be understood similarly in terms of a polarization
process and a back reaction of the system. In the thermal bath
approximation, the coefficients of friction and drift are given by an
Einstein relation and the one-body distribution function satisfies a
Fokker-Planck equation. Further away from equilibrium, the collisional
dynamics of stars is described by the gravitational Landau
equation. We have derived a new kinetic equation that should be
appropriate to the ``collisional'' relaxation of point vortices.

A system of stars or vortices can also undergo a form of violent
relaxation. This is essentially a collisionless process driven by the
rapid fluctuations of the potential as a result of collective effects
(chaotic mixing). In this collisionless regime, the stars and the
vortices are described by the Vlasov-Poisson and the Euler-Poisson
systems. These equations present a similar structure and a statistical
mechanics can be developed to predict the ``most probable state''
resulting from a complex evolution driven by a mixing process. The
relaxation towards equilibrium can be incomplete and an
out-of-equilibrium study is necessary to understand what limits
relaxation and causes a kinetic confinement of the system in a
``maximum entropy bubble''. Relaxation equations have been derived
either from a heuristic Maximum Entropy Production Principle or from a
more controllable kinetic theory, in an asymptotic regime of the
dynamics (gentle relaxation) in which a quasilinear approximation can
be implemented.

It should be noted that many results presented in this paper, in
particular those corresponding to the kinetic theory of 2D vortices
developed in Secs. \ref{sec_fluc}-\ref{sec_kin}, are very recent and
need to be completed and further discussed. In particular, it appears
indispensable to carry out extensive numerical simulations to test
their relevance and determine their domains of applicability. It is
plausible that the true dynamics of stars and vortices is more complex
than the picture that has been given here. In addition, the
description of chaos in these systems has not been addressed at all in
this paper although it is presumably an essential ingredient to
understand their dynamics. We are thus far from reaching a complete
understanding of these systems with long-range interactions. We feel,
however, that the analogy between the statistical mechanics of stars
and vortices that we have investigated is correct in its mains lines
and should lead again to fruitful developements.

\section*{Appendix A: The calculation of the diffusion coefficient}
\label{sec_appA}

In this Appendix, we calculate the diffusion coefficient of a point
vortex evolving in an inhomogeneous vortex cloud, using the Kubo formula. To
evaluate the velocity autocorrelation function $C(\tau)=\langle
V(t)V(t-\tau)\rangle$, we shall assume that between $t$ and $t-\tau$,
the point vortices follow the streamlines of the equilibrium flow. This
is a reasonable approximation in the case of strong shears.

\subsection{Unidirectional flow}
\label{sec_uni}

The trajectory of a point vortex advected by a unidirectional equilibrium flow is simply:
\begin{eqnarray}
y(t-\tau)=y(t),
\label{uni1}
\end{eqnarray}  
\begin{eqnarray}
x(t-\tau)=x(t)-\langle V\rangle_{eq}(y)\tau.
\label{uni2}
\end{eqnarray}  
The velocity auto-correlation function appearing in  Eq. (\ref{dc2}) can be written
explicitly
\begin{eqnarray}
C(\tau)={N\gamma^{2}\over 4\pi^{2}} \int dx_{1}dy_{1}{x_{1}-x\over
(x_{1}-x)^{2}+(y_{1}-y)^{2}}(t) \nonumber\\
\ \ \ \ \ \times {x_{1}-x \over
(x_{1}-x)^{2}+(y_{1}-y)^{2}}(t-\tau)P_{eq}(y),
\label{uni3}
\end{eqnarray}  
where we have used Eq. (\ref{pv2}). The second term involves the quantity
\begin{eqnarray}
(x_{1}-x)(t-\tau)=x_{1}-x+(\langle V\rangle_{eq}(y_{1})-\langle V\rangle_{eq}
(y))\tau.
\label{uni4}
\end{eqnarray} 
Since the integral in Eq. (\ref{uni3}) diverges as ${\bf
r}_{1}\rightarrow {\bf r}$, we can make a {local approximation}
and expand the velocity difference in a Taylor series in $y_{1}-y$. To
first order, we have
\begin{eqnarray}
\langle V\rangle_{eq}(y_{1})-\langle V\rangle_{eq}(y)\simeq -\Sigma(y)(y_{1}-y),
\label{uni5}
\end{eqnarray} 
where $\Sigma(y)$ is the local shear of the flow.
Introducing the variables $X\equiv x_{1}-x$ and $Y\equiv y_{1}-y$, we obtain
\begin{eqnarray}
C(\tau)={N\gamma^{2}\over 4\pi^{2}}P_{eq}(y) \int dXdY {X\over X^{2}+Y^{2}}\
{X+\Sigma(y) Y\tau\over (X+\Sigma(y) Y\tau)^{2}+Y^{2}}.
\label{uni7}
\end{eqnarray}  
The integration over $X$ can be performed easily since the integrand is just a
rational function of polynomials. After straightforward calculations, we find
\begin{eqnarray}
C(\tau)={N\gamma^{2}\over 4\pi}P_{eq}(y){1\over 1+{1\over
4}\Sigma^{2}(y)\tau^{2}} \int_{0}^{+\infty}  {dY\over Y}. 
\label{uni8}
\end{eqnarray}  
The integral over $Y$ diverges logarithmically for both small and
large $Y$. The reason for this divergence has been explained in
Sec. \ref{sec_fluc}. Introducing two cut-offs at scales $d$ (the
inter-vortex distance) and $R$  (the domain size) and noting that $\ln
(R/d)\sim {1\over 2}\ln N$, we obtain
\begin{eqnarray}
C(\tau)={N\gamma^{2}\over 8\pi}\ln N {1\over 1+{1\over
4}\Sigma^{2}(y)\tau^{2}} P_{eq}(y). 
\label{uni8a}
\end{eqnarray} 
For $\tau\rightarrow +\infty$, the correlation function decreases like
$\tau^{-2}$. This is a slow decay but it is sufficient to ensure the
convergence of the diffusion coefficient (\ref{dc2}). Using
\begin{eqnarray}
\int_{0}^{t} C(\tau) d\tau={N\gamma^{2}\over 4\pi}{\ln N\over |\Sigma(y)|}
\arctan\biggl ({1\over 2}|\Sigma(y)|t\biggr )P_{eq}(y),
\label{uni10}
\end{eqnarray} 
and taking the limit $t\rightarrow +\infty$, we establish Eq. (\ref{dc6}).

\subsection{Axisymmetric  flow}
\label{sec_axi}

In an axisymmetric flow, the trajectory of a point vortex takes the simple
form:
\begin{eqnarray}
r(t-\tau)=r(t),
\label{axi1}
\end{eqnarray}  
\begin{eqnarray}
\theta(t-\tau)=\theta(t)-{\langle V\rangle_{eq}(r)\over r}\tau.
\label{axi2}
\end{eqnarray}  
As indicated in Sec. \ref{sec_diff}, we are particularly interested by the
$r(t)r(t-\tau)$ component of the correlation tensor. Let us
introduce the separation $\delta {\bf r}\equiv {\bf r}_{1}-{\bf r}$ between the
field vortex $1$ and the test vortex. In the local approximation,  $\delta {\bf
r}$ can be considered as a small quantity. Therefore, we can write
\begin{eqnarray}
\delta {\bf r}=r\delta\theta {\bf e}_{\theta}+\delta r{\bf e}_{r}\equiv X{\bf
e}_{\theta}+Y{\bf e}_{r},
\label{axi4}
\end{eqnarray}
\begin{eqnarray}
d^{2}{\bf r}_{1}=d^{2}(\delta {\bf r})=dXdY.
\label{axi5}
\end{eqnarray}
With these notations, the correlation function appearing in Eq. (\ref{dc5}) can be rewritten
\begin{eqnarray}
C(\tau)={N\gamma^{2}\over 4\pi^{2}}P_{eq}(r)\int dXdY {X\over
X^{2}+Y^{2}}(t){X\over X^{2}+Y^{2}}({t-\tau}).
\label{axi6}
\end{eqnarray}
Now,
\begin{eqnarray}
Y(t-\tau)=\delta r(t-\tau)=r_{1}(t-\tau)-r(t-\tau)\nonumber\\
=r_{1}(t)-r(t)=Y(t)=Y,
\label{axi7}
\end{eqnarray}
and
\begin{eqnarray}
X(t-\tau)=r(t-\tau)\delta\theta(t-\tau)=r(t-\tau)
(\theta_{1}(t-\tau)-\theta (t-\tau) )\nonumber\\
 = r\biggl
(\theta_{1}(t)-\theta(t)-\biggl ({\langle V\rangle_{eq}(r_{1})\over
r_{1}}-{\langle V\rangle_{eq}(r)\over r}\biggr )\tau\biggr ).
\label{axi77}
\end{eqnarray}
In the local approximation, we can expand the last term in Eq.
(\ref{axi77}) in a Taylor series in $r_{1}-r$. This yields
\begin{eqnarray}
X(t-\tau)=r (\theta_{1}(t)-\theta (t))-r {d\over dr}\biggl ({\langle
V\rangle_{eq}(r)\over r}\biggr ) (r_{1}-r) \tau =X-\Sigma(r)Y\tau,\nonumber\\
\label{axi8}
\end{eqnarray}
where $\Sigma(r)$ is the local shear of the flow. Substituting
Eqs. (\ref{axi7}) and  (\ref{axi8}) in Eq. (\ref{axi6}), we get
\begin{eqnarray}
C(\tau)={N\gamma^{2}\over 4\pi^{2}}P_{eq}(r)\int dXdY {X\over
X^{2}+Y^{2}}{X-\Sigma(r) Y \tau\over (X-\Sigma(r) Y \tau)^{2}+Y^{2}}.
\label{axi10}
\end{eqnarray}
This integral is similar to Eq. (\ref{uni7}), so we again obtain Eq. (\ref{dc6}).

\section*{Appendix B: Calculation of the memory function}
\label{sec_appB}

In this Appendix, we calculate the memory function that occurs in
Eq. (\ref{new1}). If we assume that $P=P(r,t)$, then Eq.
(\ref{new1}) simplifies to
\begin{eqnarray}
{\partial P\over\partial t}=-{1\over r}{\partial\over\partial r}(r J_{r}),
\label{ww1}
\end{eqnarray}
where 
\begin{eqnarray}
J_{r}=-N\int_{0}^{t}d\tau\int d^{2}{\bf r}_{1}{V}_{r(t)}(1\rightarrow
0)_{t}\nonumber\\
\times\biggl\lbrace {V}_{r(t-\tau)}(1\rightarrow 0)P_{1} {\partial
P\over\partial r}-{V}_{r_{1}(t-\tau)}(1\rightarrow 0)P {\partial
P_{1}\over\partial r_{1}}\biggr\rbrace_{t-\tau},
\label{ww2}
\end{eqnarray}
and where $V_{r(t)}(1\rightarrow 0)$ is the component of the vector ${\bf
V}(1\rightarrow 0)$ in the direction of ${\bf r}(t)$. If we denote by
$(r(t),\theta(t))$ and $(r_{1}(t),\theta_{1}(t))$ the polar coordinates that
specify the position of the point vortices $0$ and $1$  at time $t$, we easily
find that
\begin{eqnarray}
V_{r(t)}(1\rightarrow 0)=-{\gamma\over 2\pi}{r_{1}\sin (\theta-\theta_{1})\over
r_{1}^{2}+r^{2}-2 r r_{1}\cos (\theta-\theta_{1})}. 
\label{ww3}
\end{eqnarray}   
We shall assume that between $t$ and $t-\tau$, the point vortices follow
circular trajectories with angular velocity $\Omega(r,t)$. In that case,
$r(t-\tau)=r$ and $\theta(t-\tau)=\theta-\Omega(r,t)\tau$. Then, we obtain
\begin{eqnarray}
V_{r(t-\tau)}(1\rightarrow 0)=-{\gamma\over 2\pi}{r_{1}\sin
(\theta-\theta_{1}-\Delta\Omega\tau)\over r_{1}^{2}+r^{2}-2 r r_{1}\cos
(\theta-\theta_{1}-\Delta\Omega\tau)} 
\label{ww5}
\end{eqnarray}  
with  
\begin{eqnarray}
\Delta\Omega=\Omega(r,t)-\Omega(r_{1},t).
\label{ww4}
\end{eqnarray} 
We find similarly that $V_{r_{1}(t-\tau)}(1\rightarrow 0)={r\over
r_{1}}V_{r(t-\tau)}(1\rightarrow 0)$. Our previous assumptions also
imply that $P(r(t-\tau),t-\tau)\simeq P(r,t)$ between $t$ and
$t-\tau$. In words, this means that the correlation time is smaller
than the time scale on which the average vorticity changes
appreciably. We do not assume that it is {\it much} smaller as in
Sec. \ref{sec_short}, so this approximation is not over
restrictive. In that case, the diffusion current becomes
\begin{eqnarray}
J_{r}=-N\int_{0}^{+\infty}d\tau\int_{0}^{2\pi}d\theta_{1}\int_{0}^{+\infty}r
r_{1}dr_{1} {V}_{r(t)}(1\rightarrow 0)\nonumber\\ 
\times {V}_{r(t-\tau)}(1\rightarrow 0)
\biggl\lbrack {1\over r}P_{1}{\partial P\over\partial r}- {1\over
r_{1}}P{\partial P_{1}\over\partial r_{1}}\biggr\rbrack,
\label{ww6}
\end{eqnarray}
where the time integral has been extended  to $+\infty$. We now need to evaluate
the memory function
\begin{eqnarray}
M=\int_{0}^{+\infty}d\tau\int_{0}^{2\pi}d\theta_{1} {V}_{r(t)}(1\rightarrow 0)
{V}_{r(t-\tau)}(1\rightarrow 0).
\label{ww7}
\end{eqnarray}
Introducing the notations $\phi=\theta_{1}-\theta$ and 
\begin{eqnarray}
\lambda={2 r r_{1}\over r_{1}^{2}+r^{2}}<1,
\label{ww8}
\end{eqnarray}
we have explicitly
\begin{eqnarray}
M=\biggl ({\gamma\lambda\over 4\pi r}\biggr
)^{2}\int_{0}^{+\infty}d\tau\int_{0}^{2\pi}d\phi {\sin\phi\over
1-\lambda\cos\phi}\  {\sin(\phi+\Delta\Omega\tau)\over 1-\lambda\cos
(\phi+\Delta\Omega\tau)}.
\label{ww9}
\end{eqnarray}
This can also be written
\begin{eqnarray}
M=\biggl ({\gamma\over 4\pi r}\biggr
)^{2}\int_{0}^{+\infty}d\tau\int_{0}^{2\pi}d\phi V'(\phi)
V'(\phi+\Delta\Omega\tau), 
\label{ww10}
\end{eqnarray}
where 
\begin{eqnarray}
V(\phi)=\ln (1-\lambda\cos\phi).
\label{ww11}
\end{eqnarray}
We now write the function $V(\phi)$ in the form of a Fourier series,
\begin{eqnarray}
V(\phi)=\sum_{n=-\infty}^{+\infty}a_{n}e^{i n\phi}\qquad {\rm with}\qquad
a_{n}={1\over 2\pi}\int_{-\pi}^{\pi} V(\phi)e^{-i n \phi}d\phi.
\label{ww12}
\end{eqnarray} 
The memory function becomes
\begin{eqnarray}
M=-{1\over 2}\biggl ({\gamma\over 4\pi r}\biggr
)^{2}\int_{-\infty}^{+\infty}d\tau\int_{0}^{2\pi}d\phi
\sum_{n,m=-\infty}^{+\infty}n m a_{n}a_{m}e^{i(n+m)\phi}e^{i m \Delta
\Omega\tau}.
\label{ww14}
\end{eqnarray}
Carrying out the integrations on $\phi$ and $\tau$ using the integral
representation of the delta function
\begin{eqnarray}
\delta (x)={1\over 2\pi}\int_{-\infty}^{+\infty} e^{-i\rho x}d\rho,
\label{wwr14}
\end{eqnarray}
we are left with
\begin{eqnarray}
M=-{\gamma^{2}\over 8 r^{2}} \sum_{n,m=-\infty}^{+\infty}n m
a_{n}a_{m}\delta_{n,-m} \delta(m\Delta\Omega)={\gamma^{2}\over 8 r^{2}}
\delta(\Delta\Omega)  \sum_{n=-\infty}^{+\infty} |n| a_{n}^{2} . 
\label{ww15}
\end{eqnarray}
It remains for us to evaluate the series that appears in the last
expression of the memory function. Using the identities
\begin{eqnarray}
\int_{0}^{\pi}\ln (1-\lambda\cos\phi)\cos(n\phi)d\phi=-{\pi\over n}\biggl
({1\over\lambda}-\sqrt{{1\over\lambda^{2}}-1}\biggr )^{n} \qquad (n>0), 
\label{ww18}
\end{eqnarray}
\begin{eqnarray}
\int_{0}^{\pi}\ln (1-\lambda\cos\phi)d\phi=\pi\ln\biggl ({1\over
2}+{\sqrt{1-\lambda^{2}}\over 2}\biggr ),
\label{ww19}
\end{eqnarray}
and the definition (\ref{ww8}) of $\lambda$, we find that $a_{0}<\infty$ and,
for $n>0$,
\begin{eqnarray}
a_{n}=-{1\over n}\biggl ({(r_{1}^{2}+r^{2})-|r_{1}^{2}-r^{2}|\over 2 r
r_{1}}\biggr )^{n}=-{1\over n}\biggl ({r_{<}\over r_{>}}\biggr )^{n},
\label{ww20}
\end{eqnarray}
where $r_{>}$ (resp. $r_{<}$) is the biggest (resp. smallest) of $r$ and
$r_{1}$. Therefore, the value of the series is
\begin{eqnarray}
\sum_{n=-\infty}^{+\infty} |n| a_{n}^{2}= 2\sum_{n=1}^{+\infty} n a_{n}^{2}=
2\sum_{n=1}^{+\infty} {1\over n}\biggl ({r_{<}\over r_{>}}\biggr )^{2n}=-2
\ln\biggl \lbrack 1-\biggl ({r_{<}\over r_{>}}\biggr )^{2}\biggr \rbrack .
\label{ww21}
\end{eqnarray}
The memory function takes the form
\begin{eqnarray}
M=-{\gamma^{2}\over 4 r^{2}} \delta(\Delta\Omega)\ln\biggl \lbrack 1-\biggl
({r_{<}\over r_{>}}\biggr )^{2}\biggr \rbrack    
\label{ww22}
\end{eqnarray}
and the diffusion current in the axisymmetric case can be written
\begin{eqnarray}
J_{r}={N\gamma^{2}\over 4 r}\int_{0}^{+\infty}r_{1}dr_{1}
\delta(\Omega-\Omega_{1})\ln\biggl \lbrack 1-\biggl ({r_{<}\over r_{>}}\biggr
)^{2}\biggr \rbrack\biggl\lbrace {1\over r}P_{1}{\partial P\over\partial r}-
{1\over r_{1}}P{\partial P_{1}\over\partial r_{1}} \biggr\rbrace. \nonumber\\  
\label{ww23}
\end{eqnarray}
This leads to the kinetic equation (\ref{day1}).

%


\bigskip 


\begin{thebibliography}{99}

\bibitem{agekyan}  {\small T. Agekyan, Sov. Astron. {\bf 5}, 809 (1962).}

\bibitem{antonov}  {\small V.A. Antonov, Vest. Leningr. Gos. Univ. {\bf 7}, 135 (1962).}

\bibitem{aronson}  {\small E.B. Aronson and C.J. Hansen, ``Thermal equilibrium states of a classical system with gravitation'', Astrophys. J.  {\bf 177}, 145 (1972).}

\bibitem{balescu}  {\small R. Balescu, Statistical Mechanics of Charged Particles (Interscience, New York, 1963).}

\bibitem{barge}  {\small P. Barge and J. Sommeria, ``Did planet formation begin inside persistent gaseous vortices?'', Astron. Astrophys. {\bf 295}, L1 (1995).}

\bibitem{barre}  {\small J. Barr\'e, D. Mukamel and S. Ruffo, ``Inequivalence of ensembles in a system with long-range interactions'', Phys. Rev. Lett. 
 {\bf 87}, 030601 (2001).}

\bibitem{bertrand}  {\small P. Bertrand, Contribution \`a l'\'etude de mod\`eles math\'ematiques de plasmas non collisionnels, PhD thesis, Universit\'e de Nancy I (1972).}

\bibitem{bilic}  {\small  N. Bilic and R.D. Viollier, ``Gravitational phase transition of heavy neutrino matter'', Phys. Lett. B {\bf 408},  75 (1997).}

\bibitem{bt}  {\small J. Binney and S. Tremaine, 
Galactic Dynamics (Princeton Series in Astrophysics, 1987).} 

\bibitem{boghosian}  {\small  B.M. Boghosian, ``Thermodynamic description of the relaxation of 2D turbulence using Tsallis statistics'' Phys. Rev. E {\bf 53}, 4754 (1996).}

\bibitem{bouchaud}  {\small J.P. Bouchaud and A. Georges, ``Anomalous diffusion in disordered media: statistical mechanisms, models and physical 
applications'', Phys. Rep. {\bf 195}, 127 (1990).}

\bibitem{bouchet0}  {\small F. Bouchet, M\'ecanique statistique pour des \'ecoulements g\'eophysiques, PhD thesis, Universit\'e J. Fourier (2001).}

\bibitem{bouchet2}  {\small F. Bouchet, P.H. Chavanis and  J. Sommeria, ``Statistical mechanics of Jupiter's Great Red Spot in the shallow water model'', preprint.}

\bibitem{bouchet1}  {\small F. Bouchet and J. Sommeria, ``Emergence of intense jets and Jupiter Great Red Spot as maximum entropy structures'', 
J. Fluid. Mech. {\bf 464}, 165 (2002).}

\bibitem{bracco}  {\small A. Bracco, P.H. Chavanis, A. Provenzale and E. 
Spiegel, ``Particle aggregation in a turbulent Keplerian flow'', Phys. Fluids {\bf 11}, 2280 (1999).}

\bibitem{brands}  {\small H. Brands, P.H. Chavanis, R. Pasmanter and J. Sommeria, ``Maximum entropy versus minimum enstrophy vortices'', Phys. Fluids {\bf 11}, 3465 (1999).}

\bibitem{caglioti}  {\small E. Caglioti, P.L. Lions, C. Marchioro and M. Pulvirenti, ``A special class of stationary flows for two-dimensional Euler 
equations: a statistical mechanics description'', Commun. Math. Phys. {\bf 143}, 501 (1992).}

\bibitem{pomeau}  {\small G.F. Carnevale, J.C. McWilliams, Y. Pomeau, J.B. Weiss and W.R. Young, ``Evolution of vortex statistics in two-dimensional turbulence'',  Phys. Rev. Lett.   {\bf 66}, 2735 (1991).}

\bibitem{chandra39}  {\small S. Chandrasekhar, 
An Introduction to the Theory of Stellar Structure (Dover 1939).} 

\bibitem{c0}  {\small S. Chandrasekhar, ``A statistical theory of stellar encounters'', Astrophys. J. {\bf 94}, 511 (1941).}

\bibitem{chandra42}  {\small S. Chandrasekhar, Principles of stellar dynamics (Dover 1942).}
 
\bibitem{chandrabrownien}  {\small S. Chandrasekhar, ``Stochastic problems in physics and astronomy'', Rev. Mod. Phys. {\bf 15}, 1 (1943).}

\bibitem{chandrafric}  {\small S. Chandrasekhar, ``Dynamical friction. I. General considerations: the coefficient of dynamical friction'', Astrophys. J. {\bf 97}, 255 (1943).}

\bibitem{c3}  {\small S. Chandrasekhar, ``The statistics of the gravitational field arising from a random distribution
of stars: III. The correlations in the forces acting at two points separated by
a finite distance'', Astrophys. J. {\bf 99}, 25 (1944).}

\bibitem{c4}  {\small S. Chandrasekhar, ``The statistics of the gravitational field arising from a random distribution
of stars: IV. The stochastic variation of the force acting on a star'', Astrophys. J. {\bf 99}, 47 (1944).}

\bibitem{chandraS}  {\small S. Chandrasekhar, ``Brownian motion, dynamical friction and stellar dynamics'', Rev. Mod. Phys. {\bf 21}, 383 (1949).}

\bibitem{cn1}  {\small S. Chandrasekhar \& J. von Neumann, ``The statistics of the gravitational field arising from a
random distribution of stars I. The speed of fluctuations'', Astrophys. J.
{\bf 95}, 489 (1942).}

\bibitem{cn2}  {\small S. Chandrasekhar \& J. von Neumann ``The statistics of the gravitational field arising from a
random distribution of stars II. The speed of fluctuations; dynamical friction;
spatial correlations'', Astrophys. J.
{\bf 97}, 1 (1943).}

\bibitem{cthese}  {\small P.H. Chavanis, Contribution \`a la m\'ecanique statistique des tourbillons bidimensionnels. Analogie avec la relaxation violente des syst\`emes stellaires, Th\`ese de doctorat, Ecole Normale Sup\'erieure de Lyon (1996).}

\bibitem{drift}  {\small P.H. Chavanis, ``Systematic drift experienced by a point vortex in two-dimensional turbulence'',  Phys. Rev. E  {\bf 58}, R1199 (1998).}

\bibitem{cg}  {\small P.H. Chavanis, ``On the coarse-grained evolution of collisionless stellar systems'',  Mon. Not. R. astr. Soc.  {\bf 300}, 981 (1998).}

\bibitem{cfloride}  {\small P.H. Chavanis, 
``From Jupiter's Great Red Spot to the structure 
of galaxies: statistical mechanics of two-dimensional 
vortices and stellar systems'',  Annals of the 
New York Academy of Sciences {\bf 867}, 120 (1998).}

\bibitem{cplanetes}  {\small P.H. Chavanis, ``Trapping of dust by coherent vortices in the solar nebula'',  Astron. Astrophys.  {\bf 356}, 1089 (2000).}

\bibitem{kin1}  {\small P.H. Chavanis, ``Quasilinear theory of the 2D Euler equation'',  Phys. Rev. Lett. {\bf 84}, 5512 (2000).}

\bibitem{japon}  {\small P.H. Chavanis, ``On the analogy between two-dimensional vortices and stellar systems'', Proceedings of the IUTAM Symposium on Geometry and Statistics of Turbulence (2001), T. Kambe, T. Nakano and T. Miyauchi Eds. (Kluwer Academic Publishers).}

\bibitem{kin2}  {\small P.H. Chavanis, ``Kinetic theory of point vortices: diffusion coefficient and systematic drift'', Phys. Rev. E {\bf 64}, 026309 (2001).}

\bibitem{chavacano}  {\small P.H. Chavanis, ``Gravitational instability of finite isothermal spheres'', Astron. Astrophys. {\bf 381}, 340 (2002).}

\bibitem{relat}  {\small P.H. Chavanis, ``Gravitational instability of finite isothermal spheres in general relativity. Analogy with neutron stars'', Astron. Astrophys. {\bf 381}, 709 (2002).}

\bibitem{poly}  {\small P.H. Chavanis, ``Gravitational instability of polytropic spheres and generalized thermodynamics'', Astron. Astrophys. {\bf 386}, 732 (2002).}

\bibitem{eff}  {\small P.H. Chavanis, ``Effective velocity created by a point vortex in two-dimensional hydrodynamics'', Phys. Rev. E {\bf 65}, 056302 (2002).}

\bibitem{pt}  {\small P.H. Chavanis, ``Phase transitions in self-gravitating systems. Self-gravitating fermions and hard sphere models'', Phys. Rev. E {\bf 65}, 056123 (2002).}

\bibitem{dubrovnik} {\small P.H. Chavanis, ``Statistical mechanics of violent relaxation in stellar systems'', Proceedings of the Conference on
Multiscale Problems in Science and Technology (Springer 2002); also available on astro-ph/0212205.}

\bibitem{cape} {\small P.H. Chavanis, ``The self-gravitating Fermi gas'',  Proceedings of the Conference Dark2002: 4th International Heidelberg Conference on Dark Matter in Astro and Particle Physics, 4-9 Feb 2002, Cape Town, South African Astroparticles (Springer); also available on astro-ph/0205426.} 

\bibitem{grand}  {\small P.H. Chavanis, ``Gravitational instability of isothermal and polytropic spheres'', Astron. Astrophys. in press [astro-ph/0207080].}

\bibitem{ispolatov}  {\small P.H. Chavanis and I. Ispolatov,
``Phase diagram of self-attracting systems'', Phys. Rev. E {\bf 66},
036109 (2002).}

\bibitem{crs}  {\small P.H. Chavanis, C. Rosier and C. Sire ``Thermodynamics of self-gravitating systems'', Phys. Rev. E. {\bf 66}, 036105 (2002).}

\bibitem{csire1}  {\small P.H. Chavanis and C. Sire,
``The statistics of velocity fluctuations arising from a random
distribution of point vortices: the speed of fluctuations and the
diffusion coefficient'', Phys. Rev. E  {\bf 62}, 490 (2000).}

\bibitem{csire2}  {\small P.H. Chavanis and C. Sire,
``The spatial correlations in the velocities arising from a random distribution of point vortices'', Phys. Fluids {\bf 13}, 1904 (2001).}

\bibitem{csfluide0}  {\small P.H. Chavanis and J. Sommeria,
``Classification of self-organized vortices in two-dimensional
turbulence: the case of a bounded domain'', J. Fluid Mech.  {\bf
314}, 267 (1996).}

\bibitem{csthermo}  {\small P.H. Chavanis and J. Sommeria,
``Thermodynamical approach for small-scale parametrization in 2D turbulence'',  Phys. Rev. Lett. {\bf 78}, 3302 (1997).}


\bibitem{csfluide}  {\small P.H. Chavanis and J. Sommeria,
``Classification of robust isolated vortices in two-dimensional hydrodynamics'',  J. Fluid Mech.  {\bf 356}, 259 (1998).}

\bibitem{cs}  {\small P.H. Chavanis and J. Sommeria,
``Degenerate equilibrium states of collisionless 
stellar systems'',  Mon. Not. R. astr. Soc.  {\bf 296}, 569 (1998).}

\bibitem{shallow}  {\small P.H. Chavanis and J. Sommeria,
``Statistical mechanics of the shallow water system'', Phys. Rev. E {\bf 65}, 026302 (2002).}

\bibitem{csr}  {\small P.H. Chavanis, J. Sommeria and 
R. Robert, ``Statistical mechanics of two-dimensional 
vortices and collisionless stellar systems'', Astrophys.
J. {\bf 471}, 385 (1996).}

\bibitem{chukbar}  {\small K.V. Chukbar,``Statistics of two-dimensional vortices and the Holtsmark distribution,'' Plasma Physics Reports. {\bf 25}, 77 (1999).}

\bibitem{cohn}  {\small H. Cohn, ``Late core collapse 
in star clusters and the gravothermal instability'', 
Astrophys. J. {\bf 242}, 765 (1980).}

\bibitem{vega4}  {\small H.J. de Vega and N. Sanchez, ``Statistical mechanics of the self-gravitating gas: I. Thermodynamical limit and phase diagrams'', Nucl. Phys. B {\bf 625}, 409 (2002).} 

\bibitem{eyink}  {\small G.L. Eyink and H. Spohn, ``Negative temperature states and large-scale, long-lived vortices in two-dimensional turbulence'', J. Stat. Phys. {\bf 70}, 833 (1993).} 

\bibitem{follana}  {\small  E. Follana and V. Laliena,  ``Thermodynamics of self-gravitating systems with softened potentials'', Phys. Rev. E  {\bf 61}, 6270 (2000).}

\bibitem{godon}  {\small P. Godon and M. Livio, ``The formation and role of vortices in protoplanetary disks'', Astrophys. J.  {\bf 537}, 396 (2000).}

\bibitem{hansen}  {\small A.E. Hansen, D. Marteau and P. Tabeling, ``Two-dimensional turbulence and dispersion in a freely decaying system'', Phys. Rev. E {\bf 58}, 7261 (1998).}

\bibitem{thirring} P. Hertel, W. Thirring, ``Thermodynamic instability of a system of gravitating fermions'' in {Quanten und Felder}, edited by H.P. D\"urr (Vieweg, Braunschweig, 1971) 

\bibitem{hjorth}  {\small J. Hjorth and J. Madsen, ``Statistical mechanics of galaxies'', Mon. Not. R. astr. Soc.  {\bf 265}, 237 (1993).}

\bibitem{holtsmark}  {\small J. Holtsmark, Ann. Phys. (Leipzig)
{\bf 58}, 577 (1919).}

\bibitem{horwitz}  {\small G. Horwitz and J. Katz, ``Steepest descent technique and stellar equilibrium statistical mechanics. III. Stability of various ensembles'', Astrophys. J. {\bf 222}, 941 (1978).}

\bibitem{huang}  {\small X.P. Huang and C.F. Driscoll, ``Relaxation of 2D turbulence to a metaequilibrium near the minimum enstrophy state'', Phys. Rev. Lett.  {\bf 72}, 2187 (1994).}

\bibitem{ibragimov}  {\small I.A. Ibragimov and Yu. V. Linnik, Independant and Stationary Sequences of Random Variables (Wolters-Noordhoff, Groningen, 1971).}

\bibitem{ilb}  {\small S. Inagaki and D. Lynden-Bell, ``Self-similar solutions for post-collapse evolution of globular clusters'', Mon. Not. R. astr. Soc.  {\bf 205}, 913 (1983).}

\bibitem{jimenez}  {\small J. Jimenez, ``Algebraic probability density tails in decaying isotropic two-dimensional turbulence'', J. Fluid Mech. {\bf 313}, 223
(1996).}

\bibitem{jm}  {\small G. Joyce and D. Montgomery, ``Negative temperature states for the two-dimensional guiding-center plasma'', J. Plasma Phys. {\bf 10}, 107
(1973).}

\bibitem{kp}  {\small B.B. Kadomtsev and O.P. Pogutse, ``Collisionless relaxation in systems with Coulomb interactions'',  Phys. Rev. Lett.  
{\bf 25}, 1155 (1970).}

\bibitem{krev}  {\small H.E. Kandrup, ``Stochastic gravitational fluctuations in a self-consistent mean field theory'',  Physics Reports   
{\bf 63}, 1 (1980).}

\bibitem{Kandrup}  {\small H.E. Kandrup, ``A generalized Landau equation for a system with a self-consistent mean field: derivation from an $N$-particle Liouville equation'',  Astrophys. J.  
{\bf 244}, 316 (1981).}

\bibitem{kandrup}  {\small H.E. Kandrup, ``Dynamical friction in a mean field approximation'',  Astrophys. \& Space Sci.  
{\bf 97}, 435 (1983).}

\bibitem{katz}  {\small J. Katz, ``On the number of 
unstable modes of an equilibrium'',  Mon. Not. R. astr. Soc.  
{\bf 183}, 765 (1978).}

\bibitem{klb}  {\small J. Katz and D. Lynden-Bell, ``The gravothermal
instability in two dimensions'', Mon. Not. R. astr. Soc.  {\bf 184},
709 (1978).}

\bibitem{newkatz}  {\small J. Katz and  I. Okamoto,  ``Fluctuations in isothermal spheres'',  Mon. Not. R. astr. Soc. {\bf 317}, 163 (2000).}

\bibitem{kaz}  {\small E. Kazantzev, J. Sommeria and J. Verron, ``Subgridscale eddy parametrization by statistical mechanics in a barotropic ocean model'',  J. Phys. Ocean. {\bf 28}, 1017 (1998).}

\bibitem{kirchhoff}  {\small G. Kirchhoff, in {Lectures in Mathematical
Physics, Mechanics} (Teubner, Leipzig, 1877).}

\bibitem{kraichnan}  {\small R.H. Kraichnan,  ``Statistical dynamics of two-dimensional flow'',  J. Fluid Mech. {\bf 67}, 155 (1975).}

\bibitem{kramers}  {\small H.A. Kramers, Physica  {\bf 7}, 284 (1940).}

\bibitem{kuv}  {\small B.N.  Kuvshinov and T.J. Schep, ``Holtsmark distribution in point vortex systems,'' Phys. Rev. Lett. {\bf 84}, 650 (2000).}

\bibitem{kuzmin}  {\small G.A. Kuzmin, ``Statistical mechanics of the organization into two-dimensional coherent structures'', in 
{Structural Turbulence}, edited by M.A. Goldshtik (Acad. Naouk CCCP Novosibirsk, Institute of ThermoPhysics, 1982), pp. 103-114.}

\bibitem{kiess}  {\small C. Lancellotti and M. Kiessling, ``Self-gravitational collapse in stellar dynamics'', Astrophys. J. {\bf 549}, L93 (2001).}

\bibitem{larson}  {\small R.B. Larson, ``A method for 
computing the evolution of star clusters'',
Mon. Not. R. astr. Soc. {\bf 147}, 323 (1970).}

\bibitem{laval}  {\small J.P. Laval, P.H. Chavanis, B. Dubrulle and C. Sire, 
``Scaling laws and vortex profiles in 2D decaying turbulence'', Phys. Rev E
{\bf 63}, 065301(R) (2001).}

\bibitem{dub}  {\small J.P. Laval, B. Dubrulle and S. Nazarenko, 
``Nonlocality of interaction of scales in the dynamics of 2D incompressible fluids'', Phys. Rev. Lett.
{\bf 83}, 4061 (1999).}

\bibitem{lee}  {\small E.P. Lee, 
``Brownian motion in a stellar system'', Astrophys. J.  {\bf 151},
687 (1968).}

\bibitem{lp}  {\small T.S. Lundgren and Y.B. Pointin, ``Statistical mechanics of two-dimensional vortices'', J. Stat. Phys. {\bf 17},
323 (1977).}

\bibitem{lb}  {\small D. Lynden-Bell, ``Statistical mechanics 
of violent relaxation in stellar systems'', Mon. Not. R. astr. Soc.  {\bf 136}, 101 (1967).}

\bibitem{lbe}  {\small D. Lynden-Bell and P.P. Eggleton, 
``On the consequences of the gravothermal catastrophe'', 
Mon. Not. R. astr. Soc.  {\bf 191}, 483 (1980).}

\bibitem{lbw}  {\small D. Lynden-Bell and R. Wood, 
``The gravothermal catastrophe in isothermal spheres and 
the onset of red-giants structure for stellar systems'', 
Mon. Not. R. astr. Soc.  {\bf 138}, 495 (1968).}

\bibitem{williams}  {\small J.C. McWilliams, ``The emergence of isolated coherent vortices in turbulent flow'', J. Fluid Mech.  {\bf 146}, 21 (1984).}

\bibitem{michel}  {\small J. Michel and R. Robert, 
``Statistical mechanical theory of the great red spot of Jupiter'', 
J. Phys. Stat.  {\bf 77}, 645 (1994).}

\bibitem{miller}  {\small J. Miller, 
``Statistical mechanics of the Euler equation in two dimensions'', 
Phys. Rev. Lett.  {\bf 65}, 2137 (1990).}

\bibitem{weichman}  {\small J. Miller, P.B. Weichman and M.C. Cross 
``Statistical mechanics, Euler's equation, and Jupiter's Red Spot'', 
Phys. Rev. A {\bf 45}, 2328 (1992).}

\bibitem{min}  {\small I.A. Min, I. Mezic and A. Leonard 
``L\'evy stable distributions for velocity and velocity difference in
systems of vortex elements'', Phys. Fluids {\bf 8}, 1169 (1996).}

\bibitem{newton}  {\small P.K. Newton, {The N-Vortex Problem: Analytical Techniques},
Springer-Verlag, Applied Mathematical Sciences Vol. 145, May 2001.}

\bibitem{nezlin}  {\small M.V. Nezlin and E.N. Snezhkin, Rossby vortices, 
spiral structures, solitons (Springer-Verlag 1993).} 

\bibitem{novikov}  {\small E.A. Novikov 
``Dynamics and statistics of a system of vortices'', Sov. Phys. JETP {\bf 41}, 937 (1975).}

\bibitem{onsager}  {\small L. Onsager, ``Statistical hydrodynamics'', Nuovo Cimento Suppl. {\bf 6}, 279 
(1949).}

\bibitem{pad2}  {\small T. Padmanabhan, ``Antonov instability 
and the gravothermal catastrophe-revisited'',  Astrophys. J. Supp.  {\bf 71}, 651 (1989).}

\bibitem{pad}  {\small T. Padmanabhan, ``Statistical mechanics of gravitating systems'',  Phys. Rep.   {\bf 188}, 285 (1990).}

\bibitem{pedlosky}  {\small J. Pedlosky, {Geophysical fluid dynamics} (Springer-Verlag, 1996).}

\bibitem{penston}  {\small M.V. Penston, ``Dynamics of self-gravitating 
gaseous spheres-III. Analytical results in the free-fall of isothermal
cases.'', Mon. Not. R. astr. Soc. {\bf 144}, 425 (1969).}

\bibitem{plastino}  {\small A. Plastino and A.R. Plastino, ``Stellar polytropes and Tsallis entropy'',  Phys. Lett. A   {\bf 226}, 257 (1997).}

\bibitem{pl}  {\small Y.B. Pointin and T.S. Lundgren, ``Statistical mechanics of two-dimensional vortices in a bounded container'',  Phys. Fluids.   {\bf 19}, 1459 (1976).}

\bibitem{risken}  {\small H. Risken, {The Fokker-Planck equation} (Springer, 1989).}

\bibitem{rob}  {\small R. Robert, ``A maximum entropy principle for two-dimensional Euler equations'', J. Stat. Phys. {\bf 65}, 531 (1991).}

\bibitem{rr}  {\small R. Robert and C. Rosier, ``The modelling of small scales in 2D turbulent flows: A statistical mechanical approach'', J. Stat. Phys. {\bf 86}, 481 (1997).}

\bibitem{rs1}  {\small R. Robert and J. Sommeria, ``Statistical equilibrium states for two-dimensional flows'', J. Fluid Mech. {\bf 229}, 291 (1991).}

\bibitem{rs}  {\small R. Robert and J. Sommeria, ``Relaxation towards a statistical equilibrium state in two-dimensional perfect fluid dynamics'', Phys. Rev. Lett. {\bf 69}, 2776 (1992).}

\bibitem{salzberg}  {\small A. Salzberg and S. Prager, J. Chem. Phys.  {\bf 38}, 2587 (1963).}

\bibitem{sl}  {\small G. Severne and M. Luwel, ``Dynamical theory of collisionless relaxation'', Astrophys. \& Space Sci.  {\bf 72}, 293 (1980).}

\bibitem{sirec}  {\small C. Sire and P.H. Chavanis, ``Numerical renormalization group of vortex aggregation in 2D decaying turbulence: 
the role of three-body interactions'', Phys. Rev. E {\bf 61}, 6644
(2000).}

\bibitem{scD}  {\small C. Sire and P.H. Chavanis, ``Thermodynamics and collapse of self-gravitating Brownian particles in $D$ dimensions'', Phys. Rev. E {\bf 66}, 046133 (2002).}

\bibitem{smith}  {\small A.R. Smith and T.M. O'Neil  ``Nonaxisymmetric thermal equilibria of a cylindrically bounded guiding-center plasma or discrete vortex system'', Phys. Fluids B {\bf 2}, 2961
(1990).}

\bibitem{nore}  {\small J. Sommeria, C. Nore, T. Dumont and R. Robert, ``Th\'eorie statistique de la Tache Rouge de Jupiter'', C.R. Acad. Sci. II {\bf 312}, 999 (1991).}

\bibitem{staquet}  {\small J. Sommeria, C. Staquet and R. Robert, ``Final equilibrium state of a two-dimensional shear layer'', J. Fluid Mech. {\bf 233}, 661 (1991).}

\bibitem{stahl}  {\small B. Stahl, M.K.H. Kiessling and K. Schindler, ``Phase transitions in gravitating systems and the formation of condensed objects'', Planet. Space Sci.  {\bf 43}, 271 (1994).}

\bibitem{tanga}  {\small P. Tanga, A. Babiano, B. Dubrulle and A. Provenzale, ``Forming planetesimals in vortices'', Icarus  {\bf 121}, 158 (1996).}

\bibitem{henon}  {\small S. Tremaine, M. H\'enon and D. Lynden-Bell, ``H-functions and mixing in violent relaxation'', Mon. Not. R. astr. Soc. {\bf 219} 285 (1986).}

\bibitem{Tsallis}  {\small C. Tsallis, ``Possible generalization of Boltzmann-Gibbs statistics'', J. Stat. Phys. {\bf 52} 479 (1988).}

\bibitem{tw}  {\small B. Turkington and N. Whitaker, ``Statistical equilibrium computations of coherent structures in turbulent shear layers'', SIAM J. Sci. Comput. {\bf 17}, 1414 (1996).}

\bibitem{weiss2}  {\small J.B. Weiss and J.C. McWilliams, ``Temporal scaling behavior of decaying two-dimensional turbulence'', Phys.
Fluids A {\bf 5}, 608 (1993).}

\bibitem{weiss}  {\small J.B. Weiss, A. Provenzale and J.C. McWilliams, ``Lagrangian dynamics in high-dimensional point vortex systems'', Phys.
Fluids {\bf 10}, 1929 (1998).}

\bibitem{wp}  {\small C.R. Willis and R.H. Picard, ``Time-dependent projection-operator approach to master equations for coupled systems'',  Phys. Rev. A  {\bf 9}, 1343
(1974).}

\bibitem{millerB}  {\small  V.P. Youngkins and B.N. Miller, ``Gravitational phase transitions in a one-dimensional spherical system'', Phys. Rev. E  {\bf 62}, 4582 (2000).}





\end{thebibliography}
\end{document}